\documentclass[a4paper,12pt]{article}
\def\parn              {  \par\noindent }

\def\parmedskip        {  \par\medskip  }

\def\parsmallskipn      {  \par\smallskip\noindent  }
\def\al{\alpha}
\def\be{\beta}
\def\ga{\gamma} 
\def\ep{\epsilon}
\def\lam{\lambda}

 \def\calB{{\cal B}}  
  
 \def\calH{{\cal H}} 
\def\calJ{{\cal J}}  
\def\calM{{\cal M}} \def\calN{{\cal N}} \def\calO{{\cal O}}

\def\calV{{\cal V}}  
  
\def\lamtil{\tilde{\lambda}}

\def\aldot{{\dot{\alpha}}}
\def\bedot{{\dot{\beta}}}

\def\onedot{{\dot{1}}}
\def\twodot{{\dot{2}}}

\def\cbar{\bar{c}}
\def\dbar{\bar{d}}

\def\Zbar{\bar{Z}}

\def\xibar{\bar{\xi}}

\def\del        {  \partial }
\def\half       {  {1\over 2}  }
\def\rootof#1   {  \left( #1 \right)^{1/2}  } 
\def\trace      {  \mbox{Tr}\,  }

\def\abs#1      {  \vert #1 \vert  }
\def\ie         {{\it i.e.}\,\,}
\def\evalat#1   {  \left\vert_{#1} \right. }

\def\comma          {\, ,}
\def\period         {\, .}
\def\lsim      {\lower .65ex \hbox{\ $\stackrel{<}{\sim}$\ } }
\def\gsim      {\lower .65ex \hbox{\ $\stackrel{>}{\sim}$\ } }

\def\bra#1{{\langle #1 | } }
\def\ket#1{{| #1 \rangle } }

\def\com#1#2{{ \left[#1, #2\right] } } 

\def\matel#1#2#3  {{\langle #1 | #2 | #3 \rangle } }

\def\lrvec#1    {\hbox{$\stackrel{\leftrightarrow}{#1}$}}
\def\lvec#1     {\hbox{$\stackrel{\leftarrow}{#1}$}}

\def\vecii#1#2      {  \left(\begin{array}{c}#1\\#2\end{array}\right)  }
\def\veciii#1#2#3   {  \left(\begin{array}{c}#1\\#2\\#3\end{array}
                     \right)  }
\def\veciv#1#2#3#4  {  \left(\begin{array}{c}#1\\#2\\#3\\#4
                                 \end{array}\right)  }
\def\vecfv#1#2#3#4#5 {  \left(\begin{array}{c}#1\\#2\\#3\\#4\\#5
                                 \end{array}\right)  }

\def\matrixii#1#2#3#4            {  \left(\begin{array}{cc}#1&#2\\#3&#4
                                       \end{array}\right) }
\def\matrixiii#1#2#3#4#5#6#7#8#9 {  \left(\begin{array}{ccc}#1&#2&#3\\
                                     #4&#5&#6\\#7&#8&#9\end{array}
                               \right)  }
\def\mativ#1#2#3#4               {  \left(\begin{array}{cccc}
                                       #1\\#2\\#3\\#4\end{array}\right) }
\def\matv#1#2#3#4#5              {  \left(\begin{array}{ccccc}
                                     #1\\#2\\#3\\#4\\#5\end{array}
                              \right)  }
\def\eqabegin         {  \begin{eqnarray}  }
\def\eqaend           {  \end{eqnarray}  }
\def\nn               {  \nonumber  }
\def\bracetwo#1#2     {  \left\{ \begin{array}{l} #1 \\ #2 \end{array}
                         \right.  }
\def\bracetwocases#1#2#3#4  {   \left\{ \begin{array}{ll} #1 &
                                 \qquad #2 \\
                                 #3 & \qquad #4 \end{array} \right.  }
\def\bracebegin#1     {  \left\{ \begin{array}{#1}   }
\def\braceend         {  \end{array}\right.   }
\def\boxit#1#2      {  \vbox{\hrule\hbox{ \hskip -4.1pt \vrule\kern3pt 
                     \vbox
                    {  \hsize #1 \strut\kern3pt #2 \kern3pt\strut  }
                       \kern3pt  \vrule} \hrule  } }
\def\centerbox#1#2  {  \mbox{  }\par\bigskip  \hfil \boxit{#1}{#2} \hfil
                       \par\bigskip\noindent }
\def\rightbox#1#2   {  \hfill\boxit{#1}{#2}  }
\def\leftbox#1#2    {  \boxit{#1}{#2}  }
\def\fullbox#1      {  \boxit{\textwidth}{#1}  }

\newcommand{\nullify}[1]{}

\def\mpg#1#2{\begin{minipage}[t]{#1} #2  \end{minipage} }

\def\bfall{\boldmath\bf }
\def\nxt{\parsmallskipn}

\def\epsfig#1#2#3{
{\lower #3 \hbox{
 \mpg{#1}{\begin{center} \includegraphics[width=#1,clip]{#2.eps} \\
 Fig. #2\end{center} }}}}

\def\papertitlepage{\baselineskip 3.5ex \thispagestyle{empty}}
\def\Title#1{\baselineskip 1cm \vspace{1.5cm}\begin{center}
 {\Large\bf #1} \end{center} 
\vspace{0.5cm}}
\def\Authors#1{\begin{center} {\it #1} \end{center}}
\def\Abstract{\vspace{1.0cm}\begin{center} {\large\bf Abstract} 
           \end{center} \par\bigskip}

\renewcommand{\thefootnote}{\fnsymbol{footnote}}

\def\bfall{\boldmath\bf}
\usepackage{ifpdf}
\ifpdf
\pdfoutput=1
\usepackage{graphicx} 
\usepackage[setpagesize=false,pagebackref=false, linktocpage, bookmarksopen=true, colorlinks=true, linkcolor=darkblue,citecolor=darkblue,urlcolor=darkblue]{hyperref}
\usepackage{hyperref}
\newcommand{\arXiv}[2]{\href{http://arxiv.org/abs/#1}{{\tt arXiv:#2}}}
\newcommand{\hep}[2]{\href{http://arxiv.org/abs/#1}{{\tt #2}}}

\else
\usepackage{graphicx}
\newcommand{\arXiv}[2]{[{\tt arXiv:#2}]}
\newcommand{\hep}[2]{[{\tt #2}]}

\fi
\usepackage{graphicx}
\usepackage{latexsym}
\usepackage{amsmath}
\usepackage{amssymb}
\usepackage[usenames]{color}
\usepackage{ascmac}
\usepackage{mathrsfs}
\usepackage{cite}
\usepackage{fancybox}
\usepackage{simplewick}
\definecolor{darkgreen}{rgb}{0,0.5,0}
\definecolor{darkblue}{cmyk}{0.9,0.9,0,0}
\definecolor{darkred}{rgb}{0.6,0,0.3}
\definecolor{refkey}{rgb}{0.5,0.5,0}
\definecolor{labelkey}{rgb}{0.5,0.5,0}
\definecolor{citekey}{rgb}{0.5,0.5,0}
\definecolor{MyRed}{cmyk}{0,1,1,0.15}
\definecolor{MyBlue}{cmyk}{1,1,0,0.25}

\def\Xbar{\bar{X}}

\def\ads3{Euclidean $AdS_3$}

\newcommand{\beqa}{\begin{eqnarray}}
\newcommand{\eeqa}{\end{eqnarray}}
\newcommand{\bea}{\begin{array}}
\newcommand{\eea}{\end{array}}
\newcommand{\beqn}{\begin{equation}}
\newcommand{\eeqn}{\end{equation}}

\def\costh2{\cos{\theta_0 \over 2}}
\def\sinth2{\sin{\theta_0 \over 2}}

\def\SL2C{{\rm SL(2,C)}}

\def\AdS3{$AdS_3$}
\def\EAdS3{$E\!AdS_3$}

\def\Zbar{{\bar{Z}}}
\def\Xbar{{\bar{X}}}

\def\eqref#1{(\ref{#1})} 

\def\n{\mathfrak{n}}

\def\psu224{{\rm psu}(2,2|4) }
\def\u224{{\rm u}(2,2|4) }
\def\SU2{{\rm SU(2)}}
\def\Zket{\ket{Z}}
\def\Zbarket{\ket{\Zbar}}

\def\bfone{{\bf 1}}
\def\singpsu{{}_{\rm psu}\bra{\bfone_{12}}}
\def\Zbra{\bra{Z}}

\def\barZbarbra{\overline{\bra{\bar{Z}}}}
\def\mutil{\tilde{\mu}}
\def\lamtil{\tilde{\lambda}}
\def\bfn{{\bf n}}

\def\rma{{\rm a}}
\def\rmb{{\rm b}}
\def\rmc{{\rm c}}
\def\rmd{{\rm d}}
\begin{document}
\papertitlepage
\vspace*{-1cm}
\hfill \begin{minipage}{4.2cm} RUP-15-12  \parn 
UT-Komaba 15-3 \parn June, 2015
               \end{minipage}
\vspace{-0.8cm}
\Title{On the singlet projector  and the monodromy relation for $\boldsymbol{{\rm psu}(2,2|4)}$ spin chains  
and 
reduction to subsectors  } 
\Authors{\baselineskip 3ex
{\sc Yoichi Kazama\footnote[2]{\textcolor{darkblue}{\tt yoichi.kazama@gmail.com}}, 
 Shota Komatsu\footnote[3]{\textcolor{darkblue}{\tt skomatsu@perimeterinstitute.ca}} and Takuya Nishimura\footnote[4]{\textcolor{darkblue}{\tt tnishimura@hep1.c.u-tokyo.ac.jp}} 
\\ }
\vskip 3ex
${}^\dagger$  Research Center for Mathematical Physics,  
Rikkyo University, \\  Toshima-ku, Tokyo  171-8501  
Japan 
  \parmedskip\noindent
${}^\ddagger$ Perimeter Institute for Theoretical Physics, \\
Waterloo, Ontario N2L 2Y5, Canada \parmedskip\noindent
 ${}^\S$    Institute of Physics, University of Tokyo, \\
\vspace{-0.42cm}
 Komaba, Meguro-ku, Tokyo 153-8902 Japan 
  }
\vspace{-1cm}
\renewcommand{\thefootnote}{\arabic{footnote}}
\numberwithin{equation}{section}
\numberwithin{figure}{section}
\numberwithin{table}{section}
\parskip=0.9ex
\baselineskip 3.3ex
\Abstract
As a step toward uncovering the relation between the weak and the strong coupling regimes of the $\calN=4$ super Yang-Mills theory beyond the spectral level,   we have developed in a previous paper [arXiv:1410.8533] a novel  group theoretic interpretation of the Wick contraction of the fields, which allowed us to compute a much more general class of three-point functions in the SU(2) sector, as in the case of strong coupling [arXiv:1312.3727],  directly in terms of the determinant representation of the partial domain wall partition function. Furthermore, we derived  a non-trivial identity for the three point functions with monodromy operators inserted, being  the discrete counterpart  of the global monodromy condition which played such a crucial  role in the computation at strong coupling. 
In this companion paper, we shall extend our study to the entire ${\rm psu}(2,2|4)$ sector and obtain  several important generalizations. They include  in particular  (i) the manifestly conformally covariant construction, from the basic principle,   of the  singlet-projection operator  for performing the Wick contraction  and (ii)  the derivation of the  monodromy relation for the case of the so-called ``harmonic R-matrix",  as well as for the usual fundamental R-matrtix.
The former case, which is new and has features rather different from the 
latter, is expected to have  important  applications.  We also describe how  the form of the monodromy relation is modified  as ${\rm psu}(2,2|4)$ is reduced to  its  subsectors. 

\newpage
\baselineskip 3.5ex
\renewcommand{\thefootnote}{\arabic{footnote}}
\thispagestyle{empty}
\enlargethispage{2\baselineskip}
\renewcommand{\contentsname}{\hrule {\small \flushleft{Contents}}}
{\footnotesize \tableofcontents}
\nxt\hrule
\newpage
\section{Introduction}
Although the idea and the use  of AdS/CFT \cite{AdSCFT1, AdSCFT2, AdSCFT3} have been expanded 
into  enormous  varieties of directions, still the understanding of the essence of the dynamical  mechanism of this remarkable duality remains as  one of the most important unsolved problems.  For such a fundamental task, it is best to 
study deeply the prototypical example which has been studied most vigorously, 
 namely  the duality between the $\calN=4$ super Yang-Mills theory in 4 dimensions  and the type IIB string theory in AdS${}_5 \times $ S${}^5$. 

In this context, undoubtedly  the pivotal work which pointed the way to analyze  this  strong/weak duality was the discovery of Minahan and Zarembo\cite{MZ } that the one-loop dilatation operator for the gauge invariant composite operators made out of six scalars of super Yang-Mills theory takes exactly the form of 
 the Hamiltonian of the integrable SO(6) spin chain system. This made (a part of) the integrability structure of the super Yang-Mills theory quite explicit, at least as far  as the 
spectrum of the composite operators are concerned. Two years later, 
the power and the relevance of integrability was revealed also on the strong coupling side by the works\cite{BPR, KMMZ}. These  works opened up the way to use  integrability as an extremely  powerful tool,   not relying on  the structure of supersymmetry and suitable for studying the dynamical aspects. Various results obtained along this way 
 up to around 2010 are summarized in the review \cite{Beisert:2010jr}. 

Subsequent  developments can be classified into several categories. 
 One is the  more  sophisticated way of computing the spectrum of 
 composite operators, even at  finite coupling. The reader should be referred to the most advanced approach \cite{Gromov:2013pga}, \cite{Gromov:2014caa} and the references therein.  Advancements in another category  are 
 the computation and understanding of other observables,  such as  the scattering amplitudes (for reviews, see  \cite{Dixon:2011xs,Roiban:2010kk,Drummond:2010km,Alday:2010kn}  ) 
and the Wilson loops (see for example \cite{Drukker:2012de,Correa:2012hh} and references therein).  

In the realm of the gauge-invariant  composite operators,  properties  beyond the spectrum level have  been vigorously pursued.  In particular,  study of the three-point functions, the main interest behind the present  work,  have been advanced  using the power of ``integrability".  Below let us briefly summarize the highlights of the recent developments in  this category both at weak and strong couplings. 

 At weak coupling, a systematic  procedure called ``tailoring" has been developed \cite{Tailoring1, Tailoring2, Tailoring3, quantum3pt, Tailoring4}, which 
essentially reduces the computations to those of certain scalar products 
 of spin chain states. To actually bring them into a tractable form 
a technical improvement was needed\cite{Foda},  and then a special class of  three-point functions for non-BPS operators have  been expressed explicitly in terms of Slavnov determinants \cite{Slavnov}. Furthermore,  the  semi-classical  limit of such three-point functions with large charges were successfully evaluated in a remarkably  compact form \cite{Tailoring3, Kostov1, Kostov2, Fixing}. 

On the other hand, at strong coupling, the lack of the method for quantization 
of strings  in curved space-time only permits one to deal with the semi-classical limit of large charges. Even in such a situation, the lack of the knowledge of the  appropriate vertex operators and of the saddle point configuration made the  complete computation quite difficult\cite{JW, KK1}. Finally, applying the idea of the  state-operator correspondence and with the use of the finite-gap integration method, the computation was rendered possible for the GKP string\cite{KK2} and for  the more interesting case of three-point functions of rather general class in the SU(2) sector, which includes the very special ones computed by the tailoring technique at weak coupling\cite{KK3}. 

Now in order to understand the connection between the weak and the strong 
 coupling regimes, it is important to find the common structures in the 
two regimes.  Since the super Yang-Mills theory at weak coupling is completely  well-defined, it would be better to start by re-analyzing  the three-point functions on this side of the duality. 

 In a previous communication\cite{KKN2}, we have developed two novel viewpoints and applied them explicitly to a class of three point functions in the SU(2) sector which are much more general than had been  treated by the tailoring procedure. Let us summarize  these two ideas as (I) and (II) below:
\begin{enumerate}
	\item[(I)] 
 One is the group theoretic reinterpretation
 of the Wick contraction of basic  fields  as a singlet projection of a tensor product of two fields. 
\begin{enumerate}
	\item[(Ia)] In the case of the SU(2) sector, one can apply this idea with respect to the more refined SU(2)${}_L \times$ SU(2)${}_R$ structure  present in that sector. This feature can be succinctly referred to  as ``double spin-chain"  and it leads to the factorization of the left and the right sector and simplifies various formulas. This formalism allowed us to study a class of three-point
 functions of operators built upon more general spin-chain vacua than the 
 special configuration discussed so far in the literature. This formulation 
makes the correspondence with the strong coupling computation done in 
 \cite{KK3} quite apparent. 
\item[(Ib)] Another conspicuous advantage of the new interpretation of the 
 Wick contraction is that one can avoid  the scalar products of off-shell states which appear in the  tailoring prescription. Emergence of such an object  required the trick\cite{Foda} to 
turn one of the off-shell states into an on-shell state in order to write it 
in terms of the Slavnov determinant. In contrast, in our formulation  one can 
 directly obtain the expression in terms of the partial domain wall partition functions, and for a certain class of correlators it can be readily expressed as a determinant even for two off-shell states. 
\end{enumerate}
\end{enumerate}

\begin{enumerate}
	\item [(II)]  The second  new idea  formulated  explicitly  in \cite{KKN2} is so-called the monodromy relation,  which can be obtained 
 by inserting the  monodromy matrices $\Omega$'s  inside the two-point or the three-point functions  and  using  the unitarity and the crossing relations.  This produces a relation  between correlation  functions of different operators and hence acts  like the Schwinger-Dyson equation.  In particular, in the special limit where the spectral parameter $u$ goes to  $\infty$, it reduces to the Ward identity for SU(2)${}_{\rm L}$ and SU(2)${}_{\rm R}$. Moreover, for the 
 three-point function in the semiclassical limit, where the  operators carry  large quantum numbers, it takes the classical relation $\Omega_1\Omega_2 \Omega_3=1$, which is precisely of the form of the monodromy relation that follows from  the integrability of the string theory, which played such a crucial role in the computation at strong coupling\cite{JW, KK3}.  
Thus, its  super Yang-Mills counterpart  should also  be considered  as a major part of the concept of  ``integrability"  beyond the spectral level. 
\end{enumerate}

Now the main purpose of the present paper is to extend these two main ideas explicitly to the full psu(2,2$|$4) sector and discuss how various general formulas are modified and simplified when we reduce them to the various subsectors. As for the monodromy relation (II), for the sake of clarity of presentation, we shall mainly concentrate on the case of the two-point functions. However, the extension to the three-point functions of our interest is straightforward, as was demonstrated in the case of SU(2) sector in \cite{KKN2}, and the form of the result will also be briefly presented.

As far as  the basic ideas (I) and (II) sketched above are concerned,  similar ideas on the Wick contraction and the monodromy relations have also been 
discussed independently by \cite{spinvertex}\footnote{They did not discuss, however, the explicit advantage described in (Ia) and (Ib) gained by the new interpretation of the Wick contraction.}. Their work was based largely on the work by \cite{ADGN},  which observed and utilized  certain similarity of the Wick-contracting operator to the string field theory (SFT) vertex in the spirit of the string bit formulation. In this fashion, the work of \cite{spinvertex} discussed already the full $\psu224$ sector making use of the similar vertex, as well as the same oscillator representations and some associated basic formulas, as \cite{ADGN}. 

However, as far as the result (I) for the ${\rm psu}(2,2|4)$ is concerned, the exponential form of the vertex written down by \cite{ADGN} was guessed  by an analogy with the delta-function overlap in SFT and unfortunately was not  ${\rm psu}(2,2|4)$ singlet.  The work of \cite{spinvertex},  which was based  on \cite{ADGN}, modified certain parts of the exponent and 
checked that it is a singlet projector a posteriori. However, there are two 
points that one wishes to improve on.  

One is the understanding of why the singlet projector is of a simple  exponential  form, which was assumed in the work of \cite{ADGN} and hence in \cite{spinvertex}.  Such a form may be 
 natural as the oscillator description of the $\delta$-function overlap familiar 
 in SFT context is indeed exponential. However, the analogy should be taken with care. For one thing the discrete indices 
of the oscillators for the string case are the Fourier mode numbers, whereas the similar indices in the super Yang-Mills case designate the location along the 
 spin chain. Furthermore, in the case of the string the parts to be identified are rather homogeneous and hence it is natural to employ the (oscillator representation of) a delta-function to connect them. On the other hand, in the case of 
 the ${\rm psu}(2,2|4)$ spin chain the  adjacent ``string bits" can be quite different and the analogy to the delta function overlap is not intuitively obvious. 
In fact it is a simple exercise to construct the singlet state in the case of 
 the spin $j$ representation of SU(2) and confirm  that it does {\it not} take an exponential form. 

Therefore the surest way to obtain  the desired  vertex which effects the Wick contraction is to construct the most general singlet projector systematically  in the space of tensor product of two spin chains. We shall show that 
the singlet projector exists for ${\rm su}(2,2|4)$ (as well as its restriction 
 ${\rm psu}(2,2|4)$)  but not for ${\rm u}(2,2|4)$ and, 
strictly speaking, for each sector of the representation of the ${\rm su}(2,2|4)$ with a definite central charge $C$, the singlet vertex is not of a simple exponential form. However, provided that one is interested only in a sector with one definite value of the central charge\footnote{This is the case for the $\calN=4$ super Yang-Mills, since the basic fields all carry  $C=0$.}, one is allowed to use the simple exponential form, which is much more tractable. (We shall further elaborate on this  later.) In this connection, we shall also explain in an appendix how 
the simple non-exponential singlet projector constructed for the SU(2) subsector in our  previous work  can be obtained from the general exponential 
projector for ${\rm psu}(2,2|4)$. 

The second point is that one wishes to  improve the situation that the singlet projector of \cite{spinvertex} is not manifestly conformally invariant, 
which is not  useful for the treatment of the computation 
 of the correlation functions of the local composite operators. 
In the present work we shall construct the version of the singlet projector 
 which is manifestly conformally invariant and hence much simpler  to use.  

To explain what  we mean by this, it is instructive to recall the following basic facts. 
 In constructing the representations of the superalgebra ${\rm u}(2,2|4)$,  there are basically two different schemes, depending on which maximal  bosonic subgroups of the supergroup ${\rm U}(2,2|4)$ to make use of:
\begin{align}
&(E):\quad {\rm U}(2,2|4) \supset {\rm U(1)}_E \times {\rm SU(2)}_L \times {\rm SU(2)}_R  \times {\rm SU(4)} \comma \label{Escheme}\\
&(D):\quad {\rm U}(2,2|4) \supset {\rm U(1)}_D \times {\rm SL}(2,C) \times \overline{
{\rm SL}(2,C)}  \times {\rm SU(4)} \label{Dscheme}
\end{align}
Their difference resides in the choice of the subgroups in the $SO(4,2)$ part. 
The one, which we shall call E-scheme, makes use of the compact subgroups with  the AdS energy $E$ being  diagonal. On the other hand, in the scheme to be called  D-scheme, the dilatation generator $D$ is diagonal and 
the rest of the subgroup chosen in $SO(4,2)$ is  the non-compact  Lorentz group ${\rm SL}(2,C) \times \overline{
{\rm SL}(2,C)}$. Therefore, the D-scheme is manifestly  conformal covariant.  It is well-known and fully discussed in \cite{DolanOsborn} that these two 
 schemes are connected by a non-unitary similarity transformation generated by the operator $U=e^{(\pi/4)(P_0-K_0)}$ such that $U^{-1} DU =iE$. 

In the treatment of \cite{ADGN},  and hence \cite{spinvertex}, the oscillators
 appropriate for the E-scheme are used as basic building blocks for the 
 generators of $\u224$ and the relevant vertex operators. Since  the D-scheme is more natural  for the main purpose  of computing  the correlation functions of the basic super Yang-Mills fields,   they transformed various quantities  to that scheme by  the similarity transformation using the operator $U$. However,  since $U$ does not map an individual 
component group,  such as SU(2)${}_{\rm L}$,  in the E-scheme to a definite component group, such as SL(2,C),  in the D-scheme\footnote{Obviously, any similarity transformation, unitary or non-unitary,  does not change  the structure of the group.},  the mapping does not make the description manifestly conformally covariant. 

In our construction, to be described fully in section 2, we will stick to the D-scheme throughout, by using the oscillators which transform covariantly under the  maximal subgroups shown in (\ref{Dscheme}). This will make the entire description quite transparent without the need of the operator $U$. 

 Let us next  turn to the ${\rm psu}(2,2|4)$ version of the monodromy relation (II).  There are two natural 
 types of monodromy matrices depending on the choice of the  auxiliary space. One is the simpler and the  fundamental one,  for which 
 the auxiliary space is taken to be ${\mathbb C}^{4|4}$. For this case, the derivation of 
 the monodromy relation is a straightforward  extension of the one for the 
 SU(2) sector given in our previous work\cite{KKN2} and agrees  with 
 the description given in \cite{spinvertex}.  Another type is the monodromy relation  associated with the so-called harmonic R-matrix, for which the structure of the auxiliary space is the same as that of the physical quantum space\cite{Kulish:1981gi, Reshetikhin:1983vw,Reshetikhin:1985vd,Ogievetsky:1986hu }. This case may be useful for obtaining local 
 conserved quantities as well as for the study of scattering amplitudes\cite{HarmonicR1,HarmonicR2,spectralreg3,spectralreg4,CK,CDK,BdR1,BdR2}. In the present work, we shall derive the monodromy 
 relation for this more complicated case as well, which was not discussed  in \cite{spinvertex}.  As was demonstrated in \cite{KKN2}, the monodromy 
 relation for the three-point functions  can be straightforwardly derived once 
 that for the two-point functions  is established, in this article we shall concentrate on the case of two-point functions.

 Now the  monodromy relations for the entire $\psu224$ sector is practically too complicated to analyze at present.  In this sense, it is of interest to look first at such relations for simpler subsectors. This has  already been done for the SU(2) sector in our previous work. In the present work, we shall first sketch 
 how this result can be rederived by the reduction of the relations for the $\psu224$ sector and then apply  similar techniques to the non-compact   SL(2)  subsector with much more detailed expositions. Under such reductions we shall see 
that certain non-trivial shifts in the spectral parameters are produced. It should be stated that all the discussions in this paper are at the tree level. It will be an important future task to extend some of the basic concepts to the loop level.

Having explained the essence of  the new findings of the present work, let us 
briefly summarize the organization of the rest of this article. 

In section 2, we start with a review (section 2.1),  where we present the representation of the generators of $\u224$ in terms of the oscillators, which transform  covariantly under 
 the maximal subgroups ${\rm U(1)}_{\rm D} \times {\rm SL(2,C)} 
 \times \overline {{\rm SL(2,C)}}$ of ${\rm SO(4,2)}$ (the ${\rm D}$-scheme choice discussed  above) and 
 ${\rm U(1)}_{\rm J} \times {\rm SU(2)}_{\rm L} \times  {\rm SU(2)}_{\rm R} $ of ${\rm SO(6)}$ respectively.  
With this set-up, in section 2.2 we solve the conditions 
 for the most general singlet state in the tensor product of two Hilbert spaces. 
This gives a state the form of which is  not quite an exponential in the tensor product of oscillators. We shall then explain that nevertheless for the application to the super Yang-Mills fields with $C=0$, one can promote it to a simple 
 exponential form. As a check, we compute the relevant 2-point functions of basic super Yang-Mills fields using this singlet state. (The demonstration that it
 reduces to a simple non-exponential form  for  the SU(2) subsector obtained 
in our previous work will be given in  Appendix B.)

In section 3, we derive  explicitly the formulas for the monodromy relations for the correlation functions in the $\psu224$ spin chain systems, 
 first  in the case of the fundamental R-matrix and then in the case of 
 the harmonic R-matrix, which is more involved (some of the details are relegated to  Appendix C.). 

In section 4, we explain how the monodromy relations for the $\psu224$ can be reduced to the ones for the subsectors. In particular, we study the case of 
 the compact SU(2) subsector and the non-compact SL(2)  subsector  and see that the reduction produces certain shifts in the spectral parameters. 

In the final discussion section (section 5), we shall summarize the essential ideas and methods employed  to obtain the new results in this work  and 
discuss how they should be utilized to try to capture  the principles through which  to relate the super Yang-Mills theory and the string theory in AdS spaces.

As already indicated, three  appendices (including Appendix A where we list 
 all the generators of ${\rm u}(2,2|4)$ in the D-scheme notation for convenience) are provided to supplement the discussions given in the main text. 
 
\section{Conformally covariant oscillator description of  {\bfall $ {\rm psu}(2,2|4)$} and the singlet projector for the contraction of basic  fields }
We begin by constructing  the singlet projector for the full  $\psu224$ sector 
from the first principle, 
with which one can efficiently perform the Wick contraction between 
 the basic fields of the $\mathcal{N}=4$ super Yang-Mills fields. As already emphasized
 in our previous work \cite{KKN2}, the use of this object is quite natural and 
 versatile  in computing  fairly  general class of correlation functions of gauge-invariant composite operators made out of SYM fields,  at least at the tree level and  possibly at the higher loop levels. 
\subsection{Oscillator representation of the generators of u(2,2$|$4) in the D-scheme}
In the case of the  SU(2) subsector discussed in our previous paper, 
the construction of the singlet projector was nothing but the elementary problem of forming  a singlet state out of two spin $1/2$ particles, once we regard  the SU(2) spin chain as a double-chain associated with the two 
distinct SU(2) groups\footnote{It should be clear that these two groups, 
belonging to SU(4),  are 
quite different from the $\SU2_{\rm L}\times \SU2_{\rm R}$ groups which  will appear as a part of the maximal subgroups of SO(2,4) in the E-scheme described below in (\ref{eq:E-scheme}).},    $\SU2_{\rm L}$ and $\SU2_{\rm R}$, acting on the chain. In the case of the full $\psu224$ spin chain, however, 
the structure of the algebra and its representation are sufficiently involved 
to render the general construction non-trivial.  Luckily, as our aim is to 
be able to perform the Wick contraction of only the basic SYM fields,  we may restrict ourselves to the singleton representation,  which can be 
 realized by a  minimal set of oscillators\cite{Gunaydin:1984fk, Gunaydin:1998sw }. However, before introducing the oscillators, we must 
recall that there are basically two different bases for the  representations 
of the $\u224$ algebra, depending on which maximal subgroups of the 
 conformal group SO(2,4) are used and the properties of the oscillators depend on such bases. Let us describe and compare them in 
 some detail below following \cite{ADGN}. 
\subsubsection{E-scheme and the D-scheme}
As already mentioned in the introduction, we shall call these two schemes  E-scheme  and D-scheme, where ``E" and ``D" stand  for the energy and the dilatation respectively, for which 
the subgroups taken are shown below:
\begin{align}
&({\rm E}) : \quad \mathrm{SO}(2,4) \supset \mathrm{SO}(2)_{E} \times \mathrm{SU}(2)_L \times \mathrm{SU}(2)_R \comma \label{eq:E-scheme} \\
&({\rm D}) : \quad \mathrm{SO}(2,4) \supset \mathrm{SO}(1,1)_{D} \times \mathrm{SL}(2,\mathbb{C}) \times \overline{\mathrm{SL}(2,\mathbb{C})} \period \label{eq:D-scheme}
\end{align}

For the E-scheme,  the maximal  subgroups are all compact, including the 
${\rm SO(2)}_E$ factor, the eigenvalue of which is identified with  the AdS energy\footnote{Strictly speaking, one considers its universal cover.}. Thus, in the context of AdS/CFT, this scheme is 
 useful in describing the states  and their spectra on the gravity/string  side. 
On the other hand, for the D-scheme, the maximal subgroups are all 
 non-compact, consisting of the dilatation and the Lorentz groups. As the 
 interpretation of SO(2,4) as  the conformal group in four dimensions is manifest in this scheme, D-scheme is more natural in discussing the correlation 
 functions in the SYM theory. 
Accordingly, the set of oscillators used in  these two schemes are 
 different, each set transforming covariantly under the respective maximal subgroups. 

Before introducing them and discussing  their difference, it is useful to first recall 
 how the SO(2,4) algebra and its representations are  described according to these two schemes.  From the point of view of the conformal algebra 
 in four dimensions, the commutation relations of  the generators of SO(2,4) 
 are given by 
\begin{align}
[ M_{\mu \nu}, M_{\rho \sigma}]&= -i(\eta_{\mu \rho}M_{\nu \sigma}-\eta_{\nu \rho}M_{\mu \sigma}+\eta_{\mu \sigma}M_{\rho \nu}-\eta_{\nu \sigma}M_{\rho \mu}) \comma \\
[ M_{\mu \nu} , P_{\rho} ]&= -i(\eta_{\mu \rho}P_{\nu}-\eta_{\nu \rho}P_{\mu}) \comma \\
[ M_{\mu \nu} , K_{\rho} ]&= -i(\eta_{\mu \rho}K_{\nu}-\eta_{\nu \rho}K_{\mu}) \comma \\
[ D, M_{\mu \nu}]&=[P_{\mu},P_{\nu}]=[K_{\mu},K_{\nu}]=0 \comma \\
[-i D, P_{\mu}]&=P_{\mu} \comma \ [-iD,K_{\mu}]=-K_{\mu} \comma
\label{DPKcom} \\
[P_{\mu}, K_{\nu}]&=2i(\eta_{\mu \nu}D+M_{\mu \nu}) \comma
\end{align}
where $\mu \comma \nu =0,1,2,3$ and  the metric signature is taken 
 to be $\eta_{\mu \nu}= \mathrm{diag}(-1,1,1,1)$. 
$M_{\mu\nu}$, $P_\mu$, $K_\mu$ and $D$ are, respectively, the Lorentz,   the momentum, the special conformal and the dilatation generators. 
This set of commutation relations can be expressed more compactly as
\begin{align}
&[ J_{KL }, J_{MN}]= -i(\eta_{K M}J_{LN}-\eta_{LM}J_{KN}+\eta_{KN}J_{ML}-\eta_{LN}J_{MK}) \comma \\
&J_{\mu \nu}:=M_{\mu \nu} \comma \ J_{\mu-1}:=\frac{1}{2}(P_{\mu}+K_{\mu}) \comma \  J_{\mu4}:= \frac{1}{2}(P_{\mu}-K_{\mu}) \comma \ J_{-14}:=D \comma 
\end{align}
for which the structure of SO(2,4) is manifest. In this representation, 
the range of six-dimensional indices and the metric are taken to be 
$M,N=-1,0,1\cdots,3,4$ and $\eta_{MN}:=\mathrm{diag}(-1,-1,1,1,1,1)$.

Now consider this algebra from the point of view of the E-scheme. It is 
 easy to find that the generators of the compact maximal subgroups 
${\rm U(1)}_E$, $\SU2_L$ and $\SU2_R$ are given respectively by 
\begin{align}
E&:=J_{0-1}=\frac{1}{2}(P_0+K_0) \comma \label{eq:E-subalgebra1} \\
L_m &:= \frac{1}{2}\left( \frac{1}{2}\epsilon_{mnl}M_{nl}+M_{m4}\right) \comma \label{eq:E-subalgebra2} \\
 R_m &:= \frac{1}{2}\left( \frac{1}{2}\epsilon_{mnl}M_{nl}-M_{m4}\right) \period \label{eq:E-subalgebra3}
\end{align}
where $m,n,l =1,2,3$. Obviously, $L_m$ and $R_m$ commute with $E$ and  hence carry zero energy. 
The rest of the generators of SO(2,4) carry either positive or negative energy 
and thus  the generators of the entire algebra are decomposed   in the following fashion:
\begin{align} 
\begin{split}
&{\rm so}(2,4) = \mathcal{E}^+ \oplus \mathcal{E}^0 \oplus \mathcal{E}^- \comma \\
[ E,\mathcal{E}^{\pm} ]=\pm \mathcal{E}^{\pm} \comma & \quad  [ E, \mathcal{E}^0]=0 \comma \quad  [ \mathcal{E}^0 ,\mathcal{E}^{\pm} ]\subset \pm \mathcal{E}^{\pm} \comma \quad  [ \mathcal{E}^+ ,\mathcal{E}^- ]\subset \mathcal{E}^0   \period
\end{split}
\end{align}
Let $\ket{e, j_L, j_R}$ be the simultaneous eigenstate of the energy $E$ 
 and the third components $L_3$ and $R_3$ of $\SU2_L$ and $\SU2_R$ 
respectively with the eigenvalues denoted by $e, j_L$ and $j_R$. Namely 
\begin{align}
E \ket{e, j_L, j_R} &= e\ket{e, j_L, j_R}\comma \quad 
L_3 \ket{e, j_L, j_R} = j_L \ket{e, j_L, j_R} \comma \quad 
R_3 \ket{e, j_L, j_R} = j_R \ket{e, j_L, j_R} \period
\end{align}
The physically relevant unitary positive energy representations are built upon the lowest weight state among the set \{ $|e,j_L,j_R \rangle$\}, which is annihilated by all the energy-lowering generators belonging to $\mathcal{E}^-$. 
We denote them by $L^{ij}$, where the $i$ and $j$ are actually the 
 spinor indices of $\SU2_L$ and $\SU2_R$ respectively and run from 1 to 2. 
Therefore we have four annihilation operators in total and the lowest weight 
 state is characterized by 
\begin{align}
L^{ij} \ket{e,j_L, j_R} =0 \period 
\end{align}
By acting onto this vacuum  the four  raising operators belonging to $\mathcal{E}^+$,  which we denote by  $L_{ij}$, one obtains unitary representations in the 
 E-scheme. 

Next consider the algebra ${\rm so}(2,4)$  from the D-scheme point of view. In this scheme, the generators of the maximal subgroups  are 
given by 
\begin{align}
D&=J_{-14} \comma \label{eq:D-subalgebra1} \\
\mathcal{M}_m&:= \frac{1}{2}\left( \frac{1}{2}\epsilon_{mnl}M_{nl}+iM_{0m}\right) \comma \label{eq:D-subalgebra2} \\
\mathcal{N}_m&:= \frac{1}{2}\left( \frac{1}{2}\epsilon_{mnl}M_{nl}-iM_{0m}\right) \period \label{eq:D-subalgebra3}
\end{align}
Here,  $\mathcal{M}_{m}$ and $\mathcal{N}_m$ denote  the generators  of the Lorentz group SL$(2,\mathbb{C}) \times\overline{\mathrm{SL}(2,\mathbb{C})}$.  In this scheme, as is apparent from the commutation
 relations (\ref{DPKcom}),  $P_{\mu}$ and $K_{\mu}$  are, respectively 
  the raising or lowering operators. 
  Hence, the decomposition of the conformal algebra $\mathrm{so}(2,4)$ 
is of the structure 
\begin{align}
\begin{split}
{\rm so}(2,4)&= \mathcal{D}^+ \oplus \mathcal{D}^0 \oplus \mathcal{D}^- \comma \\
 [ -iD,\mathcal{D}^{\pm} ]=\pm \mathcal{D}^{\pm} \comma   \  [ -iD, \mathcal{D}^0]=0 &  \comma \ [ \mathcal{D}^0 ,\mathcal{D}^{\pm} ]\subset \pm \mathcal{D}^{\pm} \comma \ [ \mathcal{D}^+ ,\mathcal{D}^- ]\subset \mathcal{D}^0  \comma 
\end{split}
\end{align} 
where $P_{\mu}\in \mathcal{D}^+$, $K_{\mu} \in \mathcal{D}^-$ and $D, \mathcal{M}_m \comma \mathcal{N}_m \in \mathcal{D}^0$. 

From the point of view of CFT in four dimensions, which is directly expressed in the D-scheme, the multiplets of operators are built upon the conformal primaries  placed at the origin $x^\mu =0$. They carry definite dilatation charges, 
belong to the definite Lorentz representations, 
 and are annihilated by the lowering operators $K_\mu$. Using the state-operator correspondence, such a  primary state,  denoted by  $|\Delta, j_{\mathcal{M}},\bar{j}_{\mathcal{N}}\rangle$  with  $\Delta$ and $(j_{\calM}, \bar{j}_{\calN})$  being the dilatation charge and the Lorentz spins, is 
 characterized by 
\begin{align}
\begin{split}
&-iD|\Delta, j_{\mathcal{M}},j_{\mathcal{N}}\rangle=\Delta |\Delta, j_{\mathcal{M}},j_{\mathcal{N}}\rangle \comma \ \ K_{\mu}|\Delta, j_{\mathcal{M}},j_{\mathcal{N}}\rangle=0 \comma \\
&\mathcal{M}_3|\Delta, j_{\mathcal{M}},j_{\mathcal{N}}\rangle=j_{\mathcal{M}}|\Delta, j_{\mathcal{M}},j_{\mathcal{N}}\rangle \comma \ \ \mathcal{N}_3|\Delta, j_{\mathcal{M}},j_{\mathcal{N}}\rangle=j_{\mathcal{N}}|\Delta, j_{\mathcal{M}},j_{\mathcal{N}}\rangle \period \label{eq:D-rep}
\end{split}
\end{align} 
Then the module is built up by the descendants generated by the multiplicative actions of the raising operators $P_\mu$. It should be emphasized that such 
 a representation relevant for discussing  the correlation functions is non-unitary,  since the anti-hermitian operator $-iD$ has real eigenvalues. 

Now let us give a brief description of the relation between the E-scheme 
 used in \cite{spinvertex, ADGN} and the D-scheme to be employed exclusively in  this work. It is well-known by the work of \cite{DolanOsborn} that 
there exists a non-unitary similarity transformation between the generators of 
 these two schemes. The correspondence between $E=J_{-1,0}$ and $-iD
=-i J_{-1,4}$ indicates that such a  transformation should rotate the non-compact  $0$-th direction into  the compact $4$-th direction and indeed it is effected by the operator 
\begin{align}
U=\exp \left( \frac{\pi}{2} M_{04}\right)=\exp \left( \frac{\pi}{4}(P_0-K_0) \right) \period
\end{align}
Explicit transformations are given by 
\begin{align}
&U^{-1}(-iD)U= E \comma\quad  U^{-1}L_mU=\mathcal{M}_m \comma
\quad  U^{-1}R_mU=\mathcal{N}_m \comma \label{eq:E-D1} \\
&U^{-1}P_{\mu}U\in \mathcal{E}^+ \comma\quad  U^{-1}K_{\mu}U\in \mathcal{E}^- \comma \label{eq:E-D2} \\
&|\Delta, j_{\mathcal{M}},j_{\mathcal{N}}\rangle = U|e,j_L,j_R \rangle \comma \quad  \mathrm{with} \ \Delta=e \comma \  j_{\mathcal{M}}=j_L \comma \  j_{\mathcal{N}}=j_R \period \label{eq:E-D3}
\end{align}
As already mentioned, for the purpose of discussing the CFT correlation 
 functions, the D-scheme is much more natural and if one starts from 
 the E-scheme description as in \cite{spinvertex, ADGN}, one must 
 necessarily manipulate with the operator $U$ in the intermediate step. 
Also, in the oscillator representations of the generators, to be elaborated 
below, the D-scheme oscillators always keep the conformal covariance manifest as opposed to those in the E-scheme. We shall make the comparison more explicit later. 
\subsubsection{Oscillator representation  in the D-scheme} 
Having argued the advantage of the D-scheme for our purpose, let us 
 now introduce appropriate oscillators for this scheme and express the generators as their  quadratic combinations. 

For this purpose, it is useful to rewrite first the  generators of 
the SO(2,4) algebra using  the dotted and the undotted spinor indices of the Lorentz group. 
We will adopt the following conventions for the conversions of vectors and the tensors:
\begin{align}
P_{\alpha \dot{\beta}}:=-\frac{1}{2}(\sigma^{\mu})_{\alpha \dot{\beta}}P_{\mu} \comma \ K^{\dot{\alpha} \beta}:=+\frac{1}{2}(\bar{\sigma}^{\mu})^{\dot{\alpha} \beta}K_{\mu} \comma \label{eq:spinor1} \\
M_{\alpha}^{\ \beta}:= \frac{i}{2} (\sigma^{\mu \nu})_{\alpha}^{\ \beta}M_{\mu \nu} \comma \ \bar{M}^{\dot{\alpha}}_{\ \dot{\beta}}:= \frac{i}{2} (\bar{\sigma}^{\mu \nu})^{\dot{\alpha}}_{\ \dot{\beta}} M_{\mu \nu} \comma  \label{eq:spinor2} 
\end{align} 
where the Lorentz  sigma matrices are defined in terms of  the Paul matrices in the following way
\begin{align}
&(\sigma^{\mu})_{\alpha \dot{\beta}}=(-1, \sigma^i)_{\alpha \dot{\beta}} \comma \
(\bar{\sigma}^{\mu})^{\dot{\alpha} \beta}=\epsilon^{\dot{\alpha}\dot{\gamma}}\epsilon^{\beta \delta}(\sigma^{\mu})_{\delta \dot{\gamma}} =(-1, -\sigma^i)^{\dot{\alpha} \beta} \comma \\
&(\sigma^{\mu \nu})_{\alpha}^{\ \beta}=(\sigma^{[\mu} \bar{\sigma}^{\nu ]})_{\alpha}^{\ \beta} \comma \ \
(\bar{\sigma}^{\mu \nu})^{\dot{\alpha}}_{\ \dot{\beta}}=(\bar{\sigma}^{[\mu} \sigma^{\nu ]})^{\dot{\alpha}}_{\ \dot{\beta}} \period
\end{align}
Our convention for the epsilon tensor, with which the indices are raised 
 and lowered, will be  $\epsilon_{12}=\epsilon_{\dot{1}\dot{2}}=-\epsilon^{12}=-\epsilon^{\dot{1}\dot{2}}=1$. 

In this notation, the SO(2,4) commutation relations take the form 
\begin{align}
&[ M_{\alpha}^{\ \beta}, J_{\gamma}]= \delta_{\gamma}^{\beta} J_{\alpha} -\frac{1}{2}\delta_{\alpha}^{\ \beta} J_{\gamma} \comma
 \ \  
[ M_{\alpha}^{\ \beta}, J^{\gamma}]= -\delta^{\gamma}_{\alpha} J^{\beta} +\frac{1}{2}\delta_{\alpha}^{\ \beta} J^{\gamma} \comma \\
&[\bar{M}^{\dot{\alpha}}_{\ \dot{\beta}}, J_{\dot{\gamma}}]=-\delta^{\dot{\alpha}}_{\dot{\gamma}} J_{\dot{\beta}}+\frac{1}{2}\delta^{\dot{\alpha}}_{\ \dot{\beta}} J^{\dot{\gamma}} \comma \ \ 
[\bar{M}^{\dot{\alpha}}_{\ \dot{\beta}}, J_{\dot{\gamma}}]=\delta^{\dot{\gamma}}_{\dot{\beta}} J^{\dot{\alpha}}-\frac{1}{2}\delta^{\dot{\alpha}}_{\ \dot{\beta}} J_{\dot{\gamma}}  \comma \\
&[ D, P_{\alpha \dot{\beta}}]=iP_{\alpha \dot{\beta}} \comma \ [D, K^{\dot{\alpha} \beta}]=-iK^{\dot{\alpha} \beta} \comma \ [D,M_{\alpha}^{\ \beta}]=[D,\bar{M}^{\dot{\alpha}}_{\ \dot{\beta}}]=0 \\
&[ P_{\alpha \dot{\beta}}, K^{\dot{\gamma} \delta}]=\delta_{\alpha}^{\delta}\bar{M}^{\dot{\gamma}}_{\ \dot{\beta}}-\delta^{\dot{\gamma}}_{\dot{\beta}} M_{\alpha}^{\ \delta}+i\delta_{\alpha}^{\delta}\delta^{\dot{\gamma}}_{\dot{\beta}} D \comma
\end{align}
where $J^\ga$ and $J_{\dot{\gamma}}$ generically stand for quantities 
 with undotted and dotted spinor indices. 

With this preparation, it is now quite natural to introduce two sets of 
bosonic oscillators, undotted and dotted, which transform under ${\rm SL}(2,\mathbb{C})$ and $\overline{{\rm SL}(2,\mathbb{C})}$  respectively 
\begin{align}
[ \mu^{\alpha}, \lambda_{\beta} ]= \delta^{\alpha}_{\beta}, \ (\alpha \comma \beta =1,2) \comma \ \ [ \tilde{\mu}^{\dot{\alpha}}, \tilde{\lambda}_{\dot{\beta}}]=\delta^{\dot{\alpha}}_{\dot{\beta}} \ (\dot{\alpha} \comma \dot{\beta} =\dot{1},\dot{2}) \period
\end{align}
\nullify{
 (All the explicit (anti-)commutation relations 
 of the entire ${\rm u}(2,2|4)$ in the spinor basis  are given for convenience 
in Appendix A.)
}
  In terms of these  oscillators, the conformal generators 
can be expressed rather simply as \footnote{Recall that $-iD$ has positive real eigenvalues in our convention.}
\begin{align}
&M_{\alpha}^{\ \beta}=\lambda_{\alpha }\mu^{\beta} - \frac{1}{2} \delta^{\beta}_{\alpha} \lambda_{\gamma} \mu^{\gamma} \comma \ \bar{M}^{\dot{\alpha}}_{\ \dot{\beta}}=-\tilde{\lambda}_{\dot{\beta} }\tilde{\mu}^{\dot{\alpha}} +\frac{1}{2} \delta^{\dot{\alpha}}_{\dot{\beta}} \tilde{\lambda}_{\dot{\gamma}} \tilde{\mu}^{\dot{\gamma}} \comma \\
&P_{\alpha \dot{\beta}}=\lambda_{\alpha} \tilde{\lambda}_{\dot{\beta}} \comma \ K^{\dot{\alpha} \beta}= \tilde{\mu}^{\dot{\alpha}} \mu^{\beta} \comma \ D= \frac{i}{2}(\lambda_{\alpha}\mu^{\alpha}+\tilde{\lambda}_{\dot{\alpha}}\tilde{\mu}^{\dot{\alpha}}+2) \period
\end{align}
Essentially the same oscillator representation was given in 
\cite{1-loop, Govil:2014uwa}.
 We shall follow \cite{Govil:2014uwa}  with slight changes of signs and conventions. 

Now to construct the SU(4) R-symmetry generators, we introduce four sets of  fermionic  oscillators satisfying the anti-commutation relations 
\begin{align}
\{ \xi^a ,\bar{\xi}_b \}=\delta^a_b \comma \ (a,b=1,2,3,4) \period 
\end{align}
Then the SU(4) generators can be constructed as 
\begin{align}
& R_a^{\ b} = \bar{\xi}_a \xi^b -\frac{1}{4}\delta^b_a \bar{\xi}_c\xi^c \comma 
\end{align}
which indeed satisfy the correct commutation relations 
$[ R_a^{\ b}, R_c^{\ d}]= \delta_b^{\ c}R_a^{\ d} - \delta_a^{\ d} R_c^{\ b}$.  It is easy to check that under $R_a{}^b$ the 
oscillators  $\xibar_b$ and $\xi^a$  transform under  the fundamental and anti-fundamental representations. As the SU(4) indices of any generator will be
carried by these fundamental oscillators, this guarantees that a generator $J_c$ ($J^c$)  having  a lower (upper)  index transforms as a fundamental (anti-fundamental), \ie 
\begin{align}
[ R_{a}^{\ b} , J_c] = \delta_c^b R_a^{\ b}-\frac{1}{4}\delta_a^{\ b}J_c \comma \ \ [R_a^{\ b}, J^c]= -\delta_a^c J^b+\frac{1}{4}\delta_a^{\ b}J^c \period 
\end{align}
The remaining generators, namely the fermionic supersymmetry and superconformal generators, 
 are expressed in a very simple way where the transformation properties 
 are directly  expressed by those of the constituent oscillators: 
\begin{align}
&Q_{\alpha}^a:= \lambda_{\alpha} \xi^a \comma \ \bar{Q}_{\dot{\alpha}a}=\tilde{\lambda}_{\dot{\alpha}} \bar{\xi}_a \comma \\
&S^{\alpha}_a = \mu^{\alpha}\bar{\xi}_a \comma \ \bar{S}^{\dot{\alpha}a}=\tilde{\mu}^{\dot{\alpha}}\xi^a \period
\end{align}
\subsubsection{Central charge and hyper charge} 
The 30 bosonic and 32 fermionic generators constructed in terms of the 
oscillators above constitute  the generators of the $\psu224$. Actually, 
 they do not close under (anti-) commutation. The closure requires the 
operator called the central charge given by 
\begin{align}
&C=\frac{1}{2}(\lambda_{\alpha}\mu^{\alpha}-\tilde{\lambda}_{\dot{\alpha}}\tilde{\mu}^{\dot{\alpha}}+\bar{\xi}_a\xi^a)-1 \period
\label{defC}
\end{align}
As the name indicates, $C$ commutes with all the generators of $\psu224$ and  hence it takes a constant value for an irreducible representation. In particular, as we shall describe  shortly, for the basic fields of the $\calN=4$ SYM (\ie 
for the field strength multiplet) of our interest,  $C$ vanishes.  Thus in this sector, we can neglect this operator. Another additional operator of interest is 
 the so-called the hypercharge operator\footnote{The definition of the hypercharge is ambiguous in the sense that we can add the central charge to it. For example, in the literature \cite{1-loop}, the hypercharge is defined by $Z:=\frac{1}{2}(\lambda_{\alpha}\mu^{\alpha}-\tilde{\lambda}_{\dot{\alpha}}\tilde{\mu}^{\dot{\alpha}})$ and it plays a
  role of the chirality operator.  The relation to our definition is  $Z=C-B+1$.} given by 
\begin{align}
B=\frac{1}{2}\bar{\xi}_a\xi^a \period
\end{align}
This is essentially the fermion number operator. One notices  that $B$ 
does not appear in all the (anti-) commutation relations of psu(2,2$|$4) and thus it can be regarded as an outer automorphism of psu(2,2$|$4). 
  By adding $B$ and $C$ to psu(2,2$|$4), we obtain the closed algebra 
called u(2,2$|$4). 

The generators of u(2,2$|$4) can be expressed succinctly in terms of 
 the oscillators as 
\begin{align}
J^A_{\ B}=\bar{\zeta}^A \zeta_B \comma \ \ 
\bar{\zeta}^A= \left( \begin{array}{c}
\lambda_{\alpha} \\
i\tilde{\mu}^{\dot{\alpha}} \\
\bar{\xi}_a 
\end{array}
 \right)^A \comma \ \
\zeta_A= \left( \begin{array}{c}
\mu^{\alpha} \\
i\tilde{\lambda}_{\dot{\alpha}} \\
\xi^a  \end{array}
 \right)_A  \period \label{eq:oscilJ}
\end{align}
One can check that $\zeta$'s satisfy 
the graded commutator of the form 
\begin{align}
[ \zeta_A, \bar{\zeta}^B]=\zeta_A\bar{\zeta}^B-(-1)^{|A||B|}\bar{\zeta_B}\zeta^A=\delta_{A}^{\ B} \comma 
\end{align}
where $|A|$ is $1$ for fermions and $0$ for bosons. Hereafter, for simplicity, 
 all the commutators should be  interpreted as graded commutator as above. 
Then, the graded commutators  between the generators of u(2,2$|$4)   are neatly summarized in  the following form:
\begin{align}
[ J^A_{\ B}, J^C_{\ D} ]= \delta^C_{\ B} J^A_{\ D} - (-1)^{(|A|+|B|)(|C|+|D|)}\delta^{A}_{\ D}J^C_{\ D} \period
\end{align}
It is useful to write down the elements  $J^A_{\ B}$ of u(2,2$|$4) in a  matrix form 
in the following way:
\begin{align}
J^A_{\ B}&:=\left(
 \begin{array}{cc|c}
 J_{\alpha}^{\ \beta} & J_{\alpha \dot{\beta}} & J_{\alpha}^{\ b} \\
 J^{\dot{\alpha} \beta} & J^{\dot{\alpha}}_{\ \dot{\beta}} & J^{\dot{\alpha} b} \\
 \hline
 J_a^{\ \beta} & J_{a \dot{\beta}} & J_a^{\ b}
\end{array} \right)_{AB}
=
\left(
\begin{array}{cc|c}
 Y_{\alpha}^{\ \beta} & iP_{\alpha \dot{\beta}} & Q_{\alpha}^{b} \\
 iK^{\dot{\alpha} \beta} & Y^{\dot{\alpha}}_{\ \dot{\beta}} & i\bar{S}^{\dot{\alpha} b} \\
 \hline
 S_a^{\beta} & i\bar{Q}_{\dot{\beta}a} & W_a^{\ b}
\end{array} \right)_{AB} \comma \label{eq:u(2,2|4)} \\
&Y_{\alpha}^{\ \beta}=\lambda_{\alpha}\mu^{\beta}= M_{\alpha}^{\ \beta}+\frac{1}{2}\delta_{\alpha}^{\ \beta}(-iD+C-B) \comma \\  &Y^{\dot{\alpha}}_{\ \dot{\beta}}=-\tilde{\mu}^{\dot{\alpha}}\tilde{\lambda}_{\dot{\beta}}=\bar{M}^{\dot{\alpha}}_{\ \dot{\beta}}+\frac{1}{2}\delta^{\dot{\alpha}}_{\ \dot{\beta}}(iD+C-B) \comma \\
& W_a^{\ b}=\bar{\xi}_a\xi^b=R_a^{\ b} +\frac{1}{2}\delta_a^{\ b}B
\period \label{eq:defW}
\end{align}
From this one can see that  the central charge and the hypercharge are related to the trace and supertrace  in the following way
\begin{align}
\mathrm{tr}J:=\sum_A J^A_{\ A}=2C \comma \qquad  \mathrm{str}J:=\sum_A (-1)^AJ^A_{\ A}=2C-4B \period \label{eq:trJ}
\end{align} 
As we shall see in section 2.2, the singlet projector we shall construct 
will be valid for the su(2,2$|$4) algebra as well as for psu(2,2$|$4), where the generators $\hat{J}^A{}_B$ of the  former is obtained from u(2,2$|$4) by imposing the supertraceless 
 condition as 
\begin{align}
\hat{J}^A_{\ B}:=J^A_{\ B}- \frac{\mathrm{str}J}{8}(-1)^{|A|}\delta^A_{\ B} \period
\end{align}
In particular this gives  $\sum_A (-1)^{|A|}\hat{J}^A_{\ A}=0$, which 
 tells us that the hypercharge $B$ is completely removed 
from su(2,2$|$4). 
\subsubsection{Oscillator vacuum  and the  representations  of the 
fundamental SYM fields}
We now move on to the oscillator representation for the fundamental fields which appear in $\mathcal{N}=4$ SYM.  For this purpose, we define the  Fock vacuum $|0\rangle$ to be the state annihilated by all the annihilation operators:
\begin{align}
\mu^{\alpha}|0\rangle = \tilde{\mu}^{\dot{\alpha}} |0\rangle = \xi^a |0\rangle =0 \period
\end{align}
To be more precise, $\ket{0}$ is a tensor product of two vacua, one for the 
 bosonic oscillators and the other for the fermionic ones. Namely, 
\begin{align}
\ket{0} &= \ket{0}_B \otimes \ket{0}_F \comma  \label{vacBF}
\end{align}
Then the Fock space is built upon this vacuum by acting by the creation operators $\lambda_{\alpha}, \tilde{\lambda}_{\dot{\alpha}},  \bar{\xi}_a$. However, not all the states produced this way  correspond to the fields of $\mathcal{N}=4$ SYM.  The relevant ones are only those carrying  zero 
 central charge.  This can be explicitly checked by the expressions 
 of the basic $\calN=4$ SYM fields in terms of the oscillators given by\cite{1-loop}
\begin{align}
F_{\alpha \beta}(0)& \leftrightarrow \lambda_{\alpha}\lambda_{\beta} |0\rangle \comma \label{eq:field_oscillator1} \\
\psi_{\alpha a} (0)& \leftrightarrow \lambda_{\alpha} \bar{\xi}_a |0\rangle \comma \\
\phi_{ab} (0)&  \leftrightarrow \bar{\xi}_a \bar{\xi}_b |0\rangle \comma 
\label{scalars}\\
\bar{\psi}_{\dot{\alpha}}^a(0) &  \leftrightarrow \frac{1}{3!}\epsilon^{abcd}\tilde{\lambda}_{\dot{\alpha}} \bar{\xi}_b\bar{\xi}_c\bar{\xi}_d |0\rangle \comma \\
\bar{F}_{\dot{\alpha}\dot{\beta}}(0) &  \leftrightarrow \frac{1}{4!}\epsilon^{abcd} \tilde{\lambda}_{\dot{\alpha}} \tilde{\lambda}_{\dot{\beta}} \bar{\xi}_a  \bar{\xi}_b\bar{\xi}_c\bar{\xi}_d |0\rangle \period \label{eq:field_oscillator5}
\end{align}
From the form of $C$ given in (\ref{defC}) it is clear that they all carry  $C=0$. Also, it is easy to check that these oscillator expressions of the fields 
 carry the correct Lorentz and R-symmetry quantum numbers. 

In addition to these fundamental fields, we need to express their  derivatives. 
The field at  the general position $x$ is  obtained by the action of the translation operator $e^{iP\cdot x}$ as 
\begin{align}
|\mathcal{O}(0) \rangle \rightarrow |\mathcal{O}(x) \rangle := e^{iP\cdot x}|\mathcal{O}(0) \rangle \period
\end{align}
From the oscillator representation $P_{\alpha \dot{\beta}}= \lambda_{\alpha}\tilde{\lambda}_{\dot{\beta}}$, we see that  the derivatives of a field can be  expressed as 
\begin{align}
&\partial_{{}^(\alpha {}_(\dot{\beta}} \cdots \partial_{\gamma^) \dot{\delta}_)}  \mathcal{O}(x) \cong (i \lambda_{\alpha}\tilde{\lambda}_{\dot{\beta}}) \ldots ( i \lambda_{\gamma} \tilde{\lambda}_{\dot{\delta}})  |\mathcal{O}(x) \rangle \comma 
\end{align}
where $\partial_{\alpha \dot{\beta}} = \partial/\partial x^{\dot{\beta}\alpha}$ and $x^{\dot{\alpha}\beta}:=x^{\mu}(\bar{\sigma}_{\mu})^{\dot{\alpha}\beta}$ and 
 we have used $P\cdot x=P_{\alpha \dot{\beta}}x^{\dot{\beta}\alpha}$ and $\partial_{\alpha \dot{\beta}}e^{iP\cdot x}=iP_{\alpha \dot{\beta}}e^{iP\cdot x}$. Notice that the spinor indices $(\alpha , \gamma, \ldots)$ and $(\dot{\beta} , \dot{\delta}, \ldots)$ are symmetrized as the bosonic oscillators $\lambda$ mutually commute. Also note that  we can replace some combinations of the (covariant) derivatives by appropriate  fields without derivatives using the equations of motion and the Bianchi identities. For example, we can set $\partial_{\alpha \dot{\beta}} \partial^{\dot{\beta}\alpha}\phi \propto \Box \phi$ and $\epsilon^{\alpha \beta} \partial_{\alpha \dot{\alpha}} \psi_{\beta}^a$ to zero due to the free equations of motion\footnote{In the interacting case, it is possible to replace the combinations of covariant derivatives such as $\mathcal{D}_{\alpha \dot{\beta}} \mathcal{D}^{\dot{\beta}\alpha}\phi $ and $\epsilon^{\alpha \beta} \mathcal{D}_{\alpha \dot{\alpha}} \psi_{\beta}^a$ by the fields without derivatives using the equations of motion as well.}. As a result,  we can express fields with derivatives by expressions 
 where all the spinor indices are totally symmetrized. Therefore, the independent fields with derivatives are simply generated by acting $P_{\alpha\dot{\beta}}= \lambda_{\alpha}\tilde{\lambda}_{\dot{\beta}}$ on the oscillator representations for the fundamental fields  (\ref{eq:field_oscillator1})-(\ref{eq:field_oscillator5}). Since $P_{\alpha \dot{\beta}}$ commutes  with the central charge, these states with derivatives are still within the subspace with vanishing central charge.
\subsubsection{Various ``vacua"  and their relations}
It is an elementary exercise in quantum mechanics to construct the 
singlet state from two spin $1/2$ states by forming a suitable combination of 
 the highest and the lowest states.  It is a slightly more involved exercise  to 
extend this to the case of the general spin $j$, but the structure is similar: 
 One combines the states built upon the lowest weight states and those 
built upon the highest weight states with simple weights. Indeed, 
up to an overall constant, the singlet state is given by 
\begin{align}
\ket{\bfone_j} = \sum_{l=0}^{2j} (-1)^l \ket{-(j-l)} \otimes \ket{j-l} 
\end{align}
This indicates that for the  construction of  the singlet state for much more complicated 
 case of psu(2,2$|$4), the basic idea should be the same and one would combine 
 the Fock states built upon the lowest weight oscillator vacuum $\ket{0}$,  already introduced,  with the states built upon  the highest weight oscillator vacuum  $\ket{\bar{0}}$, which should be 
defined to be annihilated by the creation operators as
\begin{align}
\lambda_{\alpha}|\bar{0}\rangle = \tilde{\lambda}_{\dot{\alpha}}|\bar{0}\rangle =\bar{\xi}_a |\bar{0}\rangle =0 \period \label{defzerobar}
\end{align}
Just as for $\ket{0}$ given  in (\ref{vacBF}),  the  more precise definition of $\ket{\bar{0}}$ is 
\begin{align}
\ket{\bar{0}} &\equiv \ket{\bar{0}}_B \otimes \ket{\bar{0}}_F \period
\end{align}
From (\ref{defzerobar})  it immediately follows that $|\bar{0}\rangle$ is annihilated by $P_{\alpha \dot{\beta}}=\lambda_{\alpha}\tilde{\lambda}_{\dot{\beta}}$ and thus the Fock space built on $\ket{\bar{0}}$  is a highest weight module as opposed to the lowest weight module  built on  $|0\rangle$. 

There is  an essential difference between the bosonic sector and the 
 fermionic sector. For the bosonic sector, $\ket{0}_B$ and $\ket{\bar{0}}_B$ 
cannot be related by the action of a finite number of oscillators\footnote{Actually, by using the operator  $U^2=\exp [\frac{\pi}{2}(P_0-K_0)]$
 one can map $|0\rangle$ to $|\bar{0}\rangle$ and exchange  the role of the annihilation and the creation operators.}, but for 
 the fermionic sector  one can readily identify $\ket{\bar{0}}_F =
 \xibar_1 \xibar_2\xibar_3\xibar_4 \ket{0}_F$. 

It will turn out, however, that as for the fermionic oscillator Fock space 
 describing the R-symmetry quantum numbers, ``vacua" slightly different 
 from $\ket{0}$ and $\ket{\bar{0}}$ will be more useful and more physical.
To introduce them, we rename the fermionic oscillators in the following way 
so that half of the creation (annihilation) operators are switched to annihilation (creation) operators\footnote{Such a transformation is sometimes  called a particle-hole transformation.}:
\begin{align}
&c^i= \xi^i \ (i=1,2) \comma \ \ d^i= \bar{\xi}_{i+2} \ (i=1,2)  \comma \\
&\bar{c}_i=\bar{\xi}_i \ (i=1,2) \comma \ \ \bar{d}_i= \xi^{i+2} \ (i=1,2) \period
\end{align}
We define the state $\Zket$ as annihilated by the new annihilation 
 operators $c^i$ and $d^i$, while $\Zbarket$ is defined to be annihilated 
 by the new creation operators $\cbar_i$ and $\dbar_i$. 
\begin{align}
c^i |Z\rangle =d^i |Z\rangle =0 \comma  \label{defZ} \\
\bar{c}_{i} |\bar{Z} \rangle =\bar{d}_i |\bar{Z}\rangle =0 \label{defZbar}
 \period
\end{align}
As states built on the original vacuum $\ket{0}$, these new ``vacua" can be written as 
\begin{align}
\Zket &=d^1d^2 \ket{0} = \xibar_3\xibar_4 \ket{0}\comma \\
\Zbarket &=\cbar_1\cbar_2 \ket{0} = \xibar_1\xibar_2 \ket{0}
\end{align}
 and 
 they are related as $\Zbarket= -\cbar_1\cbar_2 \dbar_1 \dbar_2 
\Zket$ or $\Zket = -c^1c^2d^1d^2 \Zbarket$.
 Now if  we recall that the SO(6) scalars are represented by  $\xibar_a \xibar_b \ket{0}$ as shown in (\ref{scalars}), $\ket{Z}$ and $\Zbarket$ correspond to some physical scalars.  To be definite let us identify them as states  carrying $\SU2_L \times \SU2_R$ quantum numbers of the SU(2) sector, where the generators are given by\footnote{Of course the choice of  $\SU2_L \times \SU2_R$ in SU(4) is not unique. We are simply taking a convenient one. 
Incidentally, our normalization for $J_\pm$ is $J_\pm = J_1 \pm i J_2$. }
\begin{align}
J_+^L =  c^1d^1\comma \quad J_-^L =  \dbar_1 \cbar_1
\comma \quad J_3^L = \half (d^1 \dbar_1 -\cbar_1 c^1) \comma 
\label{SU2L}\\
J_+^R =  c^2d^2\comma \quad J_-^R = \dbar_2  \cbar_2
\comma \quad J_3^R = \half (d^2 \dbar_2 -\cbar_2 c^2)  \period
\label{SU2R}
\end{align}
Then, it is easy to see that $\Zket$ and $\Zbarket$ carry 
 the quantum numbers $(\half, \half)$ and $(-\half, -\half)$ respectively and hence can be identified with, say,   $\phi_1+i\phi_2$ and its complex conjugate. 
The (de-)excitations of these vacua in the SU(2) sector with the quantum 
 numbers $(\half, -\half)$ and $(-\half, \half)$ respectively, which are  
often  denoted by $\ket{X}$ and $\ket{-\bar{X}}$,  are given by 
\begin{align}
\ket{X} &= J_-^R \ket{Z} =J_+^L \Zbarket =\cbar_2 d^1  \ket{0} \comma \\ 
\ket{-\bar{X}} &= J^L_- \Zket = J^R_+ \Zbarket =d^2 \cbar_1  \ket{0} \period
\end{align}
Now in order to construct the singlet projector in section 2.2, it will turn out to be convenient to define the scalar states similar to the above, except that their bosonic  part of the vacuum is switched from $\ket{0}_B$ to $\ket{\bar{0}}_B$. We will place a line over the kets 
(or the corresponding bra) to denote such scalar states. For example, 
\begin{align}
\overline{\ket{Z}} &\equiv \ket{\bar{0}}_B \otimes d^1d^2 \ket{0}_F \comma \\
\overline{\ket{\bar{Z}}} &\equiv \ket{\bar{0}}_B \otimes \cbar^1\cbar^2 \ket{0}_F \period
\end{align}
In this more precise  notation, the previously defined $\ket{Z}$ and $\ket{\bar{Z}}$ are written as 
\begin{align}
\ket{Z} &= \ket{0}_B \otimes d^1d^2 \ket{0}_F \comma \\ 
\ket{\bar{Z}} &= \ket{0}_B  \otimes \cbar^1\cbar^2 \ket{0}_F \period
\end{align}
Since overlined scalar states differ only in the bosonic sector, the properties of 
 such states under the action of the fermionic oscillators are exactly the 
 same as the un-overlined ones. For example, $c^i \overline{\ket{Z}}=0$, 
 etc., just as in (\ref{defZ}) and (\ref{defZbar}). 
\begin{align}
\bar{c}_i \overline{|Z\rangle} =\bar{d}_i \overline{|Z\rangle} =0 \comma \ \ \overline{|Z\rangle} := \xi^1 \xi^2 |\bar{0}\rangle \comma \\
c^i \overline{|\bar{Z} \rangle} =d^i \overline{|\bar{Z}\rangle }=0 \comma \ \ \overline{|\bar{Z} \rangle} := \xi^3\xi^4 |\bar{0}\rangle \period
\end{align}

As we will need them later, it should be convenient to list the  properties of the bra (or dual) vacua,  which evidently  follow  from those of the ket vacua. $\bra{0}$ and $\bra{\bar{0}}$ have the properties 
\begin{align}
\langle 0|\lambda_{\alpha}=\langle 0|\tilde{\lambda}_{\dot{\alpha}}=\langle 0|\bar{\xi}_a=0 \comma \ \ \langle 0| 0\rangle =1 \comma \\
\langle \bar{0}|\mu^{\alpha}=\langle \bar{0}|\tilde{\mu}^{\dot{\alpha}}=\langle \bar{0} |\xi^a=0 \comma \ \ \langle \bar{0}| \bar{0} \rangle =1 \period
\end{align}
This means  that the dual Fock space is generated either by  the action of $(\mu^{\alpha},\tilde{\mu}^{\dot{\alpha}},\xi_a)$ on $\langle 0|$ or 
  by the action of  $(\lambda_{\alpha}, \tilde{\lambda}_{\dot{\alpha}}, \bar{\xi}^a)$ on $\langle\bar{0}|$. 
As for the properties of the scalar bra vacua under the action of the 
 fermionic oscillators, they satisfy 
\begin{align}
\langle Z| \bar{c}_i &=\langle Z|\bar{d}_i=0 \comma
\quad \langle \bar{Z}| c^i =\langle \bar{Z}|d^i=0 \comma \\
\langle Z|Z\rangle&=\langle \bar{Z}|\bar{Z}\rangle=1 \comma 
\end{align}
and exactly the same equations hold for the overlined bra states 
$\overline{ \bra{Z}}$ and $\overline{\bra{\bar{Z}}}$. 
\subsubsection{Comparison with the E-scheme formulation}
Before we start  the systematic construction of the singlet projector
 using these oscillator representations, let us end this subsection with some comments on the difference between the oscillator representation we use and the one employed in \cite{ADGN,spinvertex}. They basically work in the E-scheme, where the oscillators are covariant  under the compact subgroup shown in  (\ref{eq:E-scheme}).  This  in turn means that the representations corresponding to local composite operators are obtained indirectly by 
 the use of  the complicated operator $U=\exp[\frac{\pi}{4}(P_0-K_0)]$. 
To be a little more specific, let us display the E-scheme oscillators and the  Fock vacuum used in \cite{ADGN,spinvertex}.
The difference from ours is in the bosonic oscillators, which are given by 
\begin{align}
[a_i, \bar{a}^j]=&\delta_i^j \comma (i,j=1,2) \  [b_s,\bar{b}^t]=\delta_s^t \comma \ (s,t=1,2)  \\
&a_i|0 \rangle_E=b_s|0 \rangle_E=0 \period
\end{align}  
Then the SO(2,4) generators are expressed as  bi-linears of these oscillators 
as
\begin{align}
L^i_j = \bar{a}^i a_j -\frac{1}{2}\delta^i_j(\bar{a}^ka_k) \comma \ R^s_t = \bar{b}^s b_t -\frac{1}{2}\delta^s_t(\bar{b}^ub_u) \comma \\
E=\frac{1}{2} (\bar{a}^ia_i+\bar{b}^sb_s)+1 \comma \ L_{is}=a_ib_s \comma \ L^{is}=\bar{a}^i\bar{b}^s \comma
\end{align}
where $L^i_j$, $R^s_t$ are SU(2)$_L \times$ SU(2)$_R$ generators, $E$ is the AdS energy   and $L_{is}$, $L^{is}$ are the elements of $\mathcal{E}^-$, $\mathcal{E}^+$ respectively. Hence, the bosonic oscillators $a_i, \bar{a}^i$ transform covariantly under SU(2)$_L$ as a doublet and $b_s,\bar{b}^s$ are doublets of SU(2)$_R$. To convert them to the D-scheme oscillators, one needs to employ 
the similarity transformation using the operator $U$. 
However, as we mentioned in the introduction, any similarity transformation 
 preserves  the structure of the algebra  and hence, for example, $\SU2_L$ does not become a Lorentz group ${\rm SL}(2,\mathbb{C})$. This is reflected in  the transformation of the oscillators themselves. 
 By  using the explicit oscillator representation of $U$, 
 we easily find, for example, $U^{-1}\bar{a}^i  U =\frac{1}{\sqrt{2}}(\bar{a}^i + b_i)$ etc., which is not informative  as far as the useful re-interpretation to the D-scheme is concerned. 
Thus although the E- and the D- schemes are connected by a 
 similarity transformation $U$, the conformal covariance cannot be 
made manifest  by just such a transformation. 
Hence  for the purpose of dealing with the local composite operators, the use of 
 D-scheme is much more  transparent and indeed in what follows we shall 
 never need the operator $U$. 
\subsection{Construction of the singlet projector for $ {\rm psu}(2,2|4)$ }
\subsubsection{Singlet condition and its solution}
We shall now give a detailed construction of the singlet projector $\singpsu$  for the states in the product of a pair of  Hilbert spaces $\calH_1 \otimes \calH_2$, which satisfies the defining equation  for the singlet projector
\begin{align}
\singpsu (J^A{}_B \otimes \bfone  + \bfone \otimes J^A{}_B)=0\comma
\qquad J^A{}_B \in \psu224 \period \label{defsingpsu}
\end{align}
In order to find the most general singlet projector, we must proceed systematically. As it will become clear below, actually the desired singlet projector satisfying  the relation above can be constructed for ${\rm su}(2,2|4)$ as well as 
 for ${\rm psu(2,2|4)}$, but not for ${\rm u(2,2|4)}$. Recall that the generators $J^A{}_B$ of  ${\rm su}(2,2|4)$ are obtained from the generator of ${\rm u}(2,2|4)$, to be tentatively denoted  by $\hat{J}^A{}_B$, by making them supertraceless, \ie 
\begin{align}
J^A{}_B &= \hat{J}^A{}_B -{1\over 8} \delta^A_B (-1)^{|A|} 
({\rm Str} \hat{J} )\period
\end{align}
Because of this condition, when we interchange the order of the two conjugate oscillators making up any diagonal generator $J^A{}_A$, constant terms produced from the (anti-)commutation relations precisely cancel. This property will be of crucial importance  for the construction of the true singlet projector. 

 First,  let us begin by  identifying  the building block for the sector involving  the oscillator pair $\lam_\al$ and $\mu^\al$. Since the generators $J^A{}_B$ are 
 quadratic in the oscillators, the building block which would realize 
 the relation (\ref{defsingpsu}) in the above sector should be 
of the form 
\begin{align}
{}_{\lam\mu}\bra{\bfone_{12}}\propto (\bra{Z}\otimes  \barZbarbra) (\mu^\al)^{n_\mu} \otimes  (\lam_\be)^{n_\lam} )
\end{align}
Now consider a useful  combination of generators $\calJ^{(1)}\equiv \lam_1 \mu^1 -\lam_2\mu^2 = \mu^1 \lam_1 -\mu^2 \lam_2$,
 which belongs to ${\rm su(2,2|4)}$ and  hence  the interchange of the order of  $\lam_\al$ and $\mu^\al$ 
does not produce any constant.   When we apply $\calJ^{(1)}
\otimes \bfone$, we should use the form $\calJ^{(1)}\equiv \lam_1 \mu^1 -\lam_2\mu^2$ since $\lam_\al$ annihilates $\Zbra$. Then we easily obtain 
\begin{align}
(\bra{Z}\otimes  \barZbarbra) (\mu^\al)^{n_\mu} \otimes  (\lam_\be)^{n_\lam} )(\calJ^{(1)}
\otimes \bfone) &= n_\mu ( \delta^\al_1-\delta_2^\al) (\bra{Z}\otimes  \barZbarbra) (\mu^\al)^{n_\mu} \otimes  (\lam_\be)^{n_\lam} ) \period
\label{mulamone}
\end{align}
On the other hand, when we apply  $ \bfone \otimes \calJ^{(1)}$, since $\barZbarbra$
 is annihilated by $\mu^\al$, we should use the form $\calJ^{(1)}=\mu^1 \lam_1 -\mu^2 \lam_2$. Then, we get
\begin{align}
(\bra{Z}\otimes  \barZbarbra) (\mu^\al)^{n_\mu} \otimes  (\lam_\be)^{n_\lam} )(\bfone \otimes \calJ^{(1)}) &=
-n_\lam ( \delta_\be^1-\delta^2_\be) (\bra{Z}\otimes  \barZbarbra) (\mu^\al)^{n_\mu} \otimes  (\lam_\be)^{n_\lam} ) \period \label{mulamtwo}
\end{align}
In order for the sum of  (\ref{mulamone}) and (\ref{mulamtwo}) to vanish, 
 we must have $n_\mu=n_\lam$ and $\al =\be$.  Hence,  the form of the 
oscillator factor should actually be the combination $(\mu^\al \otimes \lam_\al)^{n_{\mu_\al}}$

We can apply the same logic to the sectors consisting of other conjugate 
 pairs, namely $(\mutil^\aldot , \lamtil_\aldot), (\cbar_i, c^i)$ and $(\dbar_j, d^j)$ and find similar conditions. In this way, we find that the necessary form for the  singlet projector for  ${\rm su}(2,2|4)$ can be written as 
\begin{align}
{}_{{\rm su}}\bra{\bfone_{12}} &= \sum_{\bfn} f(\bfn) \bra{\bfn} \\
\bra{\bfn} &\equiv \Zbra \otimes \barZbarbra 
\prod_{\al, \bedot, i, j}{ (\mu^\al\otimes \lam_\al)^{n_{\lam_\al}}
\over n_{\mu_\al} !} {(\mutil^\aldot\otimes \lamtil_\aldot)^{n_{\lamtil_\aldot}} \over n_{\mutil_\aldot}!}  
  {(c^i \otimes \cbar_i)^{n_{c_i}} \over n_{c_i}!}  {(d^j \otimes \dbar_j)^{n_{d_j}} \over n_{d_j}!} \comma \\
f(\bfn) &= f(n_{\lam_1}, n_{\lam_2}, n_{\lamtil_\onedot}, n_{\lamtil_\twodot}, \ldots) \comma 
\end{align}
where $f(\bfn)$ at this stage  is an arbitrary function and  is to be determined by the requirement of the singlet condition.  As for the sum over the powers $n_{\mu_\al}$ etc, we shall allow  them 
 to be arbitrary non-negative  integers. 

To see what conditions should be satisfied by the function $f(\bfn)$, let us 
 focus first on a simple  generator in the $(\mu,\lam)$ sector of the form 
 $J_\al{}^\be = \lam_\al \mu^\be $, where $\al \ne \be$. Since $\lam_\al$ is the annihilation operator for the bra state $\Zbra$, just as before, we easily get
\begin{align}
\sum_{\bfn} f(\bfn) \bra{\bfn} (\lam_\al \mu^\be \otimes 1) 
 &= \sum_{\bfn} f(\bfn) \bra{\bfn} (\lam_\al \otimes 1) ( \mu^\be \otimes 1) \nn\\
&= \sum_{\bfn} f(\bfn) \bra{\bfn} { n_{\lam_\al}\over n_{\lam_\al} !} (\mu^\al\otimes \lam_\al)^{n_{\lam_\al}-1}
(1\otimes \lam_\al) ( \mu^\be \otimes 1) \cdots \nn\\
&=  \sum_{\bfn} f(n_{\lam_\al}+1, \ldots) \bra{\bfn} ( \mu^\be \otimes 1) (1\otimes \lam_\al) + \cdots \period
\end{align}
In the third line  we have shifted $\n_{\lam_\al}$ by 1 and interchanged the order of the factors $1\otimes \lam_\al$ and $\mu^\be \otimes 1$. 
Now the action of $\mu^\be \otimes 1$ on $\bra{\bfn}$ is easily seen to 
produce the structure $1 \otimes \mu^\be$ with an overall minus sign,  
together with a shift of $\n_{\lam_\be}$ by minus one unit in $f(\bfn)$
 under the sum. As for the structure of the operator part, combined with the factor $(1\otimes \lam_\al)$ already produced, we get
\begin{align}
(1\otimes \mu^\be )(1\otimes \lam_\al) &= (1\otimes \mu^\be \lam_\al) = (1\otimes \lam_\al \mu^\be) = 1\otimes J_\al{}^\be, 
\label{interchange}
\end{align}
where we have interchanged the order of $\mu^\be$ and $\lam_\al$ 
 to get back $J_\al{}^\be$ without producing any constant  since we are considering the case with $\al \ne \be$. 
 Altogether 
we obtain the formula 
\begin{align}
\sum_{\bfn} f(n_{\lam_\al}, n_{\lam_\be}, \ldots) \bra{\bfn} (J_\al{}^\be  \otimes 1) 
&= - \sum_{\bfn} f(n_{\lam_\al}+1, n_{\lam_\be}-1, \ldots ) \bra{\bfn} (1\otimes J_\al{}^\be) 
\end{align}
Thus the singlet condition demands 
\begin{align}
 f(n_{\lam_\al}, n_{\lam_\be}, \ldots) = f(n_{\lam_\al}+1, n_{\lam_\be}-1, \ldots ) \period
\end{align}
The general solution of this equation is 
\begin{align}
f(n_{\lam_1}, n_{\lam_2}, \ldots) = g(n_{\lam_1}+ n_{\lam_2}, \ldots) \comma 
\end{align}
where $g$ is an arbitrary function except that $n_{\lam_\al}$'s
  must appear  as the sum $n_{\lam_1}+ n_{\lam_2}$. 

Repeating   similar analyses for all the off-diagonal\footnote{``Off-diagonal" here means  the generators  like $\lam_\al \mu^\be$ with $\al \ne \be$, etc.  so that their (anti)commutators  vanish. For them there is no difference between ${\rm u}(2,2|4)$ and ${\rm su}(2,2|4)$.}  generators of ${\rm su}(2,2|4)$,  one obtains the following list of singlet conditions. \\
For the bosonic generators, 
we get 

\begin{tabular}[t]{c c l} 
(b1)&$\lam_\al \mu^\be$ &\qquad  $f(n_{\lam_\al}, n_{\lam_\be}, \ldots) = f(n_{\lam_\al}+1, n_{\lam_\be}-1, \ldots) $   \\
(b2)&$\lam_\al \lamtil_\bedot$ & \qquad $f(n_{\lam_\al}, n_{\lamtil_\bedot},\ldots) = -f(n_{\lam_\al}+1, n_{\lamtil_\bedot}+1, \ldots) $  \\
(b3)&$\mutil^\aldot \lamtil_\bedot$ &\qquad $
f( n_{\lamtil_\aldot}, n_{\lamtil_\bedot}, \ldots) =f(n_{\lamtil_\aldot}-1, n_{\lamtil_\bedot}+1, \ldots) $  \\
(b4)&$\mutil^\aldot \mu^\al$ &\qquad  $f(n_{\lamtil_\aldot}, n_{\lam_\al}, \ldots)=-f(n_{\lamtil_\aldot}-1, n_{\lam_\al}-1, \ldots) $  \\
(b5)&$\cbar_i \dbar_j$ & \qquad $f(n_{c_i}, n_{d_j}, \ldots)=-f(n_{c_i}+1, n_{d_j}+1, \ldots) $  \\
(b6)&$\cbar_i c^j$ & \qquad $f(n_{c_i}, n_{c_j}, \ldots)=f(n_{c_i}+1, n_{c_j}-1, \ldots) $  \\
(b7)&$ d^j c^i$ & \qquad $f(n_{d_j}, n_{c_i}, \ldots)=-f(n_{d_j}-1, n_{c_i}-1, \ldots) $  \\
(b8)&$d^j \dbar_k$ &\qquad  $f(n_{d_j}, n_{d_k}, \ldots)=f(n_{d_j}-1, n_{d_k}+1, \ldots) $ \end{tabular}
\nxt
For the fermionic generators, the conditions are 

\begin{tabular}[t]{cc l} 
(f1)&  $\lam_\al c^i$ &\qquad  $f(n_{\lam_\al}, n_{c_i}, \ldots)=-f(n_{\lam_\al}+1, n_{c_i}-1, \ldots) $  \\
(f2)&$\lam_\al \dbar_j$ & \qquad $f(n_{\lam_\al}, n_{d_j}, \ldots)=f(n_{\lam_\al}+1, n_{d_j}+1, \ldots) $  \\
(f3)&$\mutil^\aldot c^i$ &\qquad $f(n_{\lamtil_\aldot}, n_{c_i}, \ldots)=f(n_{\lamtil_\aldot}-1, n_{c_i}-1, \ldots) $ \\
(f4)&$\mutil^\aldot \dbar_j$ &\qquad  $f(n_{\lamtil_\aldot}, n_{d_j}, \ldots)=-f(n_{\lamtil_\aldot}-1, n_{d_j}+1, \ldots) $ \\
(f5)&$\cbar_i \mu^\al$ & \qquad $f(n_{c_i}, n_{\lam_\al}, \ldots)=-f(n_{c_i}+1, n_{\lam_\al}-1, \ldots) $ \\
(f6)&$\cbar_i \lamtil_\aldot$ & \qquad $f(n_{c_i}, n_{\lamtil_\aldot}, \ldots)=f(n_{c_i}+1, n_{\lamtil_\aldot}+1, \ldots) $ \\
(f7)&$ d^j \mu^\al$ & \qquad $f(n_{d_j}, n_{\lam_\al}, \ldots) =f(n_{d_j}-1, n_{\lam_\al}+1, \ldots) $ \\
(f8)&$d^j \lamtil_\aldot$ &\qquad  $f(n_{d_j}, n_{\lamtil_\aldot}, \ldots)=-f(n_{d_j}-1, n_{\lamtil_\aldot}+1, \ldots) $ 
\end{tabular}

With the hint from the analysis of the bosonic $(\mu, \lam)$ sector, it is actually easy to write down the most general solution satisfying these equations. 
The answer is 
\begin{align}
f(n_{\lam_1}, n_{\lam_2}, \ldots ) &= (-1)^{n_{\lamtil_\onedot}+n_{\lamtil_\twodot} + n_{c_1}
+n_{c_2}} h(C) \\
2C &= 
(n_{\lam_1} + n_{\lam_2}) -(n_{\lamtil_\onedot} + n_{\lamtil_\twodot})
+(n_{c_1} + n_{c_2}) -(n_{d_1} + n_{d_2})
\end{align}
where $h(x)$ is an arbitrary function of one argument.
It is important to note that $C$ is precisely  the central charge of ${\rm u}(2,2|4)$. As such it can be set to a number in an irreducible representation.  
In particular,  the fundamental SYM fields of our interest belong to the sector where $C=0$ and $h(0)$ is just an overall constant, which we shall set to unity
 for simplicity. 

Now we must examine the diagonal generators, such as $\lam_\al \mu^\al$
 and $\cbar_i c^i$, etc. Because an extra  constant is produced upon interchanging the order of the oscillators, for example like $\lam_\al \mu^\al = \mu^\al  \lam_\al -2$, etc. in the process of the manipulation as in (\ref{interchange}), in general the singlet condition is not  satisfied. 
However, as we already stressed,  for the diagonal generators 
 which belong to ${\rm su}(2,2|4)$ and $\psu224$ such constants cancel. Therefore, the conditions 
 we obtained for the function $f(\bfn)$ do not change and the singlet projector for the physical SYM states is obtained as\footnote{If one takes a different value of $C$ one obtains a singlet projector for that sector. Here we focus on the  physical SYM fields for which $C=0$.}
\begin{align}
{}_{\rm psu} \bra{\bfone_{12}} &=  \Zbra \otimes \barZbarbra  \sum_{\bfn \ge 0\comma 
C=0} 
 (-1)^{n_{\lamtil_\onedot}+n_{\lamtil_\twodot} + n_{c_1}
+n_{c_2}} \nn\\
& \times 
\prod_{\al, \bedot, i, j}{ (\mu^\al\otimes \lam_\al)^{n_{\lam_\al}}
\over n_{\lam_\al} !} {(\mutil^\aldot\otimes \lamtil_\aldot)^{n_{\lamtil_\aldot}} \over n_{\lamtil_\aldot}!}  
  {(c^i \otimes \cbar_i)^{n_{c_i}} \over n_{c_i}!}  {(d^j \otimes \dbar_j)^{n_{d_j}} \over n_{d_j}!}
\end{align}
Because of the restriction $C=0$ in the sum over the $n_\ast$'s, this expression {\it does not} quite take the form of an exponential. However, we can remove the restriction $C=0$ in the sum when applying $\bra{\bfone_{12}}$ to
the physical SYM states, since the extra 
states  with $C \ne 0$ produced are orthogonal to $C=0$ states and do not 
 contribute to the inner product with the physical states. Thus, with the $C=0$  restriction removed, the singlet state above {\it can be} written as  a simple exponential given by
\begin{align}
{}_{\rm psu} \bra{\bfone_{12}} &= \Zbra \otimes \barZbarbra 
\exp \left( \lam_\al \otimes \mu^\al - \lamtil_\aldot \otimes \mutil^\aldot
+ \cbar_i \otimes c^i -\dbar_j \otimes d^j\right) \comma   \label{singproj}
\end{align}
where in the exponent the sum is implied for the repeated indices. If one wishes to perform  the Wick contraction in a manifestly symmetric fashion, one can use the form 
\begin{align}
\half \left( {}_{\rm psu} \bra{\bfone_{12}}+ {}_{\rm psu} \bra{\tilde{\bfone}_{12}}\,  \right) \comma 
\end{align}
where 
\begin{align}
 {}_{\rm psu} \bra{\tilde{\bfone}_{12}}
&=\barZbarbra\otimes \Zbra  \exp \left( \mu^\al \otimes  \lam_\al   -  \mutil^\aldot\otimes \lamtil_\aldot 
+c^i \otimes  \cbar_i  -d^j\otimes \dbar_j  \right) \period
\end{align}
Hereafter, we shall suppress for simplicity the subscript psu and write $\bra{\bfone_{12}}$ for $ {}_{\rm psu} \bra{\bfone_{12}}$. 
\subsubsection{Crossing relations for the oscillators}
Before ending this subsection, let us make an important remark on the property of the singlet projector 
(\ref{singproj}). Although we have constructed this state by demanding that it  be singlet under the generators of $\psu224$ satisfying (\ref{defsingpsu}),  it is easy to see from the process of construction  above that 
actually  the singlet projector (\ref{singproj}) effects the following ``crossing relations" 
 for the {\it individual oscillators}:
\begin{align}
\bra{\bfone_{12}} (\bar{\zeta}^A\otimes 1) &= \bra{\bfone_{12}}(1 \otimes \bar{\zeta}^A) \comma \\
\bra{\bfone_{12}} (\zeta_A\otimes 1) &= -\bra{\bfone_{12}}(1 \otimes \zeta_A)  \period   \label{osccrossing}
\end{align}
Clearly these relations themselves have no group theoretical meaning and appear to be  stronger than the singlet condition. It is remarkable that yet they follow from the requirement of the singlet condition and will be quite useful 
in the computation of the correlation function, as we shall see  in the next 
 subsection. 
\subsection{Wick contraction of the basic fields using the singlet projector} 
Let us now show that the Wick contraction of the basic 
 fields of the $\mathcal{N}=4$ super Yang-Mills theory can be computed quite easily  by using the singlet projector constructed in the previous subsection. 
This can be identified as the method of Ward identity  already introduced in \cite{ADGN}.  However, as we use the D-scheme from the outset,  our method is much more direct and simpler, without the need of rather complicated  conversion operator $U$. 

Consider first the scalar field $\phi_{ab}(x)$ belonging to the 6-dimensional 
anti-symmetric representation of SU(4), which corresponds to the state
$\xibar_{[a}\xibar_{b]} e^{iP\cdot x} \ket{0}$. Then the Wick contraction 
of two such fields   $\contraction[1ex]{}{\phi_{ab}}{(x)}{\phi} \phi_{ab}(x) \phi_{cd}(y) $ can be computed as $\bra{\bfone_{12}} 
(\xibar_{[a}\xibar_{b]} e^{iP\cdot x} \ket{0} \otimes \xibar_{[c}\xibar_{d]} e^{iP\cdot y} \ket{0})$. Since the singlet structure for the SU(4) part 
 gets extracted  as the unique factor   $\ep_{abcd}$, we obtain 
\begin{align}
\contraction[1ex]{}{\phi_{ab}}{(x)}{\phi}
\phi_{ab}(x) \phi_{cd}(y) & \propto \ep_{abcd} I(x,y) \comma 
\end{align}
where 
\begin{align}
I(x,y) &\equiv  \langle \boldsymbol{1}_{12}| (e^{iP\cdot x}|0\rangle \otimes e^{iP\cdot y}|0\rangle) \period
\end{align}
The function $I(x,y)$ 
will be seen below to be the basic building block for the contractions of all the 
super Yang-Mills fields and can be easily fixed by the singlet 
 conditions\footnote{As we shall see below, the singlet conditions 
produce Ward identities.}  with $J^A{}_B$ taken to be 
 translation and the dilatation generators in the following way. First, applying the singlet condition (\ref{defsingpsu}) taking $J^A{}_B$ to be the  translation generator, we have
\begin{align}
0 &= \bra{\bfone_{12}}( iP_\mu e^{iP\cdot x} \ket{0} \otimes 
 e^{iP \cdot y} \ket{0} ) +  \bra{\bfone_{12}}( e^{iP\cdot x} \ket{0} \otimes 
i P_\mu e^{iP \cdot y} \ket{0} )  \nn\\
&= \left( {\del \over \del x^\mu} + {\del \over \del y^\mu} \right)
 I(x,y) 
\end{align}
This gives $I(x,y) = I(x-y)$. Next, we use the dilatation operator 
given by $D = (i/2) (\lam_\al \mu^\al + \lamtil_\aldot \mutil^\aldot +2)$. Since $P\cdot x$ can be written as $\lam_\al \lamtil_\aldot x^{\aldot \al}$, 
 the action of $\lam_\al \mu^\al$ in $D$ on $e^{iP\cdot x} \ket{0}$ gives %
\begin{align}
\lam_\al \mu^\al e^{iP\cdot x} \ket{0} &= i\lam_\al \lamtil_\aldot 
x^{\aldot \al} e^{i \lam_\al \lamtil_\aldot x^{\aldot\al}}\ket{0} =
iP\cdot x e^{iP \cdot x} \ket{0}= x^\mu {\del \over  \del x^\mu} e^{iP \cdot x} \ket{0} \period
\end{align}
Evidently, the action of $ \lamtil_\aldot \mutil^\aldot $ on $e^{iP\cdot x} \ket{0}$ gives exactly the same contribution. In a similar manner, the contribution from the $D$  acting on  $e^{iP\cdot y} \ket{0}$ in the singlet 
 condition relation produces the same result with $x^\mu$ replaced by $y^\mu$. Altogether, the singlet condition with $J^A{}_B=D$ yields 
\begin{align}
\left(  x^\mu {\del \over  \del x^\mu}  +  y^\mu {\del \over  \del y^\mu} +2 \right) I(x-y) =0 \period
\end{align}
The solution is obviously 
\begin{align}
I(x-y) \propto {1\over (x-y)^2} \period
\end{align}

Let us now describe how  the contraction of the fundamental fermions, \ie 
 $\contraction[1ex]{}{\psi_{\alpha a}}{(x)}{\bar{\psi}}
\psi_{\alpha a}(x) \bar{\psi}_{\dot{\alpha}}^b(y)$ can be done 
using the singlet projector. The singlet part for 
 the R-symmetry obviously gives $\delta_a^b$ and hence we have 
\begin{align}
\contraction[1ex]{}{\psi_{\alpha a}}{(x)}{\bar{\psi}}
\psi_{\alpha a}(x) \bar{\psi}_{\dot{\alpha}}^b(y)&\propto \delta_a^{\ b}\langle \boldsymbol{1}_{12}| (e^{iP\cdot x}\lambda_{\alpha}|0\rangle_D \otimes e^{iP\cdot y}\tilde{\lambda}_{\dot{\alpha}}|0\rangle_D) \period
\end{align}
In this case, we may use the crossing relation (\ref{osccrossing}) for the 
 oscillators  to rewrite the RHS as 
\begin{align}
\langle \boldsymbol{1}_{12}| (e^{iP\cdot x}\lambda_{\alpha}|0\rangle_D \otimes e^{iP\cdot y}\tilde{\lambda}_{\dot{\alpha}}|0\rangle_D)&=\langle \boldsymbol{1}_{12}| (e^{iP\cdot x}|0\rangle_D \otimes e^{iP\cdot y}\lambda_{\alpha}\tilde{\lambda}_{\dot{\alpha}}|0\rangle_D)=-i\frac{\partial}{\partial y^{\dot{\alpha}\alpha}} I(x-y) 
\end{align}
Therefore, up to an overall  normalization, we obtain 
\begin{align}
\contraction[1ex]{}{\psi_{\alpha a}}{(x)}{\bar{\psi}}
\psi_{\alpha a}(x) \bar{\psi}_{\dot{\alpha}}^b(y)
& \propto 2i \delta_a^b {(x-y)_{\al\aldot} \over |x-y|^4}
\end{align}

Likewise, the Wick contraction for the self-dual field strength can be computed,  again using the crossing relations for the oscillators, as 
\begin{align}
\contraction[1ex]{}{F_{\alpha \beta}}{(x)}{\bar{F}_{\dot{\alpha}\dot{\beta}}}
F_{\alpha \beta}(x)\bar{F}_{ \dot{\alpha} \dot{\beta}}(y)&\propto  \langle \boldsymbol{1}_{12}| (e^{iP\cdot x}\lambda_{\alpha}\lambda_{\beta} |0\rangle_D \otimes e^{iP\cdot y}\tilde{\lambda}_{\dot{\alpha}} \tilde{\lambda}_{\dot{\beta}}|0\rangle_D) \nn\\
&=  -\frac{1}{2}\Big( \langle \boldsymbol{1}_{12}| (e^{iPx}\lambda_{\alpha}\tilde{\lambda}_{\dot{\alpha}} |0\rangle_D \otimes e^{iPy} \lambda_{\beta}\tilde{\lambda}_{\dot{\beta}}|0\rangle_D) \nn\\
&\qquad +\langle \boldsymbol{1}_{12}| (e^{iPx}\lambda_{\alpha} \tilde{\lambda}_{\dot{\beta}}|0\rangle_D \otimes e^{iPy} \lambda_{\beta}\tilde{\lambda}_{\dot{\alpha}}|0\rangle_D)\Big) \nn\\
&= \frac{1}{2}\left( \frac{\partial}{\partial x^{\dot{\alpha}\alpha}}\frac{\partial}{\partial y^{\dot{\beta}\beta}}+\frac{\partial}{\partial x^{\dot{\beta}\alpha}}\frac{\partial}{\partial y^{\dot{\alpha}\beta}}\right) I(x-y) \nn\\
&= -6 {(x-y)_{(\al \aldot} (x-y)_{\be)\bedot} \over |x-y|^6} \period
\end{align}
Normalizations of these two point functions depend of course on the choice of 
the normalization of the individual fields but once we fix one of them, then the rest can  be determined by supersymmetry. 

With the demonstrations above,  we wish to emphasize   that our method of using the conformally covariant  D-scheme is quite simple and useful  in that the properties of the singlet projector can be  directly and effectively utilized. 
\section{Monodromy relations for correlation functions in psu(2,2$|$4) spin chain system}
Having constructed the singlet projector in the conformally covariant basis, 
 we shall now generalize the so-called monodromy relations  for the correlation functions developed in our previous work\cite{KK2} for the SU(2) sector to the full psu(2,2$|$4) sector. 
Here one must first note the following 
new features. In the case of the  SU(2) sector, 
the structure of the auxiliary Hilbert space is unequivocally chosen to be identical to that of  the quantum Hilbert space, both two dimensional,  describing  the up and down ``spin" states. On the other hand, for psu(2,2$|$4) there are 
 two appropriate  choices for the auxiliary space. To see this, we should 
recall the properties of the general R-matrix, to be denoted by 
$\mathbb{R}_{ij}(u)$, from which the monodromy matrix  is constructed. 
It is a linear map acting on the tensor product of 
 two vector spaces $V_i \otimes V_j$,  \ie $\mathbb{R}_{ij} \in {\rm End}(V_i \otimes V_j) $,   and satisfying the following 
 Yang-Baxter equation:
\begin{align}
\mathbb{R}_{12}(u_1-u_2)\mathbb{R}_{13}(u_1)\mathbb{R}_{23}(u_2)=\mathbb{R}_{23}(u_2) \mathbb{R}_{13}(u_1)\mathbb{R}_{12}(u_1-u_2) \comma \label{eq:YB_eq} 
\end{align}
where complex parameters $u_i$ are the spectral parameters. From 
 such  $\mathbb{R}_{ij}(u)$ matrices, one constructs the monodromy 
 matrix $\Omega(u) =\mathbb{R}_{a1}(u)\cdots \mathbb{R}_{a\ell}(u)$,
where $a$ here is the label for the auxiliary space $V_a$ and the 
numbers $1$ through $\ell$ denote the location of the site at which 
u(2,2$|$4) spin state resides to make up a spin chain\footnote{For the discussion of concepts requiring the Yang-Baxter equation, we must consider 
u(2,2$|$4), but not  psu(2,2$|$4),  as it is the R-matrix associated with 
the former which  satisfies the Yang-Baxter equation. We shall give more detailed discussion on this point later.} . Then, out of the 
monodromy matrix, one defines the transfer matrix $T_a(u)$ 
by  taking the trace over the auxiliary space, namely 
$T(u) := \trace_a \Omega(u)$. The prime importance of the Yang-Baxter equation (\ref{eq:YB_eq}) is that it ensures the commutativity of the 
 transfer matrices at different spectral parameters, \ie $\com{T(u)}{T(v)}=0$, which in turn implies that the quantities obtained as the coefficients of
 the power expansion in the spectral parameter  all commute. In particular, 
as one of such quantities  can be identified with the Hamiltonian of the spin chain, all the coefficients  can be interpreted as conserved charges. This is 
 at the heart of the integrability. 

Now in the case of  the u(2,2$|$4) spin chain, while the quantum Hilbert space is taken to be the Fock space $\calV$ constructed by the oscillators introduced in the previous section, there are two natural choices for  the auxiliary space $V_a$,  which should form  a representation of  u(2,2$|$4) or 
its complexified version gl(4$|4$)\footnote{The most  of the discussion to follow is insensitive to whether we consider u(2,2$|$4)
 or its complexified version gl(4$|4$). Thus, when the description is 
 easier with the complexified version, we shall use gl(4$|4$) in place of 
u(2,2$|$4).}.  
 One is the fundamental representation 
 of gl(4$|4$), \ie  $V_a=\mathbb{C}^{4|4}$ and the other is the 
 choice $V_a = \calV$, \ie  the auxiliary space being  the same as the 
infinite dimensional quantum space in structure. We shall call 
 the corresponding R-matrix  as  ``fundamental"  for the former case 
 and ``harmonic"  for the latter choice. 

For the former case, 
the monodromy matrix is finite dimensional and its  components 
 are operators acting on the quantum space. These  components satisfy the
 exchange relations (or Yang-Baxter algebra) coming from the Yang-Baxter 
 equations, and  are  quite powerful in diagonalizing the transfer matrix 
in the context of  algebraic Bethe ansatz.  It should be noted that a similar 
 finite dimensional monodromy matrix can be defined classically in the 
strong coupling regime using the flat connections of the 
string sigma  model and can be  used to determine the semi-classical spectrum\cite{BKSZ}.  Further, beyond the spectral problem, the monodromy relation of this type has its counterpart in the computation of the three-point functions in the strong coupling regime\cite{JW}, \cite{KK1, KK2, KK3} as the 
triviality of the total monodromy of the form $\Omega_1\Omega_2\Omega_3=1$, where $\Omega_i$ is the local monodromy produced around the $i$-th  vertex operator  in the so-called  auxiliary linear problem. As explained in 
\cite{JW, KK3}, this seemingly weak relation is disguisingly powerful, as it 
captures the important global information governing the three-point functions. 

 Such  monodromy relations  for the fundamental R-matrix 
for gl(4$|4$) can be derived through a procedure similar to the one 
for the SU(2) case worked out in detail  in our previous work\cite{KK2} and has  been discussed in \cite{spinvertex}. Besides the purpose of completeness, we shall re-derive these relations below since we shall use the 
 definition of the Lax operator,   slightly different from the one used in \cite{spinvertex},  which is more natural in connection with the strong coupling counterpart. 

Next let us briefly describe the characteristics of the monodromy relations we shall derive for the harmonic R-matrix, which are completely new. In this case, the monodromy matrix  is no longer finite dimensional since the auxiliary space $V_a$ is the same  as the infinite dimensional quantum spin-chain Hilbert space $\calV$.  One of the  virtues of considering such a harmonic R-matrix is that,  just as in the  case of the SU(2) Heisenberg spin chain, the construction of the conserved charges including the Hamiltonian is much easier, since due to the identical  structure of $V_a$ and $\calV$  the R-matrix at specific value of the spectral parameter serves as the permutation 
 operator ${\rm P}_{an}$. Such an operator is known to be extremely useful 
in extracting the Hamiltonian (\ie the dilatation operator).  
Because of this and other features, the harmonic R-matrix and the related quantities have already found an interesting applications in  the computation of the  scattering amplitudes  from the point of view of integrability\cite{HarmonicR1,HarmonicR2,spectralreg3,spectralreg4,CK,CDK,BdR1,BdR2} and are expected to be useful in the realm of the correlation 
 functions as well. 

In any case, since the monodromy matrix, constructed out of either ``fundamental"  or ``harmonic" R-matrices,  is a generating function of an infinite number  of conserved charges, the monodromy relations can be regarded 
as a collection of ``Ward identities" associated with such  higher charges, 
which should characterize the important properties of the correlation 
 functions. 
\nxt
\underline{Remarks on the relevance of  of u(2,2$|4$) for the monodromy relation}\\ \underline{ and  psu(2,2$|4$) for the singlet projector } 
\nxt
Before we begin the discussion of the monodromy relations, let us 
give some  important clarifying remarks  on the relevance of the different 
super algebras for the two topics we discuss in this work and the role of 
 their oscillator representation. 
\begin{itemize}
	\item The monodromy relation, to be discussed below, is deeply 
 rooted in  the integrability of the theory and hence it is crucial that 
the relevant R-matrix and the Lax matrix must satisfy the  Yang-Baxter equations and the RLL=LLR equations. A method has long been known\cite{Kulish:1981gi,Reshetikhin:1983vw, Reshetikhin:1985vd,Ogievetsky:1986hu} that one can 
construct such an R-matrix and a Lax matrix from a suitable Lie super algebra.  In the present case, one can do so  for u(2,2$|4$) algebra but not 
 for su(2,2$|4$) or psu(2,2$|4$). This is a general mathematical statement and has nothing to do with a  particular  oscillator representation  nor with the super Yang-Mills theory. However, when one makes use of  the singleton 
oscillator representation, then one can easily construct the states which form 
 the fundamental Field strength multiplet of SYM theory and the R-matrix and the monodromy matrix can be constructed in terms of the generators 
bilinear in the oscillators.  Although the basic SYM fields carry a special value 
 of the central charge $C=0$  and the global symmetry of the $\mathcal{N}=4$ SYM 
 theory is psu(2,2$|4$),  still when we discuss the monodromy relations 
for the correlation functions for the composite operators made up of these 
 SYM fields,  the generators and the related quantities to be used must be those of u(2,2$|4$). 
\item On the other hand, when we use the singlet projector to perform the Wick contractions efficiently in the computation of the correlation functions, 
 the projector is a singlet for su(2,2$|4$) and  psu(2,2$|4$). This notion 
 has nothing to do with the integrability. 
In fact a singlet projector for u(2,2$|4$) does not exist at least in the oscillator representation utilized and this point gives a subtle effect in the crossing relation, to be discussed in the next subsection. 
\end{itemize}
Thus, in the  monodromy relations for the correlation functions, 
two different superalgebras are playing their respective role. The monodromy matrices to be inserted are associated with u(2,2$|4$), while the singlet 
 projector which works as an elegant device in forming  the correlation function for the physical SYM fields is valid  for psu(2,2$|4$).

\subsection{Basic monodromy relation in the case of fundamental R-matrix}
Let us begin with the case of the monodromy relations with the use of the 
 fundamental R-matrix. We shall first give the definitions and conventions 
 for the fundamental R-matrix and the associated Lax matrix, which 
are slightly different from the ones used in \cite{spinvertex}, and then
 discuss the two important relations, namely the crossing relations 
 and the inversion relations, which will lead immediately to the 
 monodromy relations of interest. 
\subsubsection{Fundamental R-matrix and Lax operator}
Consider the  fundamental  R-matrix, for which the  quantum space is $\mathcal{V}$ and the auxiliary space is taken to be $\mathbb{C}^{4|4}$. This kind of  R-matrix is often called the Lax operator and will be denoted by  $L_{an}(u)$, where $a$ and $n$ refer, respectively, to the auxiliary space and the position on the spin chain.  It satisfies the important relation called RLL=LLR relation
\begin{align}
R_{12}(u_1-u_2)L_{a_1n}(u_1)L_{a_2n}(u_2)=L_{a_2n}(u_2)L_{a_1n}(u_1)R_{12}(u_1-u_2) \comma 
\end{align}
which follows from the basic Yang-Baxter equation (\ref{eq:YB_eq}) by setting $V_1=V_2=\mathbb{C}^{4|4}$ and $V_3=\mathcal{V}_n$, where $n$ is the position of the spin.  The R-matrix $R_{12}(u)$ appearing in this 
 equation acts on the tensor product of two copies of the auxiliary space $V_1\otimes V_2$ and,  besides the RLL=LLR equation,  it also satisfies the 
original Yang-Baxter equation denoted as RRR=RRR equation:
\begin{align}
R_{12}(u_1-u_2)R_{13}(u_1)R_{23}(u_2)=R_{23}(u_2)R_{13}(u_1)R_{12}(u_1-u_2) \period \label{eq:RLL=LLR}
\end{align}
The solution of the above RRR relation turns out to be of the form 
\begin{align}
R_{ij}(u)= u+\eta(-1)^{|B|}E^A_{i\ B}\otimes E^B_{j\ A} \comma \qquad  (E^A_{\ B})^C_D\equiv \delta^A_C \delta^B_D \comma
\end{align}
where $\eta$ is an arbitrary complex parameter\footnote{
Although the Yang-Baxter equation holds for arbitrary 
 $\eta$, we will later set $\eta =\pm i$ for our interest. } and $E^A_{i\ B}$ is the fundamental representation of gl(4$|$4) acting non-trivially on $V_i\cong \mathbb{C}^{4|4}$.  To check that  the R-matrix above actually satisfies the Yang-Baxter  it is useful to note that the operator  $\Pi_{ij}:=(-1)^{|B|}E^A_{i\ B}\otimes E^B_{j\ A}$  serves as the graded permutation operator\footnote{To prove this, we should pay attention to the non-trivial  gradings between two auxiliary spaces 
\begin{align}
(E^A_{\ B}\otimes E^C_{\ D})(a\otimes b)=(-1)^{a(|C|+|D|)} (E^A_{\ B}a)\otimes (E^C_{\ D}b) \ \ a\otimes b \in \mathbb{C}^{4|4}\otimes \mathbb{C}^{4|4} \comma \\
(E^A_{\ B}\otimes E^C_{\ D})(E^E_{\ F}\otimes E^G_{\ H})=(-1)^{(|C|+|D|)(|E|+|F|)}(E^A_{\ B}E^E_{\ F})\otimes (E^C_{\ D}E^G_{\ H}) \period
\end{align}
.}. For example, $\Pi_{12}(a\otimes b\otimes c)=(-1)^{|a||b|}(b\otimes a\otimes c)$, $\Pi_{13}(a\otimes b\otimes c)=(-1)^{|a|(|b|+|c|)+|b||c|} (c\otimes b\otimes a)$ and so on. Then, the Lax operator satisfying (\ref{eq:RLL=LLR}) is given by
\begin{align}
L_{a_in}(u)= u+\eta(-1)^{|B|} E^A_{i\ B} \otimes J^A_{n\ B} \comma
\end{align}
where $J^A_{n\ B}$'s are the generators of gl(4$|$4) defined on the $n$-th site of the spin chain. It is tedious but straightforward to show that the Lax operator indeed satisfies the RLL=LLR relation,  by explicitly computing the both sides. In performing this calculation, one should remember that  there are no grading relations  between the auxiliary space and quantum spaces, which are 
 two independent spaces. Explicitly, this means 
\begin{align}
(E^A_{\ B}\otimes J^A_{\ B})(E^C_{\ D}\otimes J^C_{\ D})=(E^A_{\ B}E^C_{\ D})\otimes (J^A_{\ B}J^C_{\ D})=E^A_{\ D}\otimes (J^A_{\ B}J^B_{\ D}) \comma
\end{align}
where we have used $E^A_{\ B}E^C_{\ D}=\delta^B_{\ C}E^A_{\ D}$. 
Differently put,  the definition of the product of the Lax operators  is not as supermatrices  but as usual matrices.  Although the choice  for the mutual grading between these two spaces is a matter of convention\footnote{In \cite{spinvertex}, the authors adopt the convention where non-trivial gradings between the auxiliary space and the quantum space exist.  Namely, $(E^A_{\ B}\otimes J^A_{\ B})(E^C_{\ D}\otimes J^C_{\ D})=(-1)^{(|A|+|B|)(|C|+|D|)}(E^A_{\ B}E^C_{\ D})\otimes (J^A_{\ B}J^C_{\ D})$. Because of this, the definition of the  Lax operator they use, \ie   $L(u):=u-i/2-i(-1)^{|A|}E^A_{\ B} \otimes J^B_{\ A}$, is slightly different from ours. },   our choice is a natural one 
 from the point of view of connecting to the strong coupling regime. 
This is simply because  the monodromy matrix at  strong coupling is defined by the path ordered exponential of the integral of the flat connection and the multiplication rule for such matrices is  the ordinary one. 
With this convention, the explicit form of the Lax operator is given in terms of
 the superconformal generators by 
\begin{align}
(L(u))^A_{\ B}&=u\delta^A_{\ B}+\eta (-1)^{|B|} J^A_{\ B}= \left( 
\begin{array}{cc|c}
u+\eta Y_{\alpha}^{\ \beta} & i\eta P_{\alpha \dot{\beta}} & -\eta Q_{\alpha}^b \\
i\eta K^{\dot{\alpha} \beta} & u+\eta Y^{\dot{\alpha}}_{\ \dot{\beta}} & -i\eta \bar{S}^{\dot{\alpha} b} \\
\hline
\eta S_a^{\beta} & i\eta \bar{Q}_{\dot{\beta} a} &u-\eta W_a^b
\end{array} \right)_{AB} \period  \label{eq:gl44Lax}
\end{align}

As usual the monodromy matrix is defined as the product of the Lax operators on each site going around the spin chain of length $\ell$:
\begin{align}
\Omega_a (u) := L_{a1}(u) \cdots L_{a\ell}(u) \period
\end{align}
The monodromy matrix so defined satisfies the following relation,  since  each Lax operator satisfies the RLL=LLR relation:
\begin{align}
R_{12}(u_1-u_2)\Omega_{a_1}(u_1)\Omega_{a_2}(u_2)=\Omega_{a_2}(u_2)\Omega_{a_1}(u_1) R_{12}(u_1-u_2) \period
\end{align}
If we write out  the above equation for each component, we obtain the so-called Yang-Baxter exchange  algebra. In the rest of this subsection, when there is no confusion we drop the indices for the auxiliary space for simplicity.
\subsubsection{Monodromy relation}
Let us now derive the generic monodromy relation. This can be achieved 
by proving the following two important relations for the Lax operators, called  the crossing relation and the inversion relation.  They are respectively of the form 
\begin{align}
(\mathrm{C})&: \ \ \langle \boldsymbol{1}_{12}| L_{n}^{(1)}(u)=-\langle \boldsymbol{1}_{12}| L_{\ell -n+1}^{(2)}(\eta -u) \comma \label{eq:crossing1} \\
(\mathrm{I})&: \ \ L_{n}^{(i)}(u)L_{n}^{(i)}(v)=u(\eta -u) \comma \ \ (u+v=\eta) \comma  \label{eq:inversion1}
\end{align}  
where the superscript $(i)$ on $L^{(i)}_n$ denotes the $i$-th 
 spin chain. 
The crossing relation (C) connects the Lax operator defined on the $n$-th site of a spin chain called  1  to that defined on the $\ell-n+1$-th site of  another spin chain called 2.  To get the feeling for the crossing relation,  it suffices to recall 
 that the singlet projector $\langle \boldsymbol{1}_{12}|$ effects  the Wick contraction  between a field at the $n$-th site of 
 one spin chain and a  field  at the $\ell -n+1$-th site of another chain. 
Actually, it is easy to prove it more precisely from the defining property of 
 the singlet $\langle \boldsymbol{1}_{12}|$. As it was already emphasized 
 in section 2.2, the operator  $\langle \boldsymbol{1}_{12}|$ is a singlet 
projector   for su(2,2$|$4) or psu(2,2$|$4) but {\it not}  for u(2,2$|$4) which  is of our concern here. So the operator $\langle \boldsymbol{1}_{12}|$  
transforms the generator $J^A{}_B$ of u(2,2$|$4) acting on the 
first spin chain into the operator $ - J^A_{\ B}-(-1)^{|A|}\delta^A_{\ B}$ acting on  the second spin chain, where the constant piece $-(-1)^{|A|}\delta^A_{\ B}$ comes from the (anti-)commutator term. Applying this to the Lax 
 operator  $(L(u))^A_{\ B}=u\delta^A_{\ B}+\eta (-1)^{|B|}J^A_{\ B}$, 
 one sees that the constant  term shifts the spectral parameter by $\eta$ and 
 we get the crossing relation as shown in (\ref{eq:crossing1}). 

The proof of the inversion relation (I), which says that  the Lax operator can be inverted for a specific value of the spectral parameter,  is slightly more involved. 
The product of two  Lax operators gives 
\begin{align}
(L(u)L(v))^A_{\ B}= uv\delta^A_{\ B}+\eta(u+v)(-1)^{|B|}J^A_{\ B} +\eta^2(-1)^{|B|+|C|}J^A_{\ C}J^C_{\ B} \period \label{eq:prod_Lax}
\end{align} 
First look at the last term quadratic in the generators. In general this cannot be simplified further. However, as we are specializing in the oscillator representation, we can write  $J^A_{\ B}=\bar{\zeta}^A\zeta_B$ using the oscillators  satisfying $[\zeta_A ,\bar{\zeta}^B]=\delta_A^{\ B}$, as described in (\ref{eq:oscilJ}). Therefore 
we can reduce the  product of the generator in the following way:
\begin{align}
(-1)^{|C|}J^A_{\ C}J^C_{\ B}=\bar{\zeta}^A(\bar{\zeta}^C\zeta_C)\zeta_B=(2C-1)J^A_{\ B} \period
\end{align}
For the first equality, we have used the fact that the constant term 
$(-1)^{|C|}\delta^C_C$,  which appears from the commutation relation,  vanishes. Further, since $[ \bar{\zeta}^C\zeta_C , \bar{\zeta}^A]=1$ and the central charge is given by $2C=\mathrm{tr}J=\bar{\zeta}^C\zeta_C$, we obtain the result above. Hence  the product of the Lax operators (\ref{eq:prod_Lax}) is simplified to 
\begin{align}
(L(u)L(v))^A_{\ B}= uv\delta^A_{\ B}+\eta(u+v+\eta(2C-1))(-1)^{|B|}J^A_{\ B} \period \label{eq:LuLv}
\end{align}
Now as we repeatedly emphasized, for the Yang-Mills fields of our interest 
we can set $C=0$ and hence the coefficient in front of $(-1)^{|B|} J^A{}_B$ becomes $\eta(u+v-\eta)$. Therefore when $u+v=\eta$, the RHS of 
(\ref{eq:LuLv}) becomes $uv \delta^A_B = u (\eta -u)\delta^A_B$, which is 
precisely the RHS of the inversion equation (\ref{eq:inversion1}). 

We are now ready to present the generic  form of the monodromy relation,
 which takes the form 
\begin{align}
\langle \boldsymbol{1}_{12}|\Omega^{(1)}(u) \Omega^{(2)}(u) =
 \langle \boldsymbol{1}_{12}| F_\ell(u) \comma  \label{eq:monodrel}
\end{align}
where $F_\ell(u)$ is some function of $u$, to be given shortly. Both sides 
of this relation should be understood as acting on  a tensor product of states on two Hilbert spaces of the form $\ket{\calO_1} \otimes \ket{\calO_2}$. 
To show (\ref{eq:monodrel}), we first prove the following relation with the use of the 
 crossing relation (\ref{eq:crossing1}):
\begin{align}
&\langle \boldsymbol{1}_{12}|\Omega^{(1)}(u)=(-1)^{\ell} \langle \boldsymbol{1}_{12}|\overleftarrow{\Omega}^{(2)}(\eta-u) \comma \label{eq:cross_mono1} \\
&\overleftarrow{\Omega}^{(2)}(u):=L^{(2)}_{\ell }(u)\cdots L^{(2)}_1(u) \period
\end{align}
Focus first on the LHS of (\ref{eq:cross_mono1}) and consider moving the 
Lax operator $L^{(1)}_n(u)$ at the $n$-th site in $\Omega^{(1)}$ to the left  towards $\langle \boldsymbol{1}_{12}|$. Since the components of the 
 Lax operators on different sites commute in the graded sense, namely 
\begin{align}
(L_{n}(u))_{AB}(L_{m}(v))_{CD}=(-1)^{(|A|+|B|)(|C|+|D|)}(L_{m}(v))_{CD}(L_{n}(u))_{AB} \comma 
\end{align}
we can move $L^{(1)}_n(u)$ all the way to the left and hit  $\langle \boldsymbol{1}_{12}|$ like 
\begin{align}
\langle \boldsymbol{1}_{12}|\ldots (L_n^{(1)}(u))_{AB}\cdots =(-1)^{(|A|+|B|)(\ldots)}\langle \boldsymbol{1}_{12}|(L_n^{(1)}(u))_{AB}\ldots \cdots \period
\end{align}
We can now use the crossing relation (C) to replace the Lax operator 
 $(L^{(1)}_n(u))_{AB}$ with  $(L_{\ell -n+1}^{(2)}(\eta -u))_{AB}$
and move it back again to the original position. In this process, 
the sign factors 
which appear through the exchange of operators exactly cancel with those produced in the previous process and 
 we get 
\begin{align}
\langle \boldsymbol{1}_{12}|\cdots L_n^{(1)}(u)\cdots =-\langle \boldsymbol{1}_{12}|\cdots L_{\ell-n+1}^{(2)}(\eta -u)\cdots  \period
\end{align}
Repeating this to all the Lax operators making up  the monodromy matrices, 
 we  immediately get (\ref{eq:cross_mono1}). Now apply $\Omega^{(2)}(u)$  to the both sides of (\ref{eq:cross_mono1}) and use the relation 
 $\overleftarrow{\Omega}^{(2)}(\eta -u) \Omega^{(2)}(u) \propto 1$, 
with the overall factor which is readily computable using the inversion relation (I). In this way we obtain (\ref{eq:monodrel}) with the function 
$F_\ell(u)$ given by $F_\ell(u) = (u(u-\eta))^\ell$. 
Now if we apply (\ref{eq:monodrel})  explicitly to the state  $\ket{\psi_1} \otimes \ket{\psi_2}$, we obtain the more explicit monodromy relation
 for the two-point function
\begin{align}
(-1)^{(|C|+|B|)|\psi_1|}\left\langle (\Omega^{(1)}(u))_{AB} |\psi_1 \rangle \comma (\Omega^{(2)}(u))_{BC} |\psi_2 \rangle \right\rangle &= F_{\ell}(u)\delta_{AC} \left\langle  |\psi_1 \rangle \comma |\psi_2 \rangle \right\rangle \comma \label{eq:psu(2,2|4)_mono} \\
F_\ell(u) &=  (u(u-\eta))^\ell \period 
\end{align}
Here  $\langle \comma \rangle$ denotes  the pairing with the singlet, which gives the Wick contraction between two operators.  The sign in front arises when we pass the monodromy through the first state $|\psi_1 \rangle$. 

At this point, it is of  importance to remark that we obtain the usual Ward identities at the leading order in the expansion of the above equation around $u=\infty$. This is a direct consequence of the su(2,2$|$4) invariance of the singlet projector. 

Once the monodromy relation is obtained for the two-point functions, the one for the three-point functions can be obtained easily, just as was shown explicitly for the SU(2) sector in \cite{KKN2}. The only differences from that case are the form of the prefactor function $F_{123}(u$) and some sign factors due to the superalgebra nature of psu(2,2$|$4). Thus the monodromy relation for the three-point function takes the form
\begin{align}
\begin{split}
&\left\langle (\Omega (u))_{AB} |\psi_1 \rangle , (-1)^{|\psi_1|(|B|+|C|)} (\Omega (u))_{BC} |\psi_2 \rangle ,  (-1)^{(|\psi_1|+|\psi_2|)(|C|+|D|)} (\Omega (u))_{CD} |\psi_3 \rangle \right\rangle \\
&= F_{123}(u) \delta_{AD} \left\langle |\psi_1 \rangle , |\psi_2 \rangle , |\psi_3 \rangle  \right\rangle  \comma \qquad 
F_{123}(u)= (u(u-\eta))^{\ell_1+\ell_2+\ell_3} \period \label{eq:Fmono_3pt}
\end{split}
\end{align}
\subsection{Basic monodromy relation in the case of harmonic R-matrix}
We shall now discuss another important version of the R-matrix, called 
 the harmonic  R-matrix, to be denoted by the  bold letter {\bf R}. We shall  derive the inversion and the crossing  relations for it and finally prove  the relations for the correlation functions obtained  with the insertion of monodromy matrices constructed out of the harmonic R-matrices. 

The term ``harmonic" stems from the form of the Hamiltonian (or dilatation 
 ) density  first derived in \cite{1-loop}, which can be expressed as\footnote{The subscript $12$ signifies that the Hamiltonian is restricted to two fields 1 and 2.} $H_{12} = h(J_{12})$, where the function $h(j)=\sum_{k=1}^j 1/k$ is the so-called  harmonic number  and $J_{12}^2$ is the quadratic 
Casimir operator. This Hamiltonian is intimately related to the one in the 
SL(2) subsector and in that context was  derived also as the logarithmic derivative of the R-matrix, just as in the case of the SU(2) Heisenberg spin chain. 

The harmonic R-matrix is recently applied  in the context of the scattering amplitudes for  the $\mathcal{N}=4$ SYM theory,  as a tool to construct  the building blocks for the deformed Grassmannian formulas  characterized 
 as Yangian invariants \cite{HarmonicR1,HarmonicR2,spectralreg3,spectralreg4,CK,CDK,BdR1,BdR2}. The spectral parameter can be  naturally introduced in the deformed formulas and turned  out to serve
  as a regulator for the IR divergences.  

Since this type of R-matrix is less well-known, we shall first give a brief review of 
 the basic facts on the harmonic R-matrix following  \cite{HarmonicR2}
and then using these properties derive the crossing and inversion relations, 
which are essential, as in the case of the fundamental R-matrix,  in obtaining the monodromy relations we seek. 
\subsubsection{Review of the harmonic R-matrix}
The harmonic R-matrix $\boldsymbol{\mathrm{R}}_{12}$ acting on the tensor product of two copies of the Fock space $\mathcal{V}_1\otimes \mathcal{V}_2$ should satisfy the following RLL=LLR relation
\begin{align}
\boldsymbol{\mathrm{R}}_{12}(u_1-u_2)L_1(u_1)L_2(u_2)=L_2(u_2)L_1(u_1)\boldsymbol{\mathrm{R}}_{12}(u_1-u_2) \period \label{eq:harmonic_YB}
\end{align}
This is obtained  from the general formula (\ref{eq:YB_eq}) by setting $V_1=\mathcal{V}_1$, $V_2=\mathcal{V}_2$,  $V_3=\mathbb{C}^{4|4}$ and replacing $\mathbb{R}_{i3}(u_i)$ with the Lax operator $L_i(u_i)=u_i+\eta (-1)^{|B|}E^A_{\ B}\otimes J^A_{i\ B}$. Renaming $u=u_1-u_2$, $u_2=\tilde{u}$ and expanding the above equation in powers of $\tilde{u}$, one can show  that the harmonic R-matrix satisfies the following two types of equations:
\begin{align}
&(i)\ \ [ \boldsymbol{\mathrm{R}}_{12}(u), J^A_{1\ B}+J^A_{2\ B}]=0 \comma \label{eq:gl_inv} \\
&(ii)\ \ (-1)^{|C|}\eta (\boldsymbol{\mathrm{R}}_{12}(u)J^A_{1\ C}J^C_{2\ B}-J^A_{2\ C}J^C_{1\ B}\boldsymbol{\mathrm{R}}_{12}(u))-u(J^A_{2\ B}\boldsymbol{\mathrm{R}}_{12}(u)-\boldsymbol{\mathrm{R}}_{12}(u)J^A_{2\ B})=0 \period \label{eq:level1_inv}
\end{align}
The first expresses  the invariance of the harmonic R-matrix 
under gl(4$|$4), while the second  implies the invariance under the level 1 generators of the Yangian algebra. They together  ensure the full Yangian invariance of the harmonic R-matrix. 
As it can be explicitly verified after constructing $\boldsymbol{\mathrm{R}}_{12}$ explicitly, the product $\boldsymbol{\mathrm{R}}_{12}(u) 
\boldsymbol{\mathrm{R}}_{12}(-u)$ is proportional to the unit operator ${\bf 1}_{12}$ but the overall normalization can be arbitrary, since the 
equations $(i)$ and $(ii)$ above are both linear in ${\bf R}$. Therefore 
one can impose the following unitarity condition, or inversion relation,  to fix the overall scale:
\begin{align}
\boldsymbol{\mathrm{R}}_{12}(u)\boldsymbol{\mathrm{R}}_{12}(-u)=\boldsymbol{1}_{12} \period \label{eq:unitarity}
\end{align}
Conversely,  the solution satisfying  (\ref{eq:gl_inv})-(\ref{eq:unitarity})
 is unique, as we demonstrate later. 

To actually find the form of $\boldsymbol{\mathrm{R}}_{12}(u)$, we will first  solve  the equation (\ref{eq:gl_inv}). For this purpose, it is convenient  to introduce the following notations for  sets of oscillators:
\begin{align}
\bar{\alpha}^{\mathsf{A}}= \left( \begin{array}{c} \lambda_{\alpha} \\ \bar{c}_i \end{array} \right) \comma \ \ \alpha_{\mathsf{A}}= \left( \begin{array}{c} \mu^{\alpha} \\ c^i \end{array} \right) \comma \label{eq:oscillator_al} \\
\bar{\beta}^{\dot{\mathsf{A}}}= \left( \begin{array}{c} \tilde{\lambda}_{\dot{\alpha}} \\ \bar{d}_i \end{array} \right) \comma \ \ 
\beta_{\dot{\mathsf{A}}}= \left( \begin{array}{c} \tilde{\mu}^{\dot{\alpha}} \\ d^i \end{array} \right) \period \label{eq:oscillator_be}
\end{align}
Notice that $\alpha_{\mathsf{A}}|Z\rangle =\beta_{\dot{\mathsf{A}}}|Z\rangle =0$.
These oscillators transform covariantly under the subalgebras gl(2$|$2) $\oplus$ gl(2$|$2) $\subset$ gl(4$|$4), whose bosonic parts are given by the Lorentz and su(2)$_L \oplus$ su(2)$_R$ R-symmetry subalgebras. In other words, the indices $\mathsf{A}, \mathsf{B}, \ldots$ are associated with  the (anti-) fundamental representation of one gl(2$|$2) and the indices $\dot{\mathsf{A}}, \dot{\mathsf{B}}, \ldots$ describe the (anti-) fundamental representation of the other gl(2$|$2).
 Accordingly, the gl(4$|$4) generators are decomposed into diagonal parts and the off-diagonal parts  with respect to these two gl(2$|$2) subalgebras in the 
 following form:
\begin{align}
J^A_{\ B} \longrightarrow 
\left( \begin{array}{cc}
J^{\mathsf{A}}_{\ \mathsf{B}} & J^{\mathsf{A}}_{\ \dot{\mathsf{B}}} \\
J^{\dot{\mathsf{A}}}_{\ \mathsf{B}} & J^{\dot{\mathsf{A}}}_{\ \dot{\mathsf{B}}} \end{array} \right) \period \label{eq:J_deco}
\end{align}
The explicit form of  these generators in terms of $\alpha ,\beta$-oscillators  are given  in \cite{HarmonicR2}, but we shall not write them down here. 

Now using the oscillators above, one introduces  the following  basis of linear operators  acting on $\mathcal{V}_1\otimes \mathcal{V}_2$, which will be useful for solving the condition (\ref{eq:gl_inv}):
\begin{align}
\boldsymbol{\mathrm{Hop}}_{k,l,m,n}^{(12)}&= \ :\frac{(\bar{\alpha}_2 \alpha^1)^k}{k!}\frac{(\bar{\beta}^2 \beta_1)^l}{l!}\frac{(\bar{\alpha}_1 \alpha^2)^m}{m!} \frac{(\bar{\beta}^1 \beta_2)^n}{n!}:\nn \\
&=\frac{1}{k!l!m!n!} \bar{\alpha}_2^{\mathsf{A}_1}\cdots \bar{\alpha}_2^{\mathsf{A}_k} \bar{\beta}^2_{\dot{\mathsf{A}}_1}\cdots \bar{\beta}^2_{\dot{\mathsf{A}}_l} \bar{\alpha}_1^{\mathsf{B}_1}\cdots \bar{\alpha}_1^{\mathsf{B}_m}\bar{\beta}^1_{\dot{\mathsf{B}}_1}\cdots \bar{\beta}^1_{\dot{\mathsf{B}}_n} \nn\\
&\quad  \cdot  \beta_2^{\dot{\mathsf{B}}_n}\cdots \beta_2^{\dot{\mathsf{B}}_1}\alpha^2_{\mathsf{B}_m}\cdots \alpha^2_{\mathsf{B}_1} \beta_1^{\dot{\mathsf{A}}_l}\cdots \beta_1^{\dot{\mathsf{A}}_1}\alpha^1_{\mathsf{A}_k}\cdots \alpha^1_{\mathsf{A}_1} 
 \period \label{eq:Hop}
\end{align}
Here the symbol  $:*:$ in the first line denotes  the normal ordering of the oscillators.
The name  $\boldsymbol{\mathrm{Hop}}$ stems from the following 
 properties of this operator.  Its action transforms  $k+l$ oscillators with label $1$ to those with label $2$ and $m+n$ oscillators with label $2$ to those with label $1$. Thus it effects  a kind of hopping operation.  Note  that these operators are manifestly invariant under the diagonal part of (\ref{eq:J_deco}), namely, $J^{\mathsf{A}}_{\ \mathsf{B}}$ and $J^{\dot{\mathsf{A}}}_{\ \dot{\mathsf{B}}}$, as all the relevant indices in (\ref{eq:Hop}) are contracted. 
Therefore,  the general solution $\boldsymbol{\mathrm{R}}_{12}(u)$ of  (\ref{eq:gl_inv}) should be obtained as a linear combination of $\boldsymbol{\mathrm{Hop}}_{k,l,m,n}^{(12)}$ of the form
\begin{align}
&\boldsymbol{\mathrm{R}}_{12}(u)=\sum_{k,l,m,n} \mathcal{A}_{k,l,m,n}^{(\boldsymbol{\mathrm{N}})}(u) \boldsymbol{\mathrm{Hop}}_{k,l,m,n}^{(12)} \comma \label{eq:RHop}
\end{align}
where  $\boldsymbol{\mathrm{N}}$ stands for the total number operator 
 defined by 
\begin{align}
&\boldsymbol{\mathrm{N}}=\boldsymbol{\mathrm{N}}^{(1)}+\boldsymbol{\mathrm{N}}^{(2)} \comma \ \ \boldsymbol{\mathrm{N}}^{(i)}=\boldsymbol{\mathrm{N}}_{\alpha}^{(i)}+\boldsymbol{\mathrm{N}}_{\beta}^{(i)}=\bar{\alpha}_i^{\mathsf{A}}\alpha^i_{\mathsf{A}}+\bar{\beta}^i_{\dot{\mathsf{A}}}\beta_i^{\dot{\mathsf{A}}} \period
\end{align}
Note that the coefficients $\mathcal{A}_{k,l,m,n}^{(\boldsymbol{\mathrm{N}})}(u) $ can depend in general on the spectral parameter $u$,  the total number operator $\boldsymbol{\mathrm{N}}$ and the central charge\footnote{Notice that the central charge is given by $C^{(i)}=\boldsymbol{\mathrm{N}}_{\alpha}^{(i)}-\boldsymbol{\mathrm{N}}_{\beta}^{(i)}$. Since we are interested in the representation in which the central charge vanishes, we neglect dependence on this combination.},  since they all commute with the diagonal generators. As explained in detail in \cite{HarmonicR2}, the invariance under the remaining off-diagonal generators $J^{\mathsf{A}}_{\ \dot{\mathsf{B}}}, J^{\dot{\mathsf{A}}}_{\ \mathsf{B}}$ together with the invariance under the level-one  Yangian generators (\ref{eq:level1_inv}) uniquely fix the coefficients up to an overall coefficient $\rho (u)$  as
\begin{align}
\begin{split}
\mathcal{A}_{k,l,m,n}^{(\boldsymbol{\mathrm{N}})}&=a_{k,l,m,n}\mathcal{A}_{I}^{(\boldsymbol{\mathrm{N}})} \comma \ \ a_{k,l,m,n}:=\delta_{k+n,l+m}(-1)^{(k+l)(m+n)} \comma \\ 
\mathcal{A}_{I}^{(\boldsymbol{\mathrm{N}})}(u)&=\rho(u)(-1)^{I+\frac{\boldsymbol{\mathrm{N}}}{2}}  \mathcal{B}(I, u+\boldsymbol{\mathrm{N}}/2) \comma \ \ \mathcal{B}(x,y):= \frac{\Gamma (x+1)}{\Gamma (x-y+1) \Gamma (y+1)} \period
\end{split} \label{eq:Hop_coefficient}
\end{align}
Here  $I:=\frac{k+l+m+n}{2}$ is an integer since  $k+n=l+m$,  and   $\mathcal{B}(x,y)$ is a natural generalization of the binomial coefficient,  whose arguments can be complex. Notice that it satisfies $\mathcal{B}(x,y)=\mathcal{B}(x,x-y)$ by definition. As already mentioned before, the overall coefficient $\rho (u)$ is determined so that the unitarity condition (\ref{eq:unitarity}) is satisfied. Explicit calculation gives 
\begin{align}
\rho (u)=\Gamma (u+1) \Gamma (1-u) \period  \label{eq:formrho}
\end{align}
An  important characteristic of the harmonic R-matrix, for which the 
quantum and the auxiliary spaces are identical just as for the SU(2) Heisenberg spin chain, is that the R-matrix at $u=0$ 
 yields  precisely the permutation operator. Because of this fact,  through  the well-known manipulation, the Hamiltonian can be extracted from the R-matrix simply as a logarithmic derivative. This is summarized as\footnote{
Equivalently, the expansion of the R-matirx around $u=0$ is of 
 the form  $\boldsymbol{\mathrm{R}}_{12}(u)=\boldsymbol{\mathrm{P}}_{12}(1+u\boldsymbol{\mathrm{H}}_{12}+\cdots)$.}
\begin{align}
\boldsymbol{\mathrm{P}}_{12}&= \boldsymbol{\mathrm{R}}_{12}(0) \comma  \\
\boldsymbol{\mathrm{H}}_{12}&=\frac{d}{du} \ln \boldsymbol{\mathrm{R}}_{12}(u)|_{u=0} \period
\end{align}

\subsubsection{Monodromy relation}
Having reviewed  the basic facts on  the harmonic R-matrix and displayed its explicit form in terms of the oscillators, we now discuss the monodromy relations 
involving such  R-matrices.  As in the case of the 
fundamental R-matrix, the basic ingredients for the derivation  is  (i) the inversion relation and (ii) the crossing relation. 

The inversion relation is already given in (\ref{eq:unitarity}),  together with the computation of the factor $\rho(u)$, shown in (\ref{eq:formrho}),  needed for the normalization. 

As for the crossing relation, its form for the harmonic R-matrix turned out to be similar to (but not identical with)  the one for the Lax matrix for the case of the fundamental R-matrix 
shown  in (\ref{eq:crossing1}) and is given by 
\begin{align}
\langle \boldsymbol{1}_{12}|\boldsymbol{\mathrm{R}}_{an}^{(1)}(u)=\langle \boldsymbol{1}_{12}|\boldsymbol{\mathrm{R}}_{a\ell-n+1}^{(2)}(-u) \period \label{eq:crossing2}
\end{align}
Once this is verified, the crossing relation for the product of harmonic R-matrices is easily  given by  
\begin{align}
\langle \boldsymbol{1}_{12}| \boldsymbol{\mathrm{R}}_{a1}^{(1)}(u) \cdots \boldsymbol{\mathrm{R}}_{a\ell}^{(1)}(u) = \langle \boldsymbol{1}_{12}| \boldsymbol{\mathrm{R}}_{a\ell}^{(2)}(-u) \cdots \boldsymbol{\mathrm{R}}_{a1}^{(2)}(-u) \period \label{eq:cross_mono2}
\end{align}
Now define the monodromy matrix as 
\begin{align}
\Omega^{(i)}_{\boldsymbol{m}\boldsymbol{n}}(u):= \langle \boldsymbol{m}| \boldsymbol{\mathrm{R}}_{a1}^{(i)}(u) \cdots \boldsymbol{\mathrm{R}}_{a\ell}^{(i)}(u)|\boldsymbol{n} \rangle_a \comma 
\end{align}
where $\{ |\boldsymbol{n}\rangle \}$ is a complete set of states in  the auxiliary space satisfying $1=\sum_{\boldsymbol{n}}|\boldsymbol{n}\rangle \langle \boldsymbol{n}|$. Note that the components of the monodromy matrix take values in operators acting on the quantum space. The crossing relation 
 for them follow immediately from  (\ref{eq:cross_mono2})  by 
 taking the matrix element between the states   $\langle \boldsymbol{m}|$ and $|\boldsymbol{n} \rangle$ and is expressed as 
\begin{align}
\langle \boldsymbol{1}_{12}|\Omega^{(1)}_{\boldsymbol{m}\boldsymbol{n}}(u) =\langle \boldsymbol{1}_{12}| \overleftarrow{\Omega}^{(2)}_{\boldsymbol{m}\boldsymbol{n}}(-u) \period
\end{align}
Now contract this relation with $\overleftarrow{\Omega}^{(2)}_{\boldsymbol{n}\boldsymbol{l}}(-u) $ and sum over $\boldsymbol{n}$. Then, since 
the completeness of  the states $\{\ket{\boldsymbol{n}}\}$ in the auxiliary space implies  $\overleftarrow{\Omega}^{(2)}_{\boldsymbol{k}\boldsymbol{l}}(-u)\Omega^{(2)}_{\boldsymbol{l}\boldsymbol{m}}(u)=\delta_{\boldsymbol{k}\boldsymbol{m}}$, we obtain the basic monodromy relation
\begin{align}
\sum_n \langle \boldsymbol{1}_{12}|\Omega^{(1)}_{\boldsymbol{m}\boldsymbol{n}}(u) \Omega^{(2)}_{\boldsymbol{n}\boldsymbol{l}}(u) 
= \langle \boldsymbol{1}_{12}| \delta_{\boldsymbol{m}\boldsymbol{l}}
\period
\end{align}
As an example for the use of this relation,   contract both sides with $\ket{\calO_1}
\otimes \ket{\calO_2}$.  We then obtain the monodromy relation for 
a two-point function of the form 
\begin{align}
\sum_{\boldsymbol{l}} \left\langle \Omega^{(1)}_{\boldsymbol{k}\boldsymbol{l}}(u) |\mathcal{O}_1\rangle \comma  \Omega^{(2)}_{\boldsymbol{l}\boldsymbol{m}}(u) |\mathcal{O}_2 \rangle \right\rangle =\delta_{\boldsymbol{k}\boldsymbol{m}} \left\langle |\mathcal{O}_1\rangle \comma |\mathcal{O}_2 \rangle \right\rangle \comma 
\end{align}
where $\big\langle \comma \big\rangle$ denotes the Wick contraction via 
 the singlet projector.
 
Just as in the case of the fundamental R-matrix, the derivation of the monodormy relation for the three-point functions is straightforward. In fact, the prefactor function in this case is trivial and the result takes the simple form:
 \begin{align}
\sum_{\boldsymbol{l},\boldsymbol{m}} \left\langle \Omega^{(1)}_{\boldsymbol{k}\boldsymbol{l}}(u) |\mathcal{O}_1\rangle \comma  \Omega^{(2)}_{\boldsymbol{l}\boldsymbol{m}}(u) |\mathcal{O}_2 \rangle , \Omega^{(3)}_{\boldsymbol{m}\boldsymbol{n}}(u) |\mathcal{O}_3 \rangle \right\rangle =\delta_{\boldsymbol{k}\boldsymbol{n}} \left\langle |\mathcal{O}_1\rangle \comma |\mathcal{O}_2 \rangle ,|\mathcal{O}_3 \rangle \right\rangle  \period
\end{align} 
The reason for the absence of the sign factors  in contrast to the case of the fundamental R-matrix (\ref{eq:Fmono_3pt}) is because  all the components of the harmonic R-matrix $\langle \boldsymbol{m}| \boldsymbol{\mathrm{R}}(u)|\boldsymbol{n}\rangle$ are bosonic:  They are composed of even number of oscillators, as  seen from the definitions (\ref{eq:Hop})-(\ref{eq:Hop_coefficient}).

Let us now describe how one can prove  the basic crossing relation (\ref{eq:crossing2}). As shown in (\ref{eq:RHop}), the harmonic R-matrix 
is made up of the hopping operators (\ref{eq:Hop})  and the coefficients
(\ref{eq:Hop_coefficient}). Therefore we need to derive the crossing 
 relations for these two quantities. Since the manipulations are somewhat 
 involved, we relegate the details to Appendix C and only sketch the 
procedures and some relevant  intermediate results below. 

First, using the crossing relations for the oscillators,  it is easy to 
derive the crossing relation for an  arbitrary function of the 
number operators. The result reads 
\begin{align}
\langle \boldsymbol{1}_{12}|f(\boldsymbol{\mathrm{N}}^{(a)}+\boldsymbol{\mathrm{N}}^{(1)})=\langle \boldsymbol{1}_{12}|f(\boldsymbol{\mathrm{N}}^{(a)}-\boldsymbol{\mathrm{N}}^{(2)}) \period 
\label{eq:Hop_N}
\end{align}
In particular,  the coefficient $\mathcal{A}_{I}^{(\boldsymbol{\mathrm{N}}^{(a)}+\boldsymbol{\mathrm{N}}^{(1)})}$ becomes $\mathcal{A}_{I}^{(\boldsymbol{\mathrm{N}}^{(a)}-\boldsymbol{\mathrm{N}}^{(2)})}$ 
 under such crossing. 

Next one can show that the crossing relation for the hopping operator 
takes the form 
\begin{align}
\begin{split}
&\langle \boldsymbol{1}_{12}| \boldsymbol{\mathrm{Hop}}_{k,l,m,n}^{(a1)} =\langle \boldsymbol{1}_{12}| \mathcal{C} \circ\boldsymbol{\mathrm{Hop}}_{k,l,m,n}^{(a1)} \label{eq:cross_Hop}  \\
&:=(-1)^{l+m}\langle \boldsymbol{1}_{12}|   \sum_{p=0}^{\mathrm{min}(k,m)}\sum_{q=0}^{\mathrm{min}(l,n)}  
\mathcal{B}(\boldsymbol{\mathrm{N}}^{(a)}_{\alpha}-m+p,p) \mathcal{B}(\boldsymbol{\mathrm{N}}^{(a)}_{\beta}-n+q,q)\boldsymbol{\mathrm{Hop}}_{k-p,l-q,m-p,n-q}^{(a2)} \comma 
\end{split}
\end{align}
where $\mathcal{C} \circ\boldsymbol{\mathrm{Hop}}_{k,l,m,n}^{(a1)}$ denotes the crossed $\boldsymbol{\mathrm{Hop}}$ operator and the function $\calB(x,y)$ was defined in (\ref{eq:Hop_coefficient}). 

Combining these crossing operations, we find that 
\begin{align}
\langle \boldsymbol{1}_{12}| \boldsymbol{\mathrm{R}}^{(1)}(u) &=\langle \boldsymbol{1}_{12}| \sum_{k,l,m,n}\sum_{p,q}a_{k,l,m,n} \mathcal{A}_{I}^{(\boldsymbol{\mathrm{N}}^{(a)}-\boldsymbol{\mathrm{N}}^{(2)}+k+l-m-n)} \mathcal{C}\circ \boldsymbol{\mathrm{Hop}}_{k,l,m,n}^{(a1)} \label{eq:cross_harmonic} \\
&= \langle \boldsymbol{1}_{12}| \sum_{k,l,m,n}a_{k,l,m,n}  \tilde{\mathcal{A}}_{I}^{(\boldsymbol{\mathrm{N}})}\boldsymbol{\mathrm{Hop}}_{k,l,m,n}^{(a2)} \comma \label{eq:AHop}
\end{align}
where 
\begin{align}
\tilde{\mathcal{A}}_{I}^{(\boldsymbol{\mathrm{N}})}(u)&=\sum_{p,q}^{\infty}(-1)^{I} \mathcal{B}(\boldsymbol{\mathrm{N}}^{(a)}_{\alpha}-m,p) \mathcal{B}(\boldsymbol{\mathrm{N}}^{(a)}_{\beta}-n,q) \mathcal{A}_{I+p+q}^{(2I-\boldsymbol{\mathrm{N}}+
2\boldsymbol{\mathrm{M}})} \period
\end{align}
In the above expression of $\tilde{\mathcal{A}}_{I}^{(\boldsymbol{\mathrm{N}})}(u)$, we have renamed  $(k-p,l-q,m-p,n-q) \rightarrow (k,l,m,n)$ and defined $\boldsymbol{\mathrm{M}}:= \boldsymbol{\mathrm{N}}^{(a)}-m-n$.
Note that, under this change of  labels, $a_{k,l,m,n}=\delta_{k+n,l+m}(-1)^{(k+l)(m+n)}$ becomes $a_{k,l,m,n}(-1)^{p+q}$ and $I=\frac{k+l+m+n}{2}$ changes into  $I+p+q$.  Using the binomial identity $\mathcal{B}(\alpha+\beta,k)=\sum_{j=0}^k\mathcal{B}(\alpha,k-j)\mathcal{B}(\beta,j)$\footnote{This is a direct consequence from the binomial theorem, namely, $(1+x)^{\alpha}=\sum_{k=0}^{\infty}\mathcal{B}(\alpha,k)x^k$. The formula readily follows by considering $(1+x)^{\alpha +\beta}=(1+x)^{\alpha}\cdot (1+x)^{\beta}$.},  the summation over $p,q$ in $\tilde{\mathcal{A}}_{I}^{(\boldsymbol{\mathrm{N}})}$ can be converted into a simpler expression
\begin{align}
\begin{split}
\tilde{\mathcal{A}}_{I}^{(\boldsymbol{\mathrm{N}})}(u)&=\sum_{r=0}^{\infty}(-1)^{I}\left( \sum_{p=0}^r \mathcal{B}(\boldsymbol{\mathrm{N}}^{(a)}_{\alpha}-m,p)\mathcal{B}(\boldsymbol{\mathrm{N}}^{(a)}_{\beta}-n,r-p) \right) \mathcal{A}_{I+r}^{(2I-\boldsymbol{\mathrm{N}}+2\boldsymbol{\mathrm{M}})}(u)  \\
&=\sum_{r=0}^{\infty}(-1)^{I} \mathcal{B}(\boldsymbol{\mathrm{M}},r) \mathcal{A}_{I+r}^{(2I-\boldsymbol{\mathrm{N}}+2\boldsymbol{\mathrm{M}})}(u) \period
\end{split}  
\end{align} 

In Appendix C, we will show that this  sum giving   $\tilde{\mathcal{A}}_{I}^{(\boldsymbol{\mathrm{N}})}(u)$ can be evaluated  and leads to 
the desired equality 
\begin{align}
\tilde{\mathcal{A}}_{I}^{(\boldsymbol{\mathrm{N}})}(u)=\sum_{r=0}^{\infty}(-1)^{I} \mathcal{B}(\boldsymbol{\mathrm{M}},r) \mathcal{A}_{I+r}^{(2I-\boldsymbol{\mathrm{N}}+2\boldsymbol{\mathrm{M}})}=\mathcal{A}_{I}^{(\boldsymbol{\mathrm{N}})}(-u) \period \label{eq:sum_A}
\end{align}
Putting this result into (\ref{eq:AHop}) and summing over $k,l,m,n$, 
 we find that the RHS of (\ref{eq:AHop}) becomes $
\langle \boldsymbol{1}_{12}| \boldsymbol{\mathrm{R}}^{(2)}(-u)
$ and this proves the crossing relation (\ref{eq:crossing2}). 
\section{Reduction of monodromy relation to subsectors}
In the previous section, we have  derived the monodromy relation 
 for the full psu(2,2$|$4) sector both in the case of the fundamental R-matrix 
 and of the harmonic R-matrix. The former can be obtained 
by a rather straightforward generalization of the SU(2) case discussed 
 in our previous work\cite{KKN2} and we tried to give  a slightly more 
detailed exposition compared with the result already given in \cite{spinvertex}. On the other hand, the case for the harmonic R-matrix is new.  Although its derivation turned out to be  substantially more involved than the case of the fundamental R-matrix, the symmetric set up for which the quantum and the auxiliary spaces carry  identical representation of psu(2,2$|$4) can be of  particular value, as 
 was already indicated in the application to the scattering amplitude. 
Also the fact that the Hamiltonian can be obtained simply as  the logarithmic derivative of the harmonic R-matrix should find useful applications. 

The most important original purpose for formulating the monodromy relations, however, is their possible use as the set of powerful equations which govern, from the integrability perspective, the structures of the correlation functions. 
Although this idea has not yet been studied explicitly, to perform such an  analysis it is  natural to begin with the simplest set-ups, namely the cases of  
 important  tractable  subsectors of the full theory. In what follows, we shall consider the compact  SU(2) sector and the non-compact SL(2) sector as 
typical examples, and derive the monodromy relations for them from the point of view of a systematic reduction of the general psu(2,2$|$4) case. 
\subsection{Reduction to the SU(2) subsector} 
\subsubsection{Embedding of ${\rm su}(2)_{\rm L} \oplus {\rm su}(2)_{\rm R}$ in 
${\rm u}(2,2|4)$} 
In the case of the SU(2) sector, the monodromy relation  was already obtained in our previous work \cite{KK2}. Therefore  the purpose here is to re-derive it through 
 the reduction of the psu(2,2$|$4) case. To do this, 
 the first step is to identify  the generators of $\SU2_{\rm L}$ and $\SU2_{\rm R}$, given in (\ref{SU2L}) and (\ref{SU2R})  in terms 
 of those of u(2,2$|$4) shown in (\ref{eq:u(2,2|4)}). This is simple since $\SU2_{\rm L} \times \SU2_{\rm R}$ is contained entirely in the R-symmetry group SU(4), and hence the relevant u(2,2$|$4) generators are 
  $W_a{}^b = \xibar_a\xi^b= R_a{}^b + \half \delta_a^b B$  given in (\ref{eq:defW}).  Explicitly, we have 
\begin{align}
J_3^L = \half (W_3{}^3 -W_1{}^1) \comma \quad 
J_+^L = -W_3{}^1 \comma \quad J^L_-=-W_1{}^3 \comma \\
J_3^R = \half (W_4{}^4 -W_2{}^2) \comma \quad 
J_+^R = -W_4{}^2 \comma \quad J^R_-=-W_2{}^4  \period
\end{align}
It will be important to recognize that the following combinations,  $B_L$ and $B_R$,  of diagonal $W_a{}^a$ generators act as  central charges for 
 the group $\SU2_{\rm L} \times \SU2_{\rm R}$:
\begin{align}
B_L &= \half (W_3{}^3 + W_1{}^1) \comma \qquad B_R = \half (W_4{}^4 + W_2{}^2) \period  \label{eq:defBLBR}
\end{align}
For example,  $B_L$ together with $J_i^L$ form ${\rm U(2)}_{\rm L}
={\rm U(1)}\times {\rm SU(2)}_{\rm L}$  of which 
 $B_L$ is the U(1) part. Thus,  $\com{B_L}{J^L_i} =0$.  Obviously 
$B_L$ commutes with $\SU2_{\rm R}$. The argument for $B_R$ is entirely similar. Being the central charges, they take definite values in an irreducible 
 representation, which in our case of interest is the spin $(\half)_L \times (\half)_R$ representation. Evaluating $B_L$ and $B_R$ on any of the  states
 in this representation, say $\ket{Z} = \xibar_3\xibar_4\ket{0}$, it is easy to  obtain $B_L = B_R = \half$. 

With this in mind, let us write down the $\SU2_{\rm L}$  Lax operators as embedded in that of ${\rm gl}(4|4)$, given in (\ref{eq:gl44Lax}). We get
\begin{align}
L_{{\rm SU}(2)_L}(u) &= \matrixii{u+{i \over 2} (W_3{}^3-W_1{}^1)}{-iW_1{}^3}{-iW_3{}^1}{u-{i \over 2}  (W_3{}^3-W_1{}^1)}
\end{align}
Using  the relation $B_L = \half = \half (W_3{}^3+W_1{}^1)$, 
\ie  $W_3{}^3 +W_1{}^1=1$, this  can be re-written as 
\begin{align}
L_{{\rm SU}(2)_L}(u) &= \matrixii{u+{i \over 2} -i W_1{}^1}{-iW_1{}^3}{-iW_3{}^1}{u+{i \over 2} -iW_3{}^3}  \label{SU2LLax}
\end{align}
Let us now recall the general form of the Lax operator for ${\rm gl}(4|4)$, 
 which is $
L(u)^A{}_B = u\delta^A_B + \eta (-1)^{|B|} J^A{}_B $, 
 and its more refined form in terms of the superconformal generators 
given in (\ref{eq:gl44Lax}). The part relevant for the SU(2) sector is 
 in the lower diagonal corner given by $u\delta_a^b -\eta  W_a{}^b$. 
For the $\SU2_{\rm L}$, we can identify the indices for  $a$ and $b$ to be 1 and 3.  In the entirely similar manner, the Lax operator for 
 the $\SU2_{\rm R}$ sector is obtained from the one for the $\SU2_{\rm L}$  sector by substitution of the indices $1 \rightarrow 2$ and $3 \rightarrow 4$. Then, it is easy to see that the Lax operators for the $\SU2_{\rm L}$ 
and $\SU2_{\rm R}$ sectors are obtained from the ${\rm gl}(4|4)$ Lax operator by taking into 
 account the shift $u \rightarrow u + i/2$ in the form 
\begin{align}
L_{{\rm SU}(2)_L}(u)_{ab} &=  L(u+i/2 )_{ab} 
\comma \qquad \mbox{with}\ \eta =i\comma \quad \{a,b\} = \{1,3\} 
\comma \\
L_{{\rm SU}(2)_R}(u)_{\bar{a}\bar{b}} &=  L(u+i/2 )_{\bar{a}\bar{b}} 
\comma \qquad \mbox{with}\ \eta =i\comma \quad \{\bar{a},\bar{b}\} = \{2,4\} \period
\end{align}
It is important to note that the occurrence  of the shift of the spectral parameter is due to the emergence  of  the   extra central charges  when a  group is restricted to its subgroup and hence is a rather general phenomenon. 
\subsubsection{Inversion relation}
Let us now derive the inversion relation.  As in the previous discussion, we shall concentrate on the SU(2)$_{\rm L}$ part.  The object to consider is the 
 product of two u(2,2$|$4) Lax operators given in (\ref{eq:prod_Lax}), with the indices $A,B$ taken to be the SU(2) indices\footnote{Here and {\it until}
 the end of subsection 4.1, we shall  use roman letters $\rma$, $\rmb$, etc. to denote the genuine SU(2) indices, which take the values 1 and 2, in order to distinguish them from 
 the italic letters $a, b$, etc., which are SU(4) indices taking values from 1 to 4.} $\rma,\rmb$.  This gives 
\begin{align}
(L(u)L(v))_{\rma\rmb}= (L(u))_{\rma\rmc}(L(v))_{\rmc\rmb}+\eta^2(-1)^{\gamma} (-1)^{|\rmb|} J^\rma_{\ \gamma} J^{\gamma}_{\ \rmb} \comma 
\label{eq:LLSU(2)}
\end{align}
where $\gamma$ runs over the indices of the auxiliary space other than 
the SU(2) indices  $\rma,\rmb,\ldots$. Using the oscillator representation, the second terms can be simplified as 
\begin{align}
\begin{split}
(-1)^{\gamma}J^\rma_{\ \gamma} J^{\gamma}_{\ \rmb} &=\bar{\zeta}^\rma(\mu^{\alpha}\lambda_{\alpha}-\tilde{\lambda}_{\dot{\alpha}}\tilde{\mu}^{\dot{\alpha}}-\xi^{\bar{d}}\bar{\xi}_{\bar{d}})\zeta_\rmb \\
&= J^\rma_{\ \rmb}(2Z_L+2Z_R+2B_R) \comma \label{eq:LuLvab}
\end{split} 
\end{align}
where 
\begin{align}
Z_L:&= \half (\lam_1\mu^1 -\lamtil_{\onedot} \mutil^{\onedot}) 
 = \half (J_1{}^1+J_\onedot{}^\onedot) \comma \label{eq:ZL}\\
Z_R:&= \half (\lam_2\mu^2 -\lamtil_{\twodot} \mutil^{\twodot}) 
 = \half (J_2{}^2+J_\twodot{}^\twodot) \comma \label{eq:ZR}
\end{align}
and $B_R$ was already defined in (\ref{eq:defBLBR}). In the SU(2) sector 
the quantities $Z_L$ and $Z_R$ vanish\footnote{Actually, as it will be 
explained in the next subsection, the quantities $Z_L$ and $Z_R$ are 
 central charges for ${\rm SL(2)}_{\rm L} \times {\rm SL(2)}_{\rm R}$. }
 and $B_R=1/2$, as was already  explained. 
Therefore the factor $2Z_L+2Z_R+2B_R$ is simply unity and we simply
 obtain
\begin{align}
(-1)^{\gamma}J^\rma_{\ \gamma} J^{\gamma}_{\ \rmb} &=J^\rma{}_\rmb \period
\end{align}
Substituting this back  into (\ref{eq:LuLvab}) and rewriting the 
spectral parameters in order to express the result in terms of the Lax operator 
 for SU(2)$_{\rm L}$, we obtain the relation
\begin{align}
\begin{split}
(L(u)L(v))_{\rma\rmb}&= (L(u))_{\rma\rmc}(L(v))_{\rmc\rmb}-\eta^2J^\rma_{\ \rmb} \\
&=(L(u+\eta /2))_{\rma\rmc}(L(v+\eta /2))_{\rmc\rmb}-((u+\eta /2)(v+\eta/2)-uv)\delta_{\rma\rmb} \\
&=(L_{{\rm SU}(2)}(u))_{\rma\rmc}(L_{{\rm SU}(2)}(v))_{\rmc\rmb}-((u+ i/2)(v+i/2)-uv)\delta_{\rma\rmb}  \period
\end{split}
\end{align}
Now from the inversion relation for the 
Lax operator for u(2,2$|4)$ given in  (\ref{eq:inversion1}),  when $u+v=\eta=i$, the left hand side of the above equation becomes $uv\delta_{ab}$, and thus we obtain the inversion relation for the SU(2)$_{\rm L}$  Lax operator 
to be of the form 
\begin{align}
(L_{{\rm SU}(2)}(u)L_{{\rm SU}(2)}(v))_{\rma\rmb}&=f_{{\rm SU}}(u,v)\delta_{\rma\rmb} \comma \ \ u+v=i \comma \\
f_{{\rm SU}}(u,v)&=uv-\frac{3}{4} \label{eq:fSU}
\end{align}
To compare with the result of \cite{KKN2}, we must substitute  $u \rightarrow -u+{i \over 2}$ and $v \rightarrow u+{i \over 2}$. Then the relation 
 above takes the form
\begin{align}
(L_{{\rm SU}(2)}(-u+(i/2))L_{{\rm SU}(2)}(u+(i/2)))_{\rma\rmb} = -(u^2+1) \delta_{\rma\rmb}\comma 
\end{align}
which agrees with the equation (5.1) of \cite{KKN2} (with the inhomogeneity parameter $\theta$ set to zero). 

From this inversion relation, one can easily obtain the monodromy relation, 
 as already described in \cite{KK2}. So we shall omit this derivation for the SU(2)  sector. Instead, in the next subsection, we shall present a derivation 
 of the monodromy relation for the SL(2) sector {\it directly}  from that for 
 the psu(2,2$|4$) sector. The result is new and the method can of course be applied to the SU(2) case as well to provide an alternative  derivation of a known result given in \cite{KKN2}. 
\subsection{Reduction to the  SL(2) subsector} 
Having been warmed up with the reduction to the simplest SU(2) sector, 
we now perform the reduction to the SL(2) subsector to derive the 
explicit form of its monodromy relation,  which is new. 
\subsubsection{Embedding of  ${\rm sl}(2)_{\rm L} \oplus {\rm sl}(2)_{\rm R}$
in u(2,2$|4$) and a derivation of monodromy relation}
Let us begin by  clarifying  how  the generators of SL(2)$_{\rm L} \times$ SL(2)$_{\rm R}$ are embedded in u(2,2$|4$). First consider the simple 
``light-cone" combinations of operators given by 
\begin{align}
p_+=\frac{1}{2}(P_0-P_3) \comma \ \ k^+=-\frac{1}{2}(K_0+K_3) \comma \ \ d_+=\frac{i}{2}(M_{03}-D) \comma \\
p_-=\frac{1}{2}(P_0+P_3) \comma \ \ k^-=\frac{1}{2}(-K_0+K_3) \comma \ \ d_-=-\frac{i}{2}(M_{03}+D) \period
\end{align}
They satisfy the following simple set of commutation relations:
\begin{align}
[ d_{\pm} ,p_{\pm}]=p_{\pm} \comma \ \ [ d_{\pm} ,k^{\pm}]=-k^{\pm} \comma \ \ [ k^{\pm},p_{\pm}]=2d_{\pm} \period
\end{align}
This shows that the generators of the SL(2)$_L \times$ SL(2)$_R$ can be taken as 
\begin{align}
\mathrm{SL(2)}_L: \ \ \ S_-=-ip_+ \comma \ \ S_+=-ik^+\comma \ \ S_0=-d_+ \comma \label{eq:generator_SL(2)_L} \\
\mathrm{SL(2)}_R: \ \ \ \tilde{S}_-=-ip_- \comma \ \ \tilde{S}_+=-ik^-\comma \ \ \tilde{S}_0=-d_- \period \label{eq:generator_SL(2)_R}
\end{align}
In this notation, the commutation relations are 
\begin{align}
\com{S_0}{S_\pm} &= \pm S_{\pm} \comma \qquad 
\com{S_+}{S_-} = 2S_0 \comma \\
\com{\tilde{S}_0}{\tilde{S}_\pm} &= \pm \tilde{S}_{\pm} \comma \qquad 
\com{\tilde{S}_+}{\tilde{S}_-} = 2\tilde{S}_0  \period
\end{align}
Now from the definition of spinor notations (\ref{eq:spinor1}), (\ref{eq:spinor2}) and the form of the ${\rm u}(2,2|4)$ generators $J^A_{\ B}$ 
given in  (\ref{eq:u(2,2|4)}), one finds that,  for example,  the generators 
$\{S_0, S_\pm\}$ of SL(2)$_{\rm L}$ are embedded in  ${\rm u}(2,2|4)$ as\footnote{$1$ and $\dot{1}$ are the indices for ${\rm SL}(2,\mathbb{C}) \times \overline{{\rm SL}(2,\mathbb{C})}$.}
\begin{align}
J_1^{\ 1}=M_1^{\ 1}-\frac{i}{2}D-\frac{1}{2}B=\frac{i}{2}(M_{03}-iM_{12})-\frac{i}{2}(D-iB)=-S_0+\frac{1}{2}M_{12}-\frac{1}{2}B 
\label{eq:Jones} \\
J^{\dot{1}}_{\ \dot{1}}=M^{\dot{1}}_{\dot{1}}+\frac{i}{2}D-\frac{1}{2}B=\frac{i}{2}(-M_{03}-iM_{12})+\frac{i}{2}(D+iB)=S_0+\frac{1}{2}M_{12}-\frac{1}{2}B \comma \label{eq:Jonedots}\\
J_{1 \dot{1}}=iP_{1 \dot{1}}=ip_+=-S_- \comma \ \ J^{\dot{1}1}=iK^{\dot{1} 1}=ik^+=-S_+ \period 
\end{align}
At this point we note that, just as the extra central charges $B_L$ and $B_R$ appeared for the subgroup  SU(2)$_{\rm L} \times $SU(2)$_{\rm R}$, 
the quantities $Z_L$ and $Z_R$ previously defined in (\ref{eq:ZL}) and 
(\ref{eq:ZR})  behave  as central charges for SL(2)$_L\times$ SL(2)$_R$. 
This will be important below. 

Now in what follows, let us concentrate on the SL(2)$_L$ subsector, where  the composite operators are constructed by  multiple actions of the  covariant derivatives along the light-cone direction as 
\begin{align}
\begin{split}
\mathcal{O}(x)&= \frac{1}{n_1!n_2!\ldots}\mathrm{Tr}( \mathcal{D}_{1\dot{1}}^{n_1} Z \mathcal{D}_{1\dot{1}}^{n_2}Z\cdots) \\
 &\mapsto  \exp (i(x^+p_++x^-p_-))  \frac{(\lambda_1 \tilde{\lambda}_{\dot{1}})^{n_1}}{n_1!}|Z\rangle \otimes \frac{(\lambda_1 \tilde{\lambda}_{\dot{1}})^{n_2}}{n_2!}|Z\rangle \otimes \cdots \period
 \end{split} 
\end{align}
Here, $\Zket$ is the scalar state $\bar{\xi}_3\bar{\xi}_4 |0\rangle $
and each factor in the total tensor product signifies the operator at different 
positions of the spin chain.  
Note that on this type of states the rotation operator $M_{12}$ vanishes 
 and the hypercharge operator $B=\half \xibar_a \xi^a$ takes the definite 
value 1. Furthermore, one can easily check that $Z_L=Z_R=0$ by acting 
 them on such a state. All these relations is consistent with the vanishing of the central charge $C=Z_L+Z_R+B-1$ for the physical SYM operators. 

With these properties, the relations (\ref{eq:Jones}) and (\ref{eq:Jonedots}) simplify and we can easily embed the SL(2)$_{\rm L}$ Lax operator 
into that of u(2,2$|4$) in the following fashion:
\begin{align}
L_{{\rm SL}(2)}(u):=\left(
\begin{array}{cc}
u+iS_0 & iS_- \\
iS_+ &u-iS_0
\end{array} \right)=\left(
\begin{array}{cc}
u-\frac{i}{2}-iJ_1^{\ 1} & -i J_{1 \dot{1}} \\
-iJ^{\dot{1}1} &u-\frac{i}{2}+iJ^{\dot{1}}_{\ \dot{1}}
\end{array} \right) \period
\end{align}
Here again the shift of $u \rightarrow u-{i \over 2}$ is due to the 
effect of the central charge, as in the case of SU(2).  Now 
 comparing this with the form of the , we readily find
 that the SL(2)$_{\rm L}$ Lax operator is embedded in the 
u(2,2$|4$) Lax operator as\footnote{Similarly, the Lax operator for SL(2)$_R$ part is embedded as ($L_{\widetilde{{\rm SL}}(2)}(u))_{\rma\rmb}=(L(u-i/2))_{\rma\rmb}$ $(\rma,\rmb=2, \dot{2})$ with $\eta=-i$. In this equation, $\rma$ and $\rmb$ refer to the indices of $2\times 2$ matrix, which are part of the ${\rm SL}(2,\mathbb{C}) \times \overline{{\rm SL}(2,\mathbb{C})}$ indices.}
\begin{align}
(L_{{\rm SL}(2)}(u))_{\rma\rmb}&=(L(u-i/2))_{\rma\rmb} \comma \qquad (\rma,\rmb=1, \dot{1})
\comma \qquad  \eta=-i \period
\end{align}
(Just as we did in the discussion of  the SU(2) sector in the previous subsection,
 we shall hereafter use the roman letters $\rma, \rmb$, etc. to denote the 
Lorentz spinor indices $\{\rma , \rmb\} = \{1, \dot{1}\}$ for the ${\rm SL}(2)_{\rm L}$ sector, in order to distinguish  them from the SU(4) indices $a,b$, etc, which run over $1$ to $4$. )

To derive the monodromy relation, we first need to find the inversion relation
 for the Lax operator. Just as in the SU(2) case, the product of the 
 SL(2)$_{\rm L}$ part of the u(2,2$|4$) Lax operators give 
\begin{align}
(L(u)L(v))_{\rma\rmb}&= (L(u))_{\rma\rmc}(L(v))_{\rmc\rmb}+\eta^2 (-1)^{|\gamma|}J^\rma_{\ \gamma} J^{\gamma}_{\ \rmb}
\end{align}
where $\gamma$ runs over the all indices of the auxiliary except for 
 those of SL(2), namely $\rma,\rmb=1,\dot{1}$. Note at this stage 
the sign of the second term is opposite to the one for SU(2) case 
given in (\ref{eq:LLSU(2)}).  Now as before we can simplify this term
 quadratic in the generators in the following manner:
\begin{align}
\begin{split}
(-1)^{|\gamma|}J^\rma_{\ \gamma} J^{\gamma}_{\ \rmb}&=\left( \begin{array}{c} \lambda_1 \\ \tilde{\mu}^1 \end{array} \right)^\rma
(\mu^2 \lambda_2-\tilde{\lambda}_{\dot{2}}\tilde{\mu}^{\dot{2}}-\xi^i\bar{\xi}_i) (\mu^1 \ -\tilde{\lambda}_{\dot{1}})_\rmb \\
&= J^\rma_{\ \rmb}( 2Z_R+2B-3) \\
&=-J^\rma_{\ \rmb} \comma  \label{eq:normal_order}
\end{split}
\end{align} 
where we used $Z_R=0, B=1$. Thus we get an extra minus sign from this manipulation and hence obtains the same sign for the second term as 
 for the SU(2) case. 
Now we make an appropriate shift of the spectral parameter and get 
\begin{align}
(L(u)L(v))_{\rma\rmb}&= (L(u))_{\rma\rmc}(L(v))_{\rmc\rmb}-\eta^2J^\rma_{\ \rmb} \\
&= (L(u-\eta/2))_{\rma\rmc}(L(v-\eta/2))_{\rmc\rmb}+\frac{\eta}{2}(u+v-\eta)\delta_{\rma\rmb}+\frac{\eta^2}{4}\delta_{\rma\rmb} \\
&= (L_{{\rm SL}(2)})(u))_{\rma\rmc}(L_{{\rm SL}(2)}(v))_{\rmc\rmb}+\frac{\eta}{2}(u+v-\eta)\delta_{\rma\rmb}+\frac{\eta^2}{4}\delta_{\rma\rmb} \comma \label{eq:Lax_prod} \period
\end{align}
Thus if we set  $u+v=\eta=-i$,  then since the left hand side of the above equation becomes  $uv\delta_{\rma\rmb}$, we get
\begin{align}
 (L_{{\rm SL}(2)})(u))_{\rma\rmc}(L_{{\rm SL}(2)}(v))_{\rmc\rmb}
&= f_{{\rm SL}}(u,v)\delta_{\rma\rmb} \comma \qquad u+v =-i \\
f_{{\rm SL}}(u,v) = uv +{1\over 4} 
\period \label{eq:fSL}
\end{align}
Note that if we compare (\ref{eq:fSL}) with (\ref{eq:fSU}), we can 
recognize that the function $f(u,v)$ for the SU(2) and the SL(2) cases can be
written in a unified manner as 
\begin{align}
f(u,v) &= uv - \mathbb{S}^2 = \bracetwocases{uv-s(s+1)}{\mbox{for SU(2)}}{uv-s(s-1)}{\mbox{for SL(2)}} \comma \qquad s=\half \comma 
\end{align}
where $\mathbb{S}^2$ is the Casimir operator for the respective group. 
\subsubsection{Method of direct reduction from psu(2,2$|$4) monodromy
 relation}
We now wish to demonstrate that we can  derive the monodromy relation for the two-point function in the SL(2) sector  {\it more directly} from that for psu(2,2$|$4), with the judicious use of the product relation (\ref{eq:Lax_prod}) for the Lax operators. 

 The monodromy relation for psu(2,2$|$4) is given in (\ref{eq:psu(2,2|4)_mono}). To reduce it to the SL(2) sector, we set the indices $A$ and $C$ to be those of SL(2), say $\rma$ and $\rmc$ which take values in $\{ 1,\dot{1} \}$, and for convenience make a shift of the spectral parameter $u \rightarrow u - i/2$. 
Hereafter we shall employ  the notation $f^-(u) \equiv f(u-i/2)$ for such a  shift for  any function $f(u)$. Then from (\ref{eq:psu(2,2|4)_mono}) we get
\begin{align}
\langle {\Omega^-_\ell}(u)_{\rma B} | \psi_1 \rangle , {\Omega^-_\ell} (u)_{B \rmc} |\psi_2 \rangle \rangle=F^-_{\ell}(u)\delta_{\rma \rmc} \langle | \psi_1 \rangle ,  |\psi_2 \rangle \rangle \comma  \label{eq:two_mono}
\end{align}
or more succinctly,  before taking the inner product with $\ket{\psi_1} \otimes \ket{\psi_2}$, 
\begin{align}
\bra{{\bf 1}_{12}} {\Omega^{(1)-}_\ell }(u)_{\rma B}  {\Omega^{(2)-}_\ell} (u)_{B \rmc}
&= F^-_{\ell}(u)\delta_{\rma \rmc} \bra{{\bf 1}_{12}} \period \label{eq:monod_rel1}
\end{align}
What we wish to derive from it is the relation involving  the SL(2) monodormy matrices with only the  SL(2) indices  $\{ 1, \dot{1} \}$.  However, obviously this reduction is non-trivial since the general gl(4$|$4)  indices  $A, B$ which occur for a neighboring product of two Lax operators $(LL)_{AB}$  in  $\Omega_{\rma B}$ may take all possible gl(4$|$4) values. 

This difficulty can be overcome by noting that the formula for the crossing
relation for the individual  ${\rm u}(2,2|4)$ Lax matrices is quite simple 
 and that a certain product of the ${\rm u}(2,2|4)$ generators which 
 appear in the product of two Lax matrices can be  reduced to a single generator, as was demonstrated already in (\ref{eq:normal_order}). 
To make use of these properties, we first focus on the leftmost and the 
 rightmost Lax operators forming the two monodromy matrices and make the 
 split
\begin{align}
\Omega^{(1)-}_\ell (u)_{\rma B}  &={L^{(1)-}_1}(u)_{\rma D}  \Omega^{(1)-}_{\ell-1} (u)_{D B }  \comma \label{eq:split1}\\
\Omega^{(2)-}_\ell (u)_{B \rmc} &= \Omega^{(2)-}_{\ell-1} (u)_{B E}
{L^{(2)-}_\ell}(u)_{E\rmc} \comma  \label{eq:split2}
\end{align}
Then, the LHS of (\ref{eq:monod_rel1}) becomes 
\begin{align}
\bra{{\bf 1}_{12}} {{\Omega^{(1)-}}_\ell} (u)_{\rma B}  {{\Omega^{(2)-}}_\ell} (u)_{B \rmc}
&= \bra{{\bf 1}_{12}} {L^{(1)-}_1}(u)_{\rma D}  {{\Omega^{(1)-}}_{\ell-1}} (u)_{D B } {{\Omega^{(2)-}}_{\ell-1}} (u)_{B E}
{L^{(2)-}_\ell}(u)_{E\rmc}  \period  \label{eq:LOmOmL}
\end{align}
As the operators at different positions all commute, we can move ${\Omega^{(1)-}}_{\ell-1}  {\Omega^{(2)-}}_{\ell-1}$ in the middle to the left until they hit $\bra{{\bf 1}_{12}}$. Then we can use  the monodromy relation (\ref{eq:two_mono}) for the case of length $\ell -1$ and write (\ref{eq:LOmOmL}) as 
\begin{align}
\bra{{\bf 1}_{12}} {{\Omega^{(1)-}}_\ell} (u)_{\rma B}  {{\Omega^{(2)-}}_\ell} (u)_{B \rmc}
&=F^-_{\ell-1} \bra{{\bf 1}_{12}} {L^{(1)-}_1}(u)_{\rma B} {L^{(2)-}_\ell}(u)_{B\rmc} \period  \label{eq:monod_ellmone}
\end{align}
Now substitute the definition $L(u)^A{}_B = u \delta^A_B 
 + \eta (-1)^{|B|} J^A{}_B$ for the Lax operators on the RHS and expand. 
This gives 
\begin{align}
\bra{{\bf 1}_{12}} {L^{(1)-}_1}(u)_{\rma B} {L^{(2)-}_1}(u)_{B\rmc}  
&=\bra{{\bf 1}_{12}}  {L^{(1)-}_1}(u)_{\rma \rmb} {L^{(2)-}_\ell}(u)_{\rmb\rmc} 
+ \bra{{\bf 1}_{12}} {L^{(1)-}_1}(u)_{\rma\beta} {L^{(2)-}_\ell}(u)_{\beta \rmc} \comma 
\label{eq:splitLL}
\end{align}
where $\beta$ stands for indices other than those of  SL(2). Further the 
second term on the RHS can be written out explicitly as 
\begin{align}
 {L^{(1)-}_1}(u)_{\rma\beta} {L^{(2)-}_\ell}(u)_{\beta \rmc} 
&= \eta^2 (-1)^{|\beta|} {J_1^{(1)}}{}^\rma{}_\beta {J_\ell^{(2)}}{}^\beta{}_\rmc \comma \label{eq:LLJJ}
\end{align}
where we used $(-1)^{|\rmc|} =1$ since $\rmc$ is a bosonic index. Apply  both 
 sides now to the singlet projector $\bra{{\bf 1}_{12}}$ and 
use the crossing relation for the generator $\bra{{\bf 1}_{12}}{J_1^{(1)}}{}^\rma{}_\beta  = -\bra{{\bf 1}_{12}}{J_\ell^{(2)}}{}^\rma{}_\beta   $, which is 
 valid since  $\rma \ne \beta$. Then using the identity (\ref{eq:normal_order}) to  the RHS, (\ref{eq:LLJJ}) becomes $
\bra{{\bf 1}_{12}}  {L^{(1)-}_1}(u)_{\rma\beta} {L^{(2)-}_\ell}(u)_{\beta \rmc} = - \eta^2 \bra{{\bf 1}_{12}} {J^{(2)}_\ell}{}^\rma{}_\rmc$ and the 
equation (\ref{eq:monod_ellmone}) can be simplified to  
\begin{align}
\bra{{\bf 1}_{12}} {{\Omega^{(1)-}}_\ell} (u)_{\rma B}  {{\Omega^{(2)-}}_\ell} (u)_{B \rmc}
&=F^-_{\ell-1} \bra{{\bf 1}_{12}} 
\left(   {L^{(1)-}_1}(u)_{\rma \rmb} {L^{(2)-}_\ell}(u)_{\rmb\rmc} -\eta^2  {J^{(2)}_\ell}{}^\rma{}_\rmc \right) \period
\end{align}
Note that on the RHS all the indices have turned into  SL(2) indices. 

Now we make a slight trick to split the last term of the RHS into identical 
halves\footnote{If one wishes, one can actually use a more general split with coefficients $\al$ and $\be$ satisfying $\al+\be=1$ and follow the same logic  to be described below for the $\half+\half$ split. This will lead to 
more general forms of the SL(2) monodromy relations. Below we shall only 
 describe the simplest split for the sake of clarity.}
 as 
$ {J^{(2)}_\ell}{}^\rma{}_\rmc = \half  {J^{(2)}_\ell}{}^\rma{}_\rmc + \half  {J^{(2)}_\ell}{}^\rma{}_\rmc $ and then hit just one half  to $ \bra{{\bf 1}_{12}} $ 
to change it into $ {J^{(1)}_\ell}{}^\rma{}_\rmc$ acting on the spin chain 1. 
Since the generator here is that of ${\rm u}(2,2|4)$, in doing so there appears an extra  constant term $\propto \delta^\rma_\rmc$ coming from the commutator  of the oscillators  forming this generator. Then we get 
\begin{align}
 \bra{{\bf 1}_{12}}  {J^{(2)}_\ell}{}^\rma{}_\rmc 
&= \half \bra{{\bf 1}_{12}}   {J^{(2)}_\ell}{}^\rma{}_\rmc
- \half \bra{{\bf 1}_{12}}
\left(   {J^{(1)}_1}{}^a{}_c + \delta^\rma_\rmc 
\right) 
\end{align}
We shall now show that the  terms linear in  the generators appearing 
on the RHS can be absorbed by a judicious shifts of the spectral parameters of the expression $ {L^{(1)-}_1}(u)_{\rma\rmb} {L^{(2)-}_\ell}(u)_{\rmb\rmc}$, which 
  is the first term on the RHS of (\ref{eq:splitLL}). 
In fact, one can easily check that (\ref{eq:splitLL}) can be re-expressed 
as 
\begin{align}
\bra{{\bf 1}_{12}} {L^-_1(u+\eta/2)}_{\rma\rmb} {L^-_\ell(u-\eta/2)}_{\rmb \rmc} 
 + f^-_1(\eta) \delta_{\rma\rmc} \bra{{\bf 1}_{12}} \comma 
\end{align}
where the factor  $f_1(\eta)$ is given by $f_1(\eta) = -\eta^2/4$. Thus combining with 
(\ref{eq:monod_ellmone}), the LHS of the original ${\rm psu}(2,2|4)$ monodromy relation (\ref{eq:monod_rel1}), namely $\bra{{\bf 1}_{12}} {\Omega^{(1)-}_\ell }(u)_{\rma B}  {\Omega^{(2)-}_\ell} (u)_{B \rmc}$, becomes 
\begin{align}
\bra{{\bf 1}_{12}}\left( F^-_{\ell-1}\delta_{\rmb\rmd}  {L_1^-(u+\eta/2)}_{ab} {L^-_\ell(u-\eta/2)}_{\rmd\rmc} + f_1(u)F^-_{\ell-1}  \delta_{\rma\rmc} \right) \period
\end{align}
As the final step, we now  rewrite  the quantity $\bra{{\bf 1}_{12}} F^-_{\ell-1}  \delta_{\rmb\rmd}$ in the first term by using the 
fundamental monodromy relation  (\ref{eq:monod_rel1}) with $\ell$ replaced by $\ell-1$. This process is a reversal of the splitting procedure we started out  with and inserts the products of the monodromy matrices defined 
 on the RHS of the splitting equations (\ref{eq:split1}) and (\ref{eq:split2}). Then, the equation above becomes 
\begin{align}
&\bra{{\bf 1}_{12}} {L_1^-(u+\eta/2)}_{ab}(\Omega^{(1)-}_{\ell-1})_{\rmb C} (\Omega^{(2)-}_{\ell-1})_{C\rmd}
 {L^-_\ell(u-\eta/2)}_{\rmd \rmc} \nn\\
&+\bra{{\bf 1}_{12}} f_1(u)F^-_{\ell-1}  \delta_{\rma\rmc} \period
\label{eq:LOOL}
\end{align}
On the other hand, each  Lax operator on  the RHS of the relation (\ref{eq:monod_ellmone}) can be interpreted as a special monodromy matrix of length one and hence we can use the monodromy relation to write the RHS as 
\begin{align}
F^-_{\ell -1} \bra{{\bf 1}_{12}}L_1^{(1)-}(u)_{\rma B} L_\ell^{(2)-}(u)_{B\rmc} = F^-_{\ell-1} F^-_1\delta_{\rma\rmc} \bra{{\bf 1}_{12}} \label{eq:FF}
\end{align}
Thus equating (\ref{eq:LOOL}) and (\ref{eq:FF}) and rearranging, we obtain an important relation
\begin{align}
\bra{{\bf 1}_{12}} {L_1^-(u+\eta/2)}_{\rma\rmb}(\Omega^{(1)-}_{\ell-1})_{\rmb C} (\Omega^{(2)-}_{\ell-1})_{C\rmd}
 {L^-_\ell(u-\eta/2)}_{\rmd\rmc}  &= g_1(u) F^-_{\ell-1} \bra \delta_{\rma\rmc}{{\bf 1}_{12}} \comma
\end{align}
where the function $g_1(u)$ is given by 
\begin{align}
g_1(u) &= F_1^-(u) -f_1^-(u) = u^2  \period
\end{align}
The point to be noted here is that the part containing the unrestricted indices 
in the above relation 
is  $(\Omega^{(1)-}_{\ell-1})_{\rmb C} (\Omega^{(2)-}_{\ell-1})_{C\rmd}$, namely the monodromy matrices of length $\ell-1$, shorter  by one unit from the original  $\ell$. 

It should  now be clear that we can perform this reduction process repeatedly 
until all the indices become those of SL(2) only. Then, taking $\eta$ to be 
$-i$ and identifying the SL(2) Lax operator as 
\begin{align}
L_{{\rm SL}(2)}(u)_{\rma\rmb}  \equiv L^-(u)_{\rma\rmb} \comma 
\end{align}
upon acting on the  state $\ket{\psi_1} \otimes \ket{\psi_2}$   we obtain 
 the monodromy relation  for the genuine SL(2) monodromy matrices inserted as 
\begin{align}
\langle (\Omega_{{\rm SL}(2)}(u-i/2))_{\rma\rmb} |\psi_1 \rangle , (\Omega_{{\rm SL}(2)}(u+i/2))_{\rmb\rmc} |\psi_2 \rangle \rangle= u^{2\ell}\delta_{\rma\rmc}\langle |\psi_1 \rangle, |\psi_2 \rangle \rangle \period
\end{align}
This completes the direct derivation of the SL(2) monodromy relation from 
 that of ${\rm psu}(2,2|4)$ relations. 

\section{Discussions}
In this paper, we studied the tree-level three-point functions in the entire ${\rm psu}(2,2|4)$ sector of $\mathcal{N}=4$ super Yang-Mills theory from a group theoretic and integrability-based point of view. We in particular developed the manifestly conformally invariant construction of the singlet-projection operator and used it to express the Wick contraction. Unlike the preceding works \cite{ADGN,spinvertex}, our construction doesn't necessitate the ``$U$-operator'' which intertwines two schemes of representations of the superconformal algebra. This property greatly simplifies the analysis and allowed us to derive the monodromy relation for the harmonic R-matrix, as well as for the usual fundamental R-matrix.

The simplicity and the manifest conformal covariance of our construction will surely be of help when analyzing the weak-coupling three-point functions using integrability. So far, such analysis was performed thoroughly only for a particular class of three-point functions in ${\rm su}(2)$ \cite{Tailoring1,KKN2}, ${\rm sl}(2)$ \cite{non-compact}, ${\rm su}(3)$ \cite{su3} and ${\rm su}(1|1)$ \cite{Thiago} sectors. In the forthcoming paper \cite{KKN4}, we will use our formalism to study more general three-point functions in the ${\rm sl}(2)$ sector, which involve more than one non-BPS operators. It would be an interesting future problem to study other sectors, in particular higher-rank sectors based on our construction. 

It would also be interesting to study the loop correction in our formulation. For this purpose, a more detailed analysis of the harmonic R-matrix may be useful since the harmonic R-matrix is intimately related to the local conserved charges including the one-loop Hamiltonian. Another avenue of research is to explore the relation with the scattering amplitudes \cite{HarmonicR1,HarmonicR2,CK,CDK,BdR1,BdR2,spectralreg3,spectralreg4}. Also in that case, the harmonic R-matrix and the monodromy relation played an important role. It would be interesting if one could make a more direct connection.

Lastly, it would be important to understand the relations with the recently-proposed non-perturbative frameworks for the string vertex \cite{SFTvertex} and the three-point functions \cite{BKV}. Understanding such a non-perturbative framework from the perturbative-gauge-theory point of view will lead to deeper understanding of the AdS/CFT correspondence.
\par\bigskip\noindent
{\large\bf Acknowledgment}\par\smallskip\noindent
Y.K. would like to thank  the research center for mathematical physics at Rikkyo university for hospitality. 
 The research of  Y.K. is supported in part by the 
 Grant-in-Aid for Scientific Research (B) 
No.~20340048, while that of T.N. is supported in part 
 by JSPS Research Fellowship for Young Scientists, from the Japan 
 Ministry of Education, Culture, Sports,  Science and Technology. The research of S.K. is supported by the Perimeter Institute for Theoretical Physics. Research at Perimeter Institute is supported by the Government of Canada through Industry Canada and by the Province of Ontario through the Ministry of Economic Development and Innovation. 
\appendix
\newpage
\section{Commutation relations for u(2,2$|$4)}
\label{sec:algebra}
In this appendix, all the explicit forms  of the commutation relations for the superconformal generators are listed in the D-scheme basis. First, the algebra for the bosonic generators, namely, SO(2,4) and SU(4) generators are given by
\begin{align}
&[ M_{\alpha}^{\ \beta}, M_{\gamma}^{\ \delta}]=\delta_{\gamma}^{\ \beta} M_{\alpha}^{\ \delta}-\delta_{\alpha}^{\ \delta}M_{\gamma}^{\ \beta} \comma \ \ [\bar{M}^{\dot{\alpha}}_{\ \dot{\beta}}, \bar{M}^{\dot{\gamma}}_{\ \dot{\delta}}]=\delta^{\dot{\gamma}}_{\ \dot{\beta}} \bar{M}^{\dot{\alpha}}_{\ \dot{\delta}}-\delta^{\dot{\alpha}}_{\ \dot{\delta}} \bar{M}^{\dot{\gamma}}_{\ \dot{\beta}} \comma \\
&[ M_{\alpha}^{\ \beta}, P_{\gamma \dot{\delta}}]=\delta_{\gamma}^{\beta} P_{\alpha \dot{\delta}}-\frac{1}{2}\delta_{\alpha}^{\ \beta} P_{\gamma \dot{\delta}} \comma \ \ [\bar{M}^{\dot{\alpha}}_{\ \dot{\beta}}, P_{\gamma \dot{\delta}}]=-\delta^{\dot{\alpha}}_{\dot{\delta}}P_{\gamma \dot{\beta}}+\frac{1}{2}\delta^{\dot{\alpha}}_{\dot{\beta}}P_{\gamma \dot{\delta}} \comma \\
&[ M_{\alpha}^{\ \beta}, K^{\dot{\gamma} \delta}]=-\delta_{\alpha}^{\delta} K^{\dot{\gamma} \beta}+\frac{1}{2}\delta_{\alpha}^{\ \beta} K^{\dot{\gamma} \delta} \comma \ \ [ \bar{M}^{\dot{\alpha}}_{\ \dot{\beta}}, K^{\dot{\gamma} \delta}]=\delta_{\dot{\beta}}^{\dot{\gamma}} K^{\dot{\alpha} \delta}-\frac{1}{2}\delta^{\dot{\alpha}}_{\ \dot{\beta}} K^{\dot{\gamma} \delta} \comma \\
&[ D, P_{\alpha \dot{\beta}}]=iP_{\alpha \dot{\beta}} \comma \ [D, K^{\dot{\alpha} \beta}]=-iK^{\dot{\alpha} \beta} \comma \ [D,M_{\alpha}^{\ \beta}]=[D, \bar{M}^{\dot{\alpha}}_{\ \dot{\beta}}]=0 \comma \\
&[ P_{\alpha \dot{\beta}}, K^{\dot{\gamma} \delta}]=\delta_{\alpha}^{\delta}\bar{M}^{\dot{\gamma}}_{\ \dot{\beta}}-\delta^{\dot{\gamma}}_{\dot{\beta}} M_{\alpha}^{\ \delta}+i\delta_{\alpha}^{\delta}\delta^{\dot{\gamma}}_{\dot{\beta}} D \comma \\
&[R_a^{\ b} , R_c^{\ d}] = \delta_c^{\ b} R_a^{\ d}-\delta_a^{\ d} R_c^{\ b} \period
\end{align}
The commutators between the fermionic generators and the conformal generators $D, P, K$ are  
\begin{align}
[ D, Q_{\alpha}^a]=\frac{i}{2} Q_{\alpha}^a \comma \ &[ D, Q_{\dot{\alpha} a}]=\frac{i}{2}Q_{\dot{\alpha} a} \comma \ [D, S_a^{\alpha}]=-\frac{i}{2} S_a^{\alpha} \comma \ [D, \bar{S}^{\dot{\alpha} a}]=-\frac{i}{2} \bar{S}^{\dot{\alpha} a} \comma \\
&[ P_{\alpha \dot{\beta}},S_a^{\gamma}]=-\delta_{\alpha}^{\gamma} \bar{Q}_{\dot{\beta} a} \comma \ \  [ P_{\alpha \dot{\beta}},\bar{S}^{\dot{\gamma} a}]=-\delta_{\dot{\beta}}^{\dot{\gamma}} Q_{\alpha}^a \comma \\
& [ K^{\dot{\alpha} \beta}, Q_{\gamma}^a]=\delta^{\beta}_{\gamma} \bar{S}^{\dot{\alpha} a} \comma \ \ [ K^{\dot{\alpha} \beta}, \bar{Q}_{\dot{\gamma} a}]=\delta^{\dot{\alpha}}_{\dot{\gamma}} S^{\beta }_a \period
\end{align}
Under the action of the Lorentz generators and the R-symmetry generators, the fermionic generators transform as follows 
\begin{align}
&[  M_{\alpha}^{\ \beta}, Q_{\gamma}^a]= \delta^{\beta}_{\gamma} Q_{\alpha}^a-\frac{1}{2}\delta^{\ \beta}_{\alpha}Q_{\gamma}^a \comma \ \ [\bar{M}^{\dot{\alpha}}_{\ \dot{\beta}}, \bar{Q}_{\dot{\gamma}a}]= -\delta^{\dot{\alpha}}_{ \dot{\gamma}}\bar{Q}_{\dot{\beta}a}+\frac{1}{2} \delta^{\dot{\alpha}}_{\ \dot{\beta}}\bar{Q}_{\dot{\gamma}a} \comma \\
&[  M_{\alpha}^{\ \beta}, S^{\gamma a}]= -\delta_{\alpha}^{\gamma} S^{\beta a}+\frac{1}{2}\delta^{\ \beta}_{\alpha}S^{\gamma a} \comma  \ \ [\bar{M}^{\dot{\alpha}}_{\ \dot{\beta}}, \bar{S}^{\dot{\gamma}a}]= \delta_{\dot{\beta}}^{ \dot{\gamma}}\bar{S}^{\dot{\alpha}a}-\frac{1}{2} \delta^{\dot{\alpha}}_{\ \dot{\beta}}\bar{S}^{\dot{\gamma}a} \comma \\
& [ R_a^{\ b} ,Q_{\alpha}^c ]= -\delta_a^{\ c}Q_{\alpha}^b+\frac{1}{4}\delta_a^{\ b} Q_{\alpha}^a \comma \ \ [ R_a^{\ b}, \bar{Q}_{\dot{\alpha} c}]=\delta_c^{\ b} \bar{Q}_{\dot{\alpha} a}-\frac{1}{4}\delta_a^{\ b} \bar{Q}_{\dot{\alpha} c} \comma \\
& [ R_a^{\ b}, S_c^{\alpha} ]=\delta_c^{\ b} S_a^{\alpha} -\frac{1}{4}\delta_a^{\ b} S_c^{\alpha} \comma \ \   [ R_a^{\ b}, \bar{S}^{\dot{\alpha} c} ]=-\delta_a^{\ c} \bar{S}^{\dot{\alpha} b} +\frac{1}{4}\delta_a^{\ b} \bar{S}^{\dot{\alpha} c} \period
\end{align}
The anti-commutators for the fermionic generators are
\begin{align}
& \{ Q_{\alpha}^a , \bar{Q}_{\dot{\beta} b} \} =\delta^a_b P_{\alpha \dot{\beta}} \comma \ \ \{ \bar{S}^{\dot{\alpha} a}, S_b^{ \beta} \} =\delta^a_b K^{\dot{\alpha} \beta} \comma  \\
& \{ Q_{\alpha}^a , S_b^{\beta} \} = \delta^a_b M_{\alpha}^{\ \beta} -\frac{i}{2}\delta^a_b \delta_{\alpha}^{\ \beta} (D+iC)+\delta^{\beta}_{\alpha}R_b^{\ a} \comma \\
& \{ \bar{S}^{\dot{\alpha} a} , \bar{Q}_{\dot{\beta} b} \} = -\delta^a_b \bar{M}^{\dot{\alpha}}_{\ \dot{\beta}}-\frac{i}{2}\delta^a_b \delta^{\dot{\alpha}}_{\ \dot{\beta}} (D-iC)-\delta^{\dot{\alpha}}_{\dot{\beta}}R_b^{\ a} \period
\end{align}
Notice that the central charge $C$ appears in the anti-commutators of supercharges and superconformal charges. When we impose the condition of supertracelessness for the generators, we obtain su(2,2$|$4). If we further drop the 
 central charges on the RHS of the anti-commutators, we get  psu(2,2$|$4). 
Of course the central charge commutes with all the generators and 
the hypercharge essentially counts the fermion number $F(J)$ of the generator  $J$:
\begin{align}
[ C, J]=0 \comma \ \ [B, J]= \frac{1}{2} F(J) J \period
\end{align}
The only generators carrying  non-vanishing fermion numbers  are the supercharges and the superconformal charges. Their fermion numbers are   $F(\bar{Q}_{\dot{\alpha} a})=F(S_a^{\alpha})=1$ and $F(Q_{\alpha}^a)=F(\bar{S}^{\dot{\alpha} a})=-1$.
\section{Comment on the relation to the singlet state for the SU(2) sector of the previous paper}
The exponential form of the singlet projector for ${\rm u}(2,2|4)$ constructed in section 2.2 looks rather different from the simple non-exponential form  given in our previous work \cite{KKN2} for the SU(2) subsector. 
If we write it explicitly in terms of the scalar states in this subsector, it is given by\footnote{Here, since we are only concerned with the SU(2) sector where only the SU(4) oscillators are relevant, 
 we shall denote  $\overline{\Zbarket}$  by $\Zbarket$ for simplicity .}
\begin{align}
\ket{\bfone_{12}}_{\rm SO(4)} = \Zket \otimes \Zbarket - \ket{X} \otimes \ket{(-\Xbar)} 
 - \ket{(-\Xbar)} \otimes \ket{X} + \Zbarket \otimes \Zket  \period
\label{SU2singlet}
\end{align}
In this appendix, we briefly explain how this form is indeed obtained from the 
 exponential form. 

The SU(2) sector is only a  part of the large Hilbert space in which the 
 exponential state belongs. Also, in our previous paper, we were only considering the spin $1/2$ representation for $\SU2_L$ and $\SU2_R$. Thus to 
 get our singlet formula (\ref{SU2singlet}) from the exponential form, we must project out such a sector from the full exponential projector. 

It turns out that to do this appropriately,  we must  first write out the exponential state  for the full ${\rm SU(4)}\simeq {\rm SO(6)}$ sector which are  generated by the fermionic oscillators only. This is given by 
\begin{align}
\ket{\bfone_{12}}_{{\rm SO(6)}} &= e^A \Zket \otimes \Zbarket \\
A &= \cbar_1 \otimes c^1 +  \cbar_2 \otimes c^2  -\dbar_1 \otimes d^1 
-\dbar_2 \otimes d^2 
\end{align}
Since, for each Hilbert space component, $A$ consists of four different fermionic oscillators, the expansion of $e^A$ stops at order $A^4$. The terms coming from the odd powers of $A$, \ie $A$ and $A^3$, are fermionic.  When we take inner product with 
scalar states, they do not contribute.  Thus we only need to look at terms of order $1, A^2$ and  $A^4$.  

(i)\ At order 1, we simply get $\Zket \otimes \Zbarket$. (ii)\ The next 
simplest contribution comes  from $A^4$. Writing this out explicitly, 
we get
\begin{align}
 {1\over 4!} A^4 \Zket \otimes \Zbarket = (\cbar_1\dbar_1) (\cbar_2\dbar_2)\Zket \otimes (d^1c^1)(d^2c^2)\Zbarket \period 
\label{A4term}
\end{align}
To see the meaning of this expression clearly, it is instructive to write down the generators of  $\SU2_L \times \SU2_R$ in terms of these fermionic oscillators. They are given by 
\begin{align}
J_3^L &= \half (d^1 \dbar_1 -\cbar_1 c^1) 
\comma \qquad J_+^L = d_1c_1 \comma \qquad J_-^L = \cbar_1 \dbar_1\comma  \\
J_3^R &= \half (d^2 \dbar_2 -\cbar_2 c^2)\comma \qquad 
J_+^R= d_2c_2 \comma \qquad J_-^R = \cbar_2 \dbar_2 
\period
\end{align}
From this we see that the RHS of (\ref{A4term}) can be written as 
$ J_-^L J_-^R \Zket \otimes J_+^L J_+^R \Zbarket$. The action of these  lowering and raising operators turn $\Zket$ into $\Zbarket$ and $\Zbarket$
 into $\Zket$, so that we get the simple result 
\begin{align}
 {1\over 4!} A^4 \Zket \otimes \Zbarket =\Zbarket \otimes \Zket \period
\end{align}
(iii)\ Finally consider the $A^2$ terms. This produces 6 terms of various 
 structures. To see which terms are relevant to the SO(4) sector, it is 
useful to look at  the  $\SU2_L\times \SU2_R$ quantum numbers of the oscillators:
\begin{align}
c^1  &= (\half,0) \comma \quad \cbar_1= ( -\half,0)\comma \quad 
c^2 = ( 0,\half) \comma \quad \cbar_2 =(0,-\half)  \\
d^1  &= (\half,0) \comma \quad \dbar_1= ( -\half,0)\comma \quad 
d^2 = (0,\half) \comma \quad \dbar_2 =(0,-\half)
\end{align}
Then, we can classify the 6 terms produced at order  $A^2$ by their  quantum numbers as follows:
\begin{align}
\cbar_1 \cbar_2 \Zket \otimes c^1c^2 \Zbarket &:\quad (0,0)\otimes (0,0) \\
-\cbar_1 \dbar_1 \Zket \otimes c^1 d^1\Zbarket &:\quad -(-\half, \half)
\otimes (\half, -\half) \simeq -\ket{-\Xbar} \otimes \ket{X} \\
-\cbar_1 \dbar_2 \Zket \otimes c^1d^2 \Zbarket&:\quad -(0, 0) \otimes (0,0) \\
-\cbar_2\dbar_1 \Zket \otimes c^2d^1 \Zbarket&:\quad -(0,0) \otimes (0,0) \\
-\cbar_2\dbar_2 \Zket \otimes c^2d^2\Zbarket&:\quad -(\half ,-\half) \otimes (-\half, \half) \simeq -\ket{X} \otimes \ket{-\Xbar} 
\\
\dbar_1\dbar_2 \Zket \otimes d^1d^2 \Zbarket&:\quad 
(0,0) \otimes (0,0) 
\end{align}
The four terms with  the quantum numbers  $(0,0)\otimes (0,0)$ are orthogonal to the SO(4) scalar states of our interest and hence can be ignored in the 
singlet projector for the SU(2) sector. Thus, collecting the relevant states, 
 we find 
\begin{align}
\ket{\bfone_{12}}_{{\rm SO}(6)} &= e^A  \Zket \otimes \Zbarket \nn\\
&\ni \ket{\bfone_{12}} _{{\rm SO}(4)}  =\Zket \otimes \Zbarket - \ket{X} \otimes \ket{(-\Xbar)}  - \ket{(-\Xbar)} \otimes \ket{X} + \Zbarket \otimes \Zket, 
\end{align}
which is precisely the singlet state (\ref{SU2singlet}) constructed in our 
 previous work. 
\section{Some  details for the derivation of the crossing relation 
for the  harmonic R-matrix}
\label{sec:detail_R}
In this appendix, we provide some details of the derivation of the 
intermediate formulas which are needed for the proof of  the crossing relation 
 for the harmonic R-matrix. 
\nxt
\underline{Proof of the formula (\ref{eq:Hop_N})}
\nxt
To prove the formula (\ref{eq:Hop_N}) for the crossing of the number 
 operators for the quantum and the auxiliary spaces, we should first 
 recall the crossing property of  the oscillators given in (\ref{eq:oscillator_al}), (\ref{eq:oscillator_be}):
\begin{align}
&\langle \boldsymbol{1}_{12}| \bar{\alpha}_{(1)}=\langle \boldsymbol{1}_{12}| \bar{\alpha}_{(2)} \comma \ \ \langle \boldsymbol{1}_{12}| \alpha^{(1)}=-\langle \boldsymbol{1}_{12}| \alpha^{(2)} \comma \\
&\langle \boldsymbol{1}_{12}| \bar{\beta}_{(1)}=-\langle \boldsymbol{1}_{12}| \bar{\beta}_{(2)} \comma \ \ \langle \boldsymbol{1}_{12}| \beta^{(1)}=\langle \boldsymbol{1}_{12}| \beta^{(2)} \period 
\end{align}
Here, the subscripts (1),(2) label the two different spin chains corresponding to two operators. For simplicity  we have suppressed  the indices for the gl(2$|$2)$\oplus$gl(2$|$2) and the labels for the different sites in the spin chain.  
Now from these relations, we immediately see that, under crossing,  the number operator for the quantum space $\boldsymbol{\mathrm{N}}^{(1)}$ transforms as  $\boldsymbol{\mathrm{N}}^{(1)} \rightarrow - \boldsymbol{\mathrm{N}}^{(2)}$ \footnote{To be precise,  $\boldsymbol{\mathrm{N}}^{(1)}_{\alpha}$ transforms as  $\boldsymbol{\mathrm{N}}^{(1)}_{\alpha}\rightarrow -\boldsymbol{\mathrm{N}}^{(2)}_{\alpha}-(-1)^{|\mathsf{A}|}\delta^{\mathsf{A}}_{\ \mathsf{A}}$.  However, the constant term vanishes as the signs  are opposite for the bosonic and fermionic oscillators and hence they exactly cancel with each other in the present case. This is also true for $\boldsymbol{\mathrm{N}}^{(1)}_{\beta}$.},  while the number operator for the auxiliary space  $\boldsymbol{\mathrm{N}}^{(a)}$ does not change. 
\nxt
\underline{Proof of the formula (\ref{eq:cross_Hop})}
\nxt
This formula can be understood in the following way. For simplicity, we concentrate on the oscillators $\bar{\alpha}, \alpha$. We first transform the creation operators $ \bar{\alpha}^{\mathsf{A}}_1$ by crossing and  get
\begin{align}
\langle \boldsymbol{1}_{12}|\frac{1}{k!l!m!n!} \bar{\alpha}^{\mathsf{A}_1}_2\cdots \bar{\alpha}^{\mathsf{A}_k}_2\cdots \bar{\alpha}^{\mathsf{B}_1}_a\cdots \bar{\alpha}^{\mathsf{B}_m}_a \cdots \alpha_{\mathsf{B}_m}^1\cdots \alpha_{\mathsf{B}_1}^1\cdots \alpha_{\mathsf{A}_k}^a\cdots \alpha_{\mathsf{A}_1}^a \period
\end{align} 
Then, we wish to move the annihilation operators $ \alpha_{\mathsf{A}_1}^1$ next to the singlet projector in order to use the crossing formula for  them. This can be easily done, but  after the crossing, we need to move them back to the original position, which in turn generates extra terms since $ \alpha_{\mathsf{A}}^2$'s do not commute with the creation operators $ \bar{\alpha}^{\mathsf{A}}_2$.  As a result, the Kronecker delta $\delta^{\mathsf{A}}_{\mathsf{B}}$ appears, which contracts the indices for the oscillators of the auxiliary space $\bar{\alpha}^{\mathsf{B}}_a, \alpha_{\mathsf{A}}^a$. 
In this way  we find that the number operator of the auxiliary space $\boldsymbol{\mathrm{N}}^{(a)}_{\alpha}$ is inserted in the middle of the oscillators. Now when we move the number operator to the left most position, the number operator is shifted by a constant due to the presence of the creation operators
on the way.  This gives the expression  of the form
\begin{align}
\begin{split}
&(\boldsymbol{\mathrm{N}}^{(a)}_{\alpha}-m+1)\cdots (\boldsymbol{\mathrm{N}}^{(a)}_{\alpha}-m+p) \\
&\times \bar{\alpha}^{\mathsf{A}_1}_2\cdots \bar{\alpha}^{\mathsf{A}_{k-p}}_2\cdots \bar{\alpha}^{\mathsf{B}_1}_a\cdots \bar{\alpha}^{\mathsf{B}_{m-p}}_a \cdots \alpha_{\mathsf{B}_{m-p}}^2\cdots \alpha_{\mathsf{B}_1}^2\cdots \alpha_{\mathsf{A}_{k-p}}^a\cdots \alpha_{\mathsf{A}_1}^a \period
\end{split}
\end{align}
By carefully treating the numerical coefficients and performing  the same calculation for the oscillators $\bar{\beta}, \beta$, we find  that the crossing formula for the hopping operator is given by (\ref{eq:cross_Hop}).
\nxt
\underline{Explanation of (\ref{eq:cross_harmonic})}
\nxt
Let us make  cautionary remarks for using the crossing relations 
 for the coefficients and the hopping operator already obtained 
to derive the crossing relation for the harmonic R-matrix given in 
(\ref{eq:cross_harmonic}).  This has to do with the effects due to  the order of 
 crossing. 
 Although the hopping operators  preserve the total number of oscillators of the quantum space and as well as of the auxiliary space and commute with 
the number operator $\boldsymbol{\mathrm{N}}$, the expression 
$\boldsymbol{\mathrm{Hop}}_{k,l,m,n}^{(a2)}$  which appears {\it after} the crossing no longer commutes  with the total number operator of the form $\boldsymbol{\mathrm{N}}^{(a)}+\boldsymbol{\mathrm{N}}^{(1)}$. 
In fact, since the hopping operator $\boldsymbol{\mathrm{Hop}}_{k,l,m,n}^{(a2)}$ moves $k+l$ oscillators from the auxiliary space to the quantum space and moves $m+n$ oscillators from the quantum space to the auxiliary space, the following exchange relation holds:
\begin{align}
\boldsymbol{\mathrm{Hop}}_{k,l,m,n}^{(a2)}f(\boldsymbol{\mathrm{N}}^{(a)}+\boldsymbol{\mathrm{N}}^{(1)})=f(\boldsymbol{\mathrm{N}}^{(a)}+\boldsymbol{\mathrm{N}}^{(1)}+k+l-m-n)\boldsymbol{\mathrm{Hop}}_{k,l,m,n}^{(a2)} \period \label{eq:shift_N}
\end{align}
This effect has to be duly taken into account. More specifically, we first move 
 the hopping operator 
 $\boldsymbol{\mathrm{Hop}}_{k,l,m,n}^{(a1)}$ to the left all the way 
 until it hits the singlet projector. This operation does not shift the number operator as the labels for the quantum space are different and they commute with each other.  Now, upon hitting the singlet state we use the crossing relation
 to convert it to $\boldsymbol{\mathrm{Hop}}_{k,l,m,n}^{(a2)}$ and 
 then  we try to  move it back to the original position. In this process  we 
come across the shift for the number operator as in (\ref{eq:shift_N})\footnote{Actually, we need to exchange $\boldsymbol{\mathrm{Hop}}_{k-p,l-q,m-p,n-q}^{(a2)}$ through the coefficient $\mathcal{A}_{I}^{(\boldsymbol{\mathrm{N}})}$. But this produces the same shift as in (\ref{eq:shift_N}).}. 
After this procedure we make the crossing of the coefficients as  $\mathcal{A}_{I}^{(\boldsymbol{\mathrm{N}})} \rightarrow \mathcal{A}_{I}^{(\boldsymbol{\mathrm{N}}^{(a)}-\boldsymbol{\mathrm{N}}^{(2)})}$. 
In this way, we obtain the relation (\ref{eq:cross_harmonic}). 
\nxt
\underline{Proof of the relation (\ref{eq:sum_A})} 
\nxt
Finally, we shall provide a proof of the relation  (\ref{eq:sum_A}). Let us recall that the definition for the coefficient $ \mathcal{A}^{(\boldsymbol{\mathrm{N}})}_{I}$ is given in terms of the generalized binomial (\ref{eq:Hop_coefficient}). Hence, we have
\begin{align}
\tilde{\mathcal{A}}_{I}^{(\boldsymbol{\mathrm{N}})}=\rho(u)(-1)^{I+\frac{\boldsymbol{\mathrm{N}}}{2}+\boldsymbol{\mathrm{M}}} \sum_{r=0}^{\infty} (-1)^r \mathcal{B}(\boldsymbol{\mathrm{M}},r)  \mathcal{B}(I+r,r-\boldsymbol{\mathrm{M}}-u+\frac{\boldsymbol{\mathrm{N}}}{2})  \comma
\end{align}
where we have used the identity $\mathcal{B}(x,y)=\mathcal{B}(x,x-y)$.  From the above expression, it turns out that the proof for the relation (\ref{eq:sum_A}) is equivalent to verify the relation
\begin{align}
\sum_{r=0}^{\infty} (-1)^r \mathcal{B}(\boldsymbol{\mathrm{M}},r) \mathcal{B}(I+r,r-\boldsymbol{\mathrm{M}}+-u+\frac{\boldsymbol{\mathrm{N}}}{2}) =(-1)^{\boldsymbol{\mathrm{M}}}\mathcal{B}(I,-u+\frac{\boldsymbol{\mathrm{N}}}{2}) \period
\end{align}
For this purpose, we will consider the following more general formula 
\begin{align}
\sum_{r=0}^{\infty} (-1)^r \mathcal{B}(\gamma,r) \mathcal{B}(\alpha+r,r-\gamma +\beta) = \frac{\sin \pi(\beta- \gamma)}{\sin \pi \beta} \mathcal{B}(\alpha,\beta) \comma \label{eq:binomial_formula1}
\end{align}
where $\alpha , \gamma$ are arbitrary complex numbers and we assume $\beta $ to be generally  a non-integer complex number.  Once we can justify this relation, we easily obtain the relation we need by setting $\alpha =I$, $\beta =-u+\frac{\boldsymbol{\mathrm{N}}}{2}$ and $\gamma=\boldsymbol{\mathrm{M}}$. Using the definition for the generalized binomial and the well-known identity for the gamma function $\Gamma (x)\Gamma (1-x)=\frac{1}{\sin \pi x}$, the left hand side becomes
\begin{align}
\begin{split}
&(\mathrm{L.H.S}) =\sum_{r=0}^{\infty} (-1)^r \frac{\Gamma (\gamma+1)}{\Gamma (r+1)\Gamma (\gamma-r+1)} \frac{\Gamma (\alpha+r+1)}{\Gamma (r-\gamma+\beta+1) \Gamma (\alpha+\gamma-\beta+1)} \\
&= \frac{\Gamma (\gamma+1)\Gamma (\alpha+1)}{\Gamma (\alpha+\gamma-\beta+1)} \sum_{r=0}^{\infty} (-1)^r \frac{\Gamma (\alpha+r+1)}{\Gamma (r+1) \Gamma (\alpha+1)} \frac{\Gamma (\gamma-r-\beta) \sin \pi(r-\gamma+\beta+1)}{\Gamma(\gamma-r+1)}  \\
&= \frac{\sin \pi(\beta- \gamma)}{\sin \pi \beta} \frac{\Gamma (\gamma+1)\Gamma (\alpha+1)}{\Gamma (\alpha+\gamma-\beta+1) \Gamma (\beta+1)} \sum_{r=0}^{\infty} 
 \mathcal{B}(\alpha +r,r) \mathcal{B}(\gamma -r-\beta-1,\gamma -r) \period
\end{split}
\end{align}
The summation over the products of binomials turns out to be equal to $ \mathcal{B}(\alpha -\beta +\gamma,\gamma)$ since the following identity holds for arbitrary complex numbers $a ,b, c$
\begin{align}
\mathcal{B}( a + b +c-1, c) = \sum_{k=0}^{\infty} \mathcal{B}( a +k-1, k)\mathcal{B}( b +c-k-1, c-k)  \period \label{eq:binomial_formula2}
\end{align}
When $c$  is any positive integer, the above relation immediately follows from 
\begin{align}
\frac{1}{(1+x)^a}=\sum_{k=0}^{\infty} (-1)^k\mathcal{B}( a +k-1, k)x^k \comma \ \ \frac{1}{(1+x)^{a+b}}= \frac{1}{(1+x)^a}\cdot  \frac{1}{(1+x)^b} \period
\end{align}
As the both sides of (\ref{eq:binomial_formula2}) are analytic functions of  $c$, it turns out that the relation holds for arbitrary complex $c$ by analytic continuation. By using this identity with $a=\alpha+1, b=-\beta , c=\gamma$, we find
\begin{align}
(\mathrm{L.H.S})= \frac{\sin \pi(\beta- \gamma)}{\sin \pi \beta} \frac{\Gamma (\gamma+1)\Gamma (\alpha+1)}{\Gamma (\alpha+\gamma-\beta+1) \Gamma (\beta+1)} \mathcal{B}(\alpha -\beta +\gamma,\gamma) = \frac{\sin \pi(\beta- \gamma)}{\sin \pi \beta} \mathcal{B}(\alpha,\beta) \period
\end{align}
Therefore, we have shown (\ref{eq:binomial_formula1}), which completes the proof. 
\renewenvironment{thebibliography}{\pagebreak[3]\par\vspace{0.6em}
\begin{flushleft}{\large \bf References}\end{flushleft}
\vspace{-1.0em}
\renewcommand{\labelenumi}{[\arabic{enumi}]\ }
\begin{enumerate}\if@twocolumn\baselineskip=0.6em\itemsep -0.2em
\else\itemsep -0.2em\fi\labelsep 0.1em}{\end{enumerate} }


\end{document}